\documentclass[trackchanges,twocolumn]{aastex7}

\usepackage[utf8]{inputenc}
\usepackage{subcaption}
\usepackage{xcolor}
\usepackage{cancel}
\usepackage{amsmath}
\usepackage{comment}
\usepackage{multirow}
\usepackage{graphicx}
\newcommand{\Ms}{M$_{\odot}$}
\newcommand{\Mdot}{M$_{\odot}$~yr$^{-1}$}
\newcommand{\mdot}{$\dot{M}$}
\newcommand{\Ls}{L$_{\odot}$}

\newcommand{\mic}{$\mu$m}
\newcommand{\about}{$\sim$}
\newcommand{\bet}{$\beta$}

\shorttitle{Dust Formation in Red Supergiants}
\shortauthors{Sujit, Sarangi \& Sengupta}

\begin{document}

\title{On the Origin of a Dusty Circumstellar Medium around Red Supergiants}

\author[0009-0002-3321-4307]{Das Sujit}
\affiliation{Indian Institute of Astrophysics, 
100 Feet Rd, Koramangala, Bengaluru, Karnataka 560034, India}
\affiliation{Pondicherry University, R.V. Nagar, Kalapet, 605014, Puducherry, India}
\email[show]{sujit.das@iiap.res.in}

\author[0000-0002-9820-679X]{Arkaprabha Sarangi}
\affiliation{Indian Institute of Astrophysics, 
100 Feet Rd, Koramangala, Bengaluru, Karnataka 560034, India}
\affiliation{Pondicherry University, R.V. Nagar, Kalapet, 605014, Puducherry, India}
\email[show]{arkaprabha.sarangi@iiap.res.in}

\author[0000-0003-0191-4157]{Sutirtha Sengupta}
\affiliation{Indian Institute of Astrophysics, 
100 Feet Rd, Koramangala, Bengaluru, Karnataka 560034, India} 
\email{sutirtha.sg86@gmail.com}

\begin{abstract}

Red supergiant (RSG) stars are widely recognized as significant sources of dust, enriching the interstellar medium. However, the physical conditions that regulate dust nucleation in their winds remain poorly constrained. We investigate the formation of molecules and dust in RSG and explore how enhanced mass loss can produce a dense, dust-rich circumstellar medium (CSM) before core collapse. We couple time-dependent mass loss with non-equilibrium chemistry to model the formation of molecules and dust precursors using mass loss rates ranging from 10$^{-6}$ to 10$^{-2}$ \Mdot\ and both constant and accelerating wind profiles. Molecules such as CO, H$_2$O, SiO, HCN, CS, SO, NH$_3$, H$_2$, and O$_2$ form efficiently in the CSM, with masses varying between 10$^{-15}$--10$^{-2}$ \Ms. O-rich dust, namely silicates and alumina, dominates the dust composition. The total dust mass ranges between 10$^{-8}$ and 3$\times$10$^{-3}$ \Ms. Accelerated winds produce more dust and allow dust formation closer to the stellar surface. The resulting fluxes exhibit strong mid-infrared excesses. The 9.7 and 18 \mic\ silicate features appear in either emission or absorption depending on the optical depth of the circumstellar medium. The time-dependent mass-loss history of our SN~2023ixf progenitor models results in a gradual increase in CSM dust mass toward explosion. A clumpy CSM provides a substantially better match to the observed optical and infrared fluxes, demonstrating the importance of time-dependent mass loss, CSM structure, and wind acceleration in shaping the observable properties of red supergiant progenitors.

\end{abstract}


\section{Background} \label{sec:Intro}

Red supergiants (RSGs) represent a late evolutionary phase of massive stars with initial masses between 8--25~\Ms, characterized by extended, cool atmospheres with effective temperatures of $\sim 2500$--$4000~\mathrm{K}$ and large stellar radii, typically several hundred to $\sim 1500~R_{\odot}$ \citep{MauronJosselin2011,Levesque2005ApJ,Levesque2006ApJ,Smith2014,vanloon2005}. Observationally, many galactic and Magellanic Cloud RSGs, exhibit strong evidence of reddening and the presence of infrared (IR) excess in the spectrum. This clearly indicates to the presence of dust in the circumstellar medium (CSM). To significantly impact the spectral energy distribution (SED), the dust must be present close to the stellar surface, and is closely coupled to the evolutionary history of the RSG star. 

Pre-explosion progenitor detections of nearby hydrogen-rich SNe indicate that most Type~II core-collapse supernovae (CCSNe) arise from RSG progenitors \citep{Smartt2009, Smartt2015, Davies2017, DiazRodriguez2021, VanDyk2025}. A subclass of CCSNe are the interacting supernovae, which are characterized by the presence of a dense CSM around the progenitor \citep{fil97, smith_2017}. Some of these SNe, e.g. SN~1998S, SN~2005ip, SN~2010jl, SN~2006jc, SN~2011ht, SN~2020ywx, show a prolonged phase of interaction for several months to years post-explosion \citep{fra14, katsuda_2014, poz04,fas01, mattila_2008, fox11, baer-way_2025}. In the last couple of years, several nearby Type~II SNe with directly identified dusty infrared-bright RSG progenitors \citep{VanDyk2025}, including SN~2023ixf \citep{Kilpatrick2023, Soraisam2023, Szalai2023, Qin2024, ragosta_2026}, SN~2024ggi \citep{Xiang2024,Bostroem2026,JacobsonGalan2026}, and SN~2025pht \citep{Kilpatrick2025,VanDyk2026}. The mass of dust measured in these SN progenitors could vary between 10$^{-5}$ and 10$^{-2}$ \Ms. There are large uncertainties in the dust composition, between O-rich and C-rich dust types. 

The diversity in the mass loss history, characterized by the rate and the velocity, strongly depends on the initial mass and metallicity of the star \citep{vanLoon_2025}. Most red supergiants are observed to lose mass at moderate rates between 10$^{-9}$ to 10$^{-5}$~\Ms\ yr$^{-1}$ \citep{antoniadis_2024, wen_2024} and velocities between 10--100 km s$^{-1}$ \citep{dessart2024}. In very late stages of advanced nuclear burning, the mass loss rates may be enhanced to 10$^{-4}$~\Ms\ yr$^{-1}$ or larger \citep{YoonCantiello2010}. These phases of mass loss may be extremely short-lived (a few decades to years) or can even span over the last few centuries of the RSG's life \citep{Sengupta_Sujit_Sarangi2026}. The density structure of the CSM around RSGs reflects the mass loss histories, which are only measured after the star explodes as CCSNe \citep{Nayana2025ApJ...985...51N, chandra_2018, baer-way_2025}.

Several RSGs are known to form dust in their winds, as evident from the presence of dusty circumstellar matter around stars such as VY CMa, Antares, NML Cyg, S Per, \citep{walmswell_2012, gordon_2018, clayton_2009, kaminski_2019, cannon_2021, niu_2023,gaston_2025} in our Galaxy and, [W60] B90 \citep{Munoz-Sanchez2024} and WOH-G64 \citep{Ohnaka2024,VanLoonOhnaka2025} in the LMC. Dust is also reported as the cause of the `Great Dimming' phase in the nearby RSG star Betelgeuse \citep{montar_2021, cannon_2023}. Dust formation in stellar winds depends on the mass-loss rates, wind velocities, abundances, and the ambient radiation field \citep{verhoelst_2009, cherchneff2013b, nozawa_2014,beasor_2022}. Collisions between strong winds around massive stars create radiative shocks, which are proposed to aid dust formation \citep{lau_2022}. \cite{nozawa_2014} predict the formation of C-rich dust in the wind of a zero-metallicity RSG with an initial mass of 500 \Ms\, assuming a C-rich wind. At present, dust formation in RSG stars is being monitored by the JWST survey program GO-5441 alongside individual stars in program GO-3777, 3589.

While continuum emission from dust can sometimes be difficult to quantify in the presence of a cool stellar background, molecules often serve as valuable tracers of the chemistry in such stellar winds \citep{cherchneff2013b,Gottlieb2021}. In addition, molecules are important indicators of the gas temperature and cooling conditions, thereby pointing to environments that are conducive to dust formation. Mid-IR and submillimeter (submm) observations of Galactic supergiants, including VY CMa, IRC+10420, NML Cyg, VX Sgr, Betelgeuse, and a few others, reveal the presence of molecules such as CO, H$_2$O, SiO, HCN, CS, AlO, SO$_2$, SiS, and NH$_3$ in their stellar winds \citep{Smith_2009, kaminski_2013a, kaminski_2013b, cherchneff2013b, matsuura_2014,Shinnaga2025, ziurys_2025}. The presence of specific molecular species also provides important clues to the chemical abundances of the stellar wind. For example, the detection of O- and H-bearing molecules such as SiO, SO$_2$, and H$_2$O indicates an oxygen-rich gas, consistent with the expected composition of the outer envelopes of massive stars \citep{sukhbold_2016,Yang2023}. 
The submm spectrum of VY CMa, obtained by \texttt{Herschel Space Observatory}, is used to measure the $^{12}$C/$^{13}$C ratio in CO molecules, and the same for O isotopes. The reported $^{12}$C/$^{13}$C ratio of $5.6\pm1.8$ \citep{matsuura_2014} is anomalously low \citep{Praztzos2018}, suggestive of non-canonical internal mixing of CN-cycled material from the burning shell to the stellar envelope \citep{McCormick2023MNRAS}. Such isotopic ratios can thus be used as diagnostic tools for stellar evolutionary scenarios, based on their progenitor mass, rotation rates and internal mixing \citep{milam_2009,Maeder_2013}. However, the coexistence of dust and molecules poses a challenge for the analysis of molecular lines, given the large optical depth of dust even at mid-IR wavelengths. 

\begin{figure*}[t]
    \centering
    \includegraphics[width=0.99\textwidth]{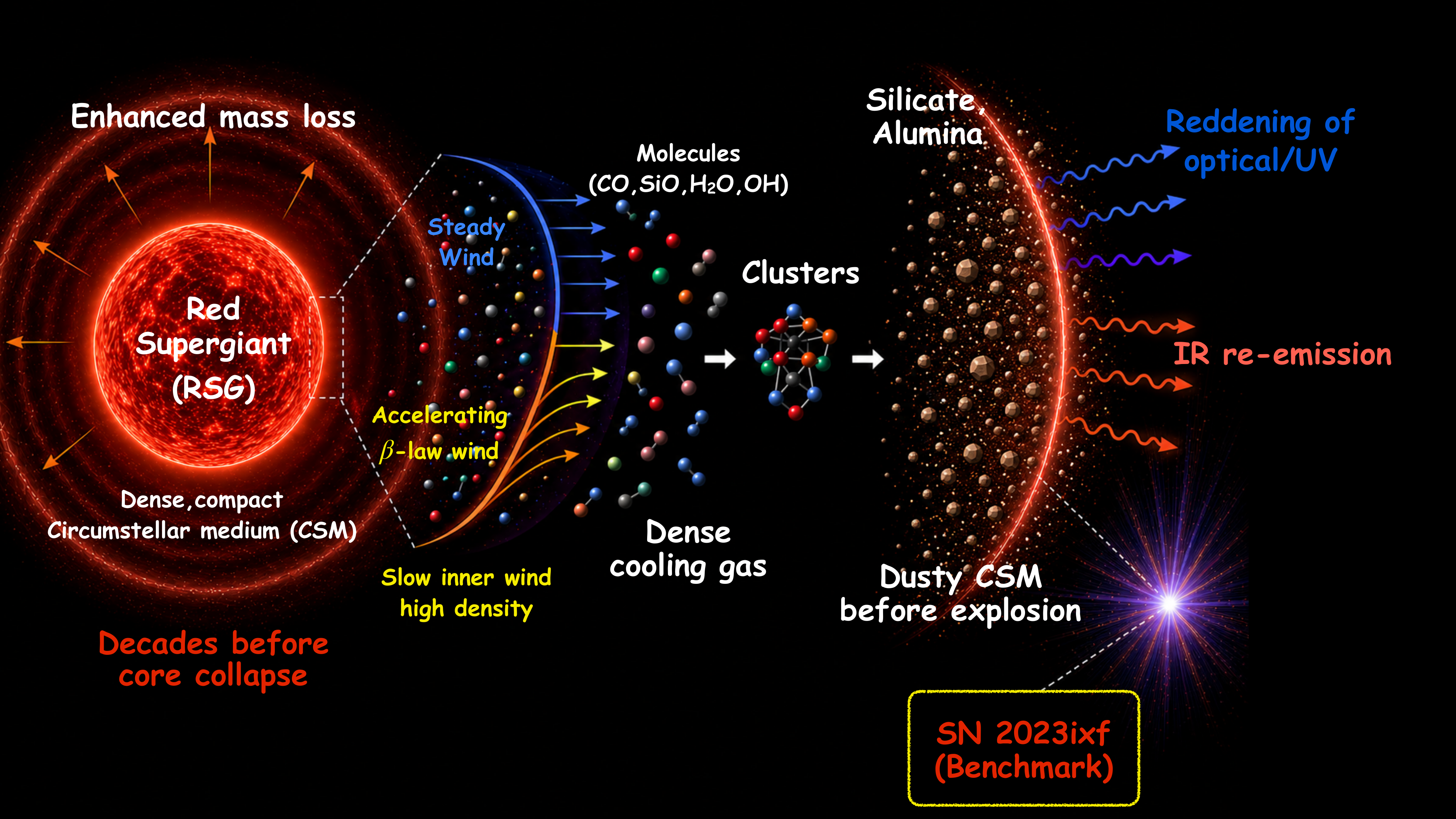}
    \caption{
    Schematic overview of the physical framework explored in this work. 
    We model (a) the nature of the RSG wind in terms of time-dependent mass loss rates and velocity profile, (b) the chemistry of molecule and dust synthesis in the winds leading to the formation of a dusty CSM, and (c) the spectral signatures of the system the star combining the stellar and the reprocessed radiation by dust. We used the observed UV/optical/IR fluxes of SN~2023ixf as a reference case study. }
    \label{fig:conceptual_overview}
\end{figure*}

\section{Rationale of this work} \label{sec:rationale}

The winds of RSG stars are known to form dust. However, the process of dust formation, as well as the quantity and composition of the dust produced, remains poorly understood. CCSN progenitors, particularly those of interacting SNe such as Type IIn, provide a valuable link between the CSM and the pre-explosion evolution of massive stars. The formation of a dense CSM in SN progenitors around solar-metallicity, non-rotating stars has been attributed to a pulsation-driven enhanced mass-loss phase during the last few centuries of the RSG's lifetime, likely occurring in stars with initial masses greater than 15~\Ms \citep{Sengupta_Sujit_Sarangi2026}. This late-phase evolution of these stars depends strongly on the initial mass, mass loss history leading up to the RSG phase and possibly also its rotational velocity \citep{Heger1997}, while dust formation in the CSM is closely tied to the mass-loss history over the final stages of evolution of the SN progenitor. Uncertainties in the mass-loss of the SN progenitor therefore directly propagate into uncertainties in the dust formation rates, dust composition, and the resulting spectral features.

Only a small fraction of the observed RSG population in nearby galaxies appears to be heavily dust-enshrouded \citep{beasor_2022, ghaziasgar_2025, sarbadhicary_2025, hassani_2026}. Furthermore, the occurrence of dusty RSGs does not always correlate with the modest mass-loss rates inferred for the majority of the RSG population \citep{antoniadis_2024, wen_2024}. 

This raises two important questions: (a) why are some RSGs heavily enshrouded by dust while others are not, and (b) how do the relatively modest mass-loss rates observed in RSGs lead to efficient dust formation in their winds. In this study, we develop dust formation models for stellar winds spanning a range of mass-loss rates from stellar evolution models \citep{Sengupta_Sujit_Sarangi2026} and the resulting CSM properties, and compute their corresponding spectral signatures. Figure~\ref{fig:conceptual_overview} presents a schematic overview of the geometry and the chronology of events, which we model in this paper to estimate the dust formation scenario for RSGs. 

The paper is arranged as follows. In Section~\ref{Sec_csm_origin} we discuss the mechanism for the formation of CSM through the stellar wind, defining the density and abundance of the gas. Section~\ref{sec_temperature} defines the gas temperature in the CSM for various mass loss rates. Thereafter, in Section~\ref{sec:chemistry} we describe the chemical processes that govern molecule and dust formation. Correspondingly, Section~\ref{sec:molecules} discusses the molecular yields for various wind models. In the following Section \ref{sec:dust_general}, we discuss the dust densities and masses formed in the RSG wind for various cases of mass loss and wind profiles. The particular case of SN~2023ixf progenitor is discussed in Section~\ref{sec:dust_SN2023ixf}. The model SEDs for all scenarios are derived in Section~\ref{sec_synthetic_spectra}, and the case of SN~2023ixf progenitor is compared to observed fluxes in Section~\ref{sec_sed_2023ixf}. We summarize the results in Section~\ref{sec_summary} and conclude with a detailed discussion in Section~\ref{sec_discussion}.

\section{Origin of the Circumstellar Matter}
\label{Sec_csm_origin}

To construct a consistent picture of dust formation in the winds of a RSG star, we need the radial profile of the CSM density, gas temperature, and abundances. As shown in Figure~\ref{fig:conceptual_overview}, these factors act as the necessary inputs for the chemistry that prevails in these stellar wind.

\subsection{Mass loss in Red-Supergiants} 

\begin{figure}[h]
    \centering
    \includegraphics[width=0.45\textwidth]{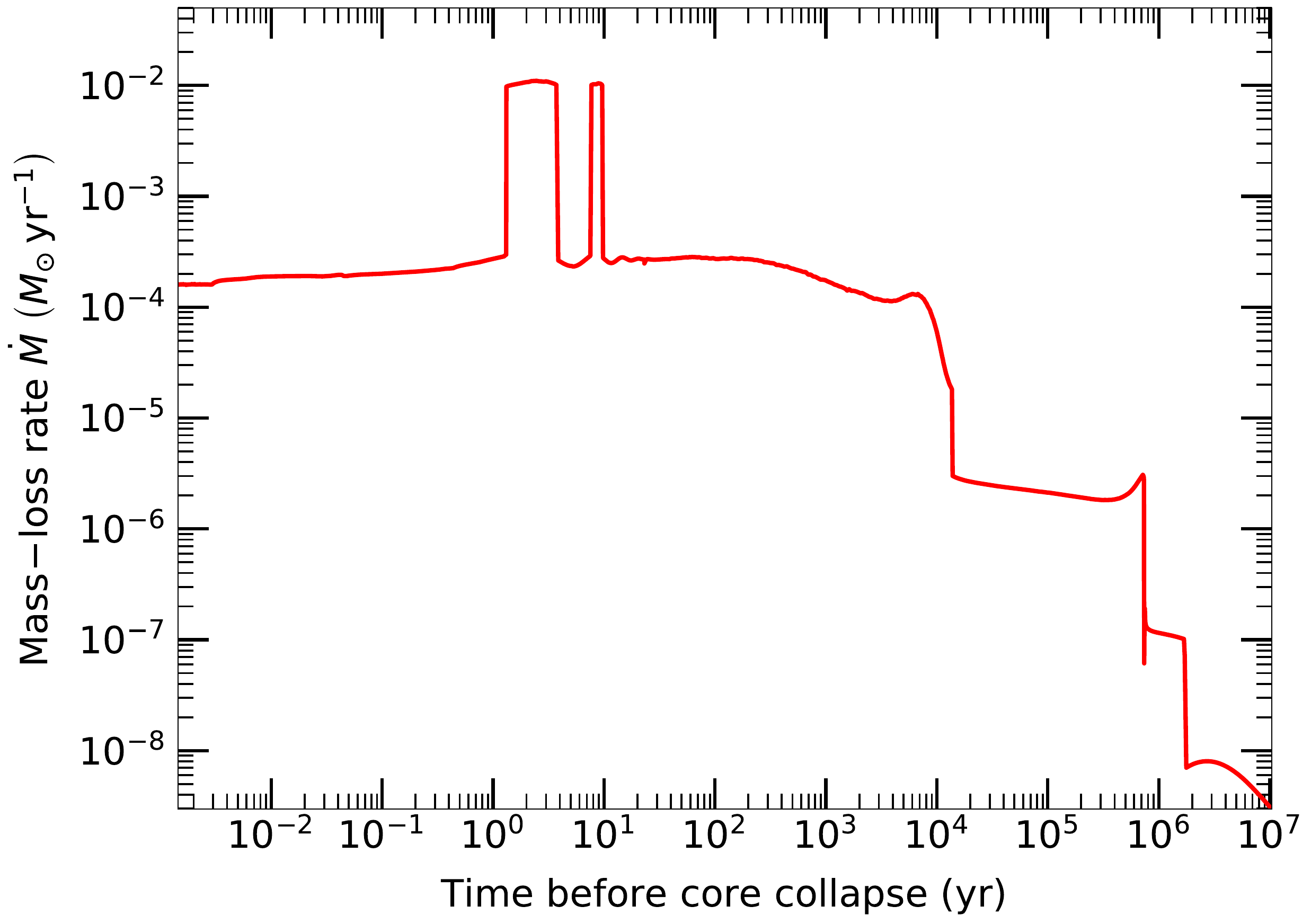}
    \caption{The time evolution of the mass loss rate, $\dot{M}$, of the RSG progenitor of SN~2023ixf
     as a function of time before CC. The mass-loss history is taken directly from our previous work \cite{Sengupta_Sujit_Sarangi2026} and spans the final $\sim10^{7}$~yr until explosion.}
    \label{fig:mass-loss_rate_vs_time}
\end{figure}

In our previous work \citep{Sengupta_Sujit_Sarangi2026}\footnote{For the inlists and implementation of mass-loss, refer to the Zenodo repository: S.
Sengupta 2025; doi: https://doi.org/10.5281/zenodo.17605468}, we developed an evolutionary sequence of models for non-rotating, solar-metallicity RSGs with pulsation-driven enhanced mass loss \citep{YoonCantiello2010,Clayton2017} to compare with the hydrogen column densities along the line of sight inferred observationally for SN~2005ip \citep{Fox2020}, SN~2023ixf \citep{Nayana2025ApJ...985...51N}, SN~2017hcc \citep{Chandra2022} and SN~2020ywx \citep{baer-way_2025}. Our models predict that stars with initial masses greater than 15~\Ms\ undergo enhanced mass loss  due to dynamical ejection of their outer layers once $\log((L/L_\odot)/(M/M_\odot)) \geq 4.15$ \citep{clayton2018a}, leading to mass loss rates in the range 10$^{-4}$--10$^{-2}$~\Mdot\ prior to CC. Figure~\ref{fig:mass-loss_rate_vs_time} shows the mass-loss history as a function of time before core collapse that best fits the circumstellar medium (CSM) and hydrogen column densities inferred for SN~2023ixf \citep{Nayana2025ApJ...985...51N}. The wind velocity is parameterized using an accelerating wind profile \citep{Moriya2018}, given by a $\beta$-law:
\begin{equation} \label{eq:betalaw}
v_w(r) = v_0 + (v_\infty-v_0)\left(1-\frac{R_*}{r}\right)^\beta,
\end{equation}
with an exponent $\beta = 1.2$ (refer Appendix B of \citealt{Sengupta_Sujit_Sarangi2026}), launch velocity $v_0=0.1$~km/s, terminal velocity $v_\infty=20$~km/s, and stellar radius $R_{*} =1.1 \times 10^{14}$ cm \citep{Sengupta_Sujit_Sarangi2026}. The velocity of the wind as a function of radial distance is presented in Figure \ref{fig:csm_density_with_velocity_profile} (left panel), and compared to a steady wind velocity set at $v_\infty$. 

We investigate dust formation across the range of mass loss rates, varying between 10$^{-6}$--10$^{-2}$ \Mdot, within the limits shown in Figure~\ref{fig:mass-loss_rate_vs_time}. For each assumed value of mass loss rate, we consider two scenarios: one in which the wind velocity remains at a constant value, $v_\infty$,  and another in which the wind accelerates according to Equation~\ref{eq:betalaw}. We then use the exact time-dependent mass loss history relevant for SN~2023ixf, shown in Figure~\ref{fig:mass-loss_rate_vs_time}, to derive the dust distribution in the CSM. Our model for SN~2023ixf \citep{Sengupta_Sujit_Sarangi2026} suggests a 18~\Ms~initial mass star, with luminosity of $\sim$ 10$^{5}$ \Ls\ and an effective temperature of $\sim$ $~3067$~K. For simplicity, we keep the stellar luminosity and temperature identical for all cases in this paper.

\subsection{CSM density}

\begin{figure*}[htbp]
    \centering

    \begin{subfigure}{0.46\textwidth}
        \centering
        \includegraphics[width=\linewidth]{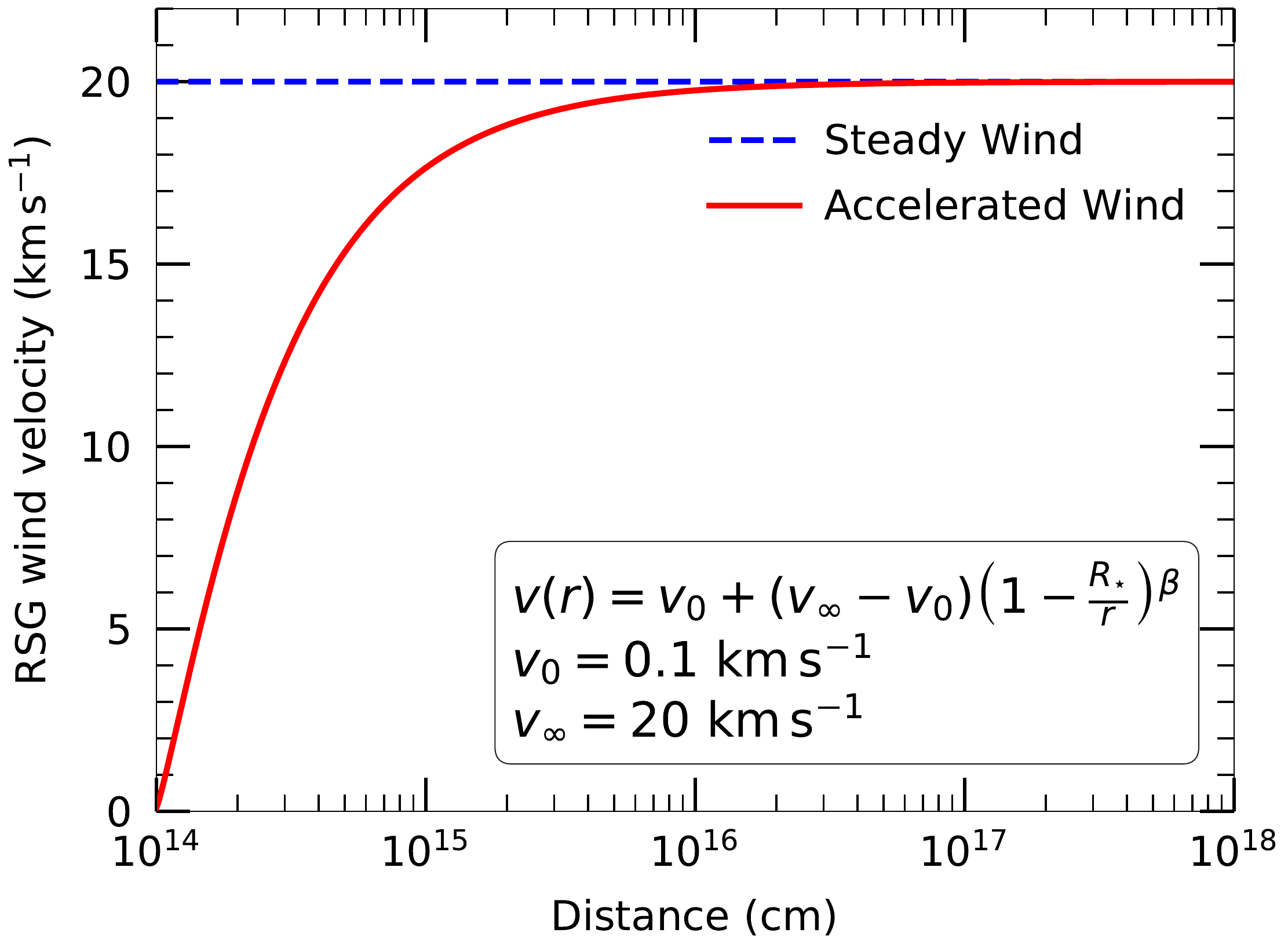}
        
    \end{subfigure}
    \hfill
    \begin{subfigure}{0.48\textwidth}
        \centering
        \includegraphics[width=\linewidth]{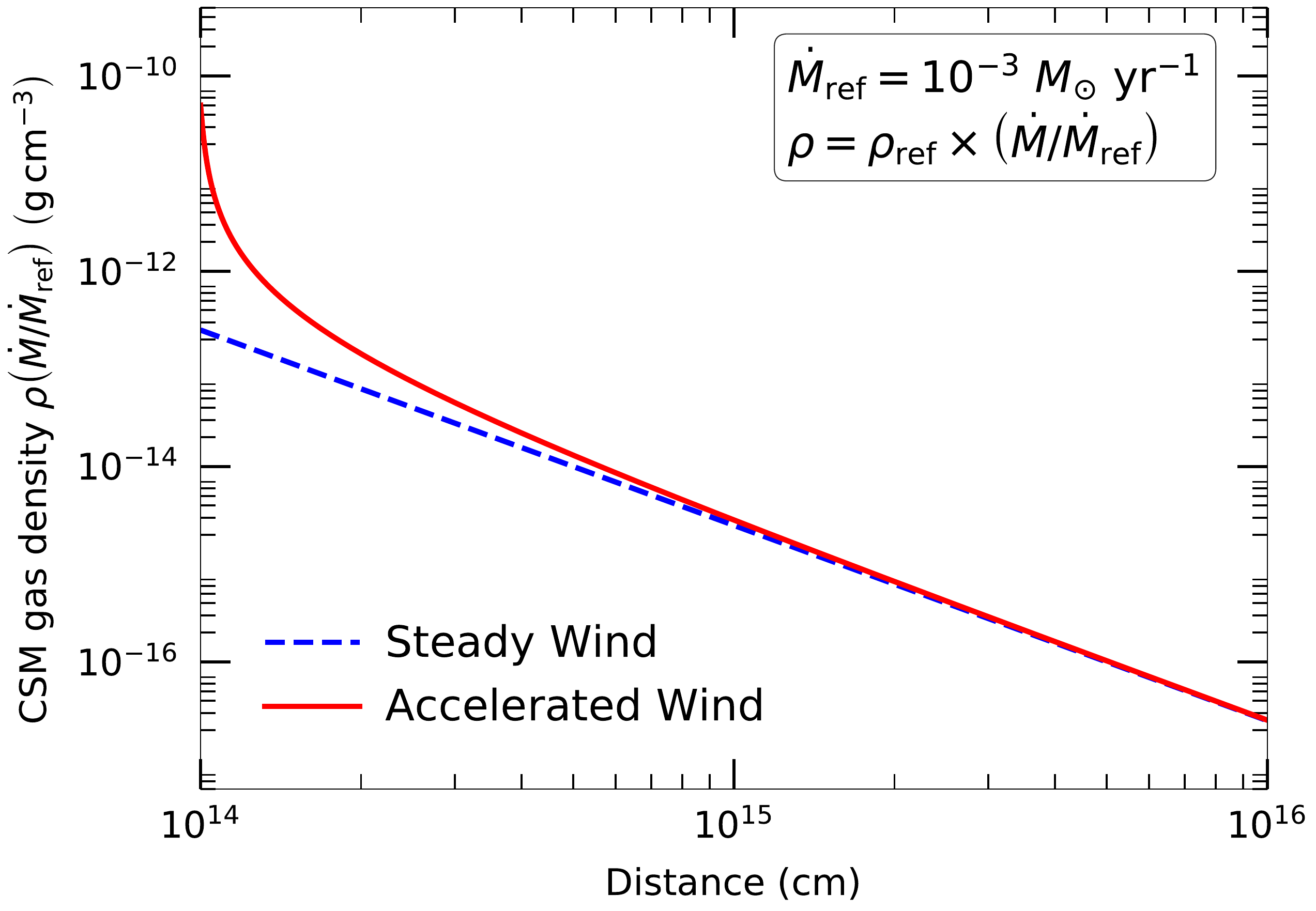}
        
    \end{subfigure}

    \caption{
    \textit{Left panel}: The radial velocity, $v(r)$, of a given gas parcel along the wind, as a function of distance from the stellar surface. The dashed blue curve represents a steady wind with constant velocity, $v_\infty =20$~km s$^{-1}$. The solid red curve shows the accelerated wind given by the $\beta$-law velocity profile :
    $v(r)=v_{0}+(v_{\infty}-v_{0})\left(1-\frac{R_{\star}}{r}\right)^{\beta}$, where $R_{\star}=10^{14}$~cm is the stellar radius and $v_0=0.1$~km s$^{-1}$ is the wind launch velocity at the stellar surface. 
    \textit{Right panel}: Typical CSM densities as a function of radial distance $r$, from the stellar surface, comparing the constant velocity wind and the accelerated $\beta$-law wind. The values on the y-axis correspond to an \mdot\ of 10$^{-3}$ \Mdot, and the densities are linearly related to \mdot.  
    }
    \label{fig:csm_density_with_velocity_profile}

\end{figure*}

For the RSG wind (as illustrated in the left panel of Figure~\ref{fig:csm_density_with_velocity_profile}), we calculate the CSM gas density at radius $r$, given by 
mass continuity, as
\begin{equation}
\label{eq_CSMdensity}
\rho(r) = \frac{\dot{M}(t_{\rm ej})}{4\pi r^{2} v_{\rm w}(r)},
\end{equation}
where $t_{\rm ej}$ denotes the time before CC at which the material currently located at
radius $r$ was ejected from the stellar surface. The constant-velocity case ($\beta = 0$) provides a useful limiting comparison, allowing us to isolate the effect of wind acceleration on the CSM density.

The right panel of Figure~\ref{fig:csm_density_with_velocity_profile} shows the resulting CSM density as a function of time before CC for a range of mass-loss rates. The nature of the wind velocity, $v_{\rm w}(r)$ (Figure~\ref{fig:csm_density_with_velocity_profile}, left panel) in Equation \ref{eq_CSMdensity}, is reflected in the density structure. Independent of the mass loss rate, the acceleration of the RSG wind close to its surface, parameterized in terms of the $\beta$-law, produces a pronounced increase in density of the inner CSM. Material ejected during the final decades to centuries prior to explosion is confined to radii of $\sim 10^{14}$--$10^{15}$~cm, reaching densities several orders of magnitude higher than those associated with earlier RSG mass-loss episodes. The accelerating wind further increases the densities within this region as the wind velocity gradually increases from the stellar surface starting from the near-zero launch speed, thereby increasing the time for which material resides in this dense region that is more conducive to dust formation. 

\subsection{Abundances in the CSM}

\begin{table}
\caption{Abundances of key elements in the stellar envelope of a 18 \Ms\ initial mass star at the pre-SN stage \citep{sukhbold_2016} that forms the CSM through mass loss. Their 18 \Ms\ run is chosen to best match the CSM properties of SN 2023ixf \citep{Sengupta_Sujit_Sarangi2026}. These abundances remain nearly unchanged for all initial mass stars between 12--20 \Ms.}
\label{tab:abundance_summary}
\centering
\begin{tabular}{l c}
\hline\hline
Species & Initial abundance \\
\hline
H        & $6.26\times10^{-1}$ \\
He       & $3.59\times10^{-1}$ \\
C        & $1.34\times10^{-3}$ \\
N        & $3.51\times10^{-3}$ \\
O        & $5.05\times10^{-3}$ \\
Mg       & $7.91\times10^{-4}$ \\
Si       & $8.29\times10^{-4}$ \\
S        & $4.23\times10^{-4}$ \\
\hline
\end{tabular}
\end{table}

The stellar wind originates from the outer H-envelope of the star. Assuming a non-rotating star, the abundances remain almost identical in the convective envelope, and therefore in the CSM. However, if a star loses the entire H-rich envelope through the wind, the abundances in the CSM will certainly alter as the inner layers of the star gets exposed. In this paper, we do not attempt to model such extreme levels of envelope stripping, as it is not expected in the case of Type II SNe progenitors of the kind studied here. 
In \cite{Sengupta_Sujit_Sarangi2026}, we predicted the mass loss and CSM structure from stellar evolutionary calculations made using the open-source stellar evolution code Modules for Experiments in Stellar Astrophysics, \texttt{MESA} \citep{Jermyn2023}. In our setup, a limited nuclear-reaction network \citep{1999ApJS..124..241T} was used until  the end of core He-burning  to reduce computational cost. Hence, we compare the elemental abundances in the outer layers of the pre-explosion models of \cite{sukhbold_2016}, which use larger reaction networks throughout their evolutionary calculations \citep{Rauscher_2002}. Table~\ref{tab:abundance_summary} lists the initial abundances of the refractory elements in the CSM considering a 18 \Ms\ initial mass star. We also verified that these abundances have a negligible dependence on the progenitor mass, in the 12--20 \Ms\ range, so these values can be safely used for all RSG models in the above mass range.


\section{Circumstellar Temperature profile}
\label{sec_temperature}


\begin{figure}[h]
    \centering
    \includegraphics[width=0.45\textwidth]{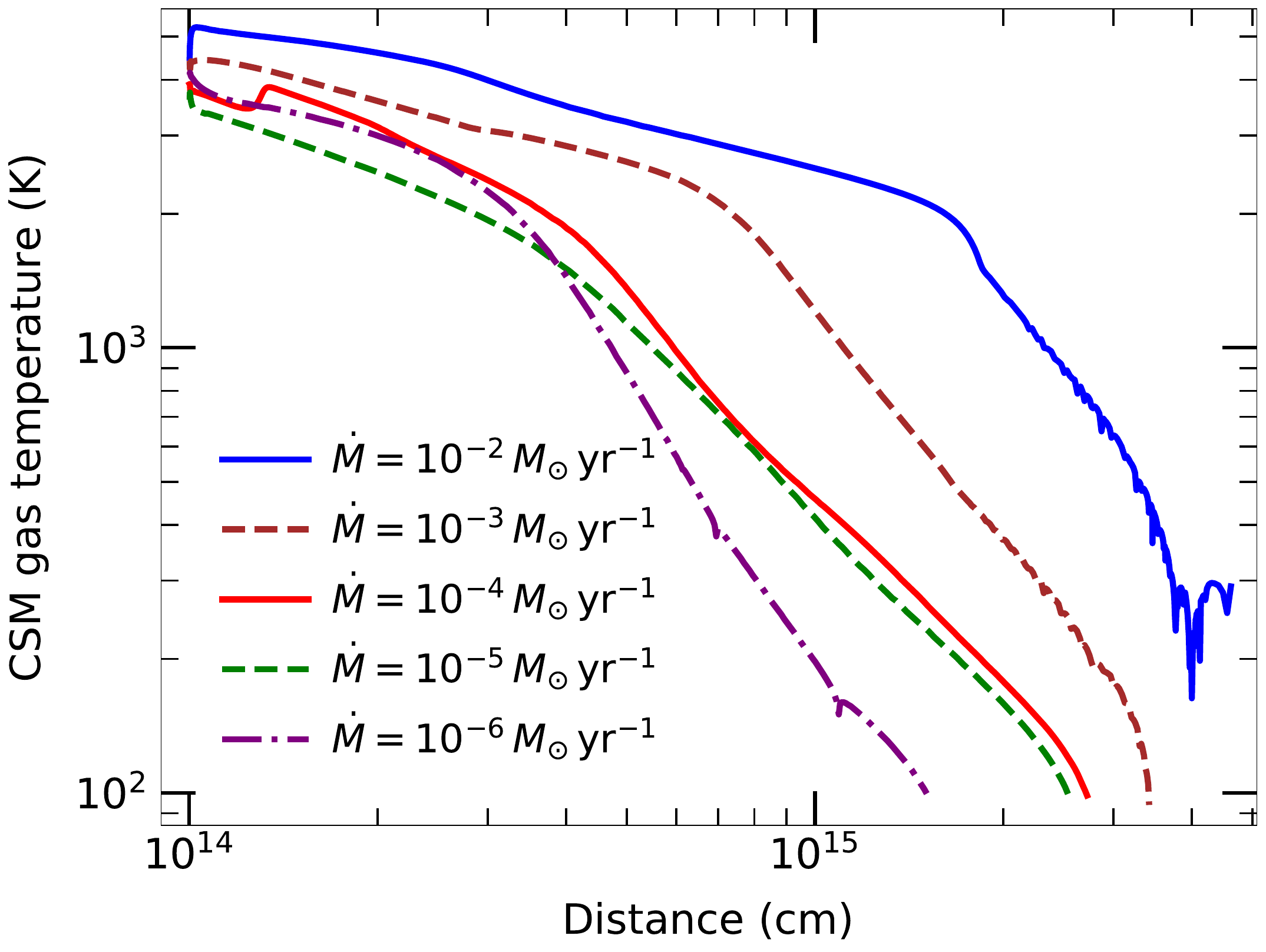}
    \caption{
    Radial gas temperature profiles of the CSM for \mdot\ varying between 10$^{-6}$--10$^{-2}$ \Mdot. Spectral synthesis code \texttt{CLOUDY} \citep{ferland_2013} is used to compute the density dependent temperature profiles. See Section \ref{sec_temperature} for details. }
    \label{fig:CSM_Temperature_profile}
\end{figure}

The thermal structure of the CSM plays a central role in
regulating dust formation and survival in RSG winds. 
Together with the density structure derived in  Section~\ref{Sec_csm_origin}, the gas temperature,
$T_{\rm gas}(r)$, determines molecular and dust chemistry, and the locations of dust condensation in the expanding wind.

We compute the circumstellar temperatures using the
spectral synthesis code \texttt{CLOUDY} \citep{Ferland2013}, adopting the density profiles derived from the
time-dependent mass-loss history and the stellar radiation field of the reference RSG model, detailed in Section~\ref{Sec_csm_origin}. 
The stellar luminosity and effective temperature define the incident radiation
field at the inner boundary, while radiative heating, line cooling, and adiabatic expansion collectively determine the radial temperature
profile throughout the wind.
The temperature structure is therefore intrinsically coupled to the
mass loss rate through the resulting density profile and its opacity.  

Figure~\ref{fig:CSM_Temperature_profile} shows the resulting gas temperature profiles as a
function of distance for mass loss rates spanning the range found in our reference RSG model sequence (as shown in Figure~\ref{fig:mass-loss_rate_vs_time}) leading upto CC i.e.
$\dot{M} \sim 10^{-6}$--$10^{-2}\,M_{\odot}\,\mathrm{yr^{-1}}$.

At all mass-loss rates, the gas temperature decreases monotonically with radius,
from several thousand Kelvin in the inner wind to a few hundred Kelvin at
distances of $\sim 10^{15}$--$10^{16}$~cm.
However, both the absolute temperatures and their radial gradient depend
sensitively on $\dot{M}$. 
Close to the star, higher mass loss rates produce denser, optically thick winds, leading to
enhanced trapping of energy, thereby resulting in higher temperatures at a given
radius. When we go further out, the density and optical depth drop, and then cooling expedites. 
Lower mass loss rates result in slightly lower temperatures close to the stellar surface mainly because of photons escaping more efficiently from the gas.

Given the non-linear behaviour of gas temperature and dependence on mass loss rates, the temperature cannot be simplified as a uniform power law as previously approximated by \cite{nozawa_2014, zubko_H2O_2004, ziurys_2009}. 

Important to note, the temperature profiles, shown in Figure \ref{fig:CSM_Temperature_profile}, are snapshot scenarios, which do not present the time evolution of individual parcels of gas. We adopt the temperature profile as a function of radius and mass loss rate, and map them for the evolution of each parcel of gas. That is used as a key input when addressing the molecular and dust chemistry described in the following Sections.

\section{Molecule and dust formation chemistry}
\label{sec:chemistry}

The formation of dust in stellar wind is governed by a large network of simultaneous gas-phase nucleation and grain condensation processes  \citep{cherchneff2013b}. The chemical evolution of a parcel of gas depends on its dynamic temperature, density, and radiation field, along with the abundances. The rapid evolution of physical conditions is best captured in a non-equilibrium setup accounting for all chemical processes simultaneously while using a small time-step for the simulation. 

Based on the abundances in the stratified ejecta of SNe, the relevant chemical network was developed by \cite{sar13, sarangi_2022b}. The molecule and dust formation in H-rich wind in AGB stars was modeled by \cite{gob16}, and was adapted for the CSM in Type IIn SNe by \cite{sarangi_2022a}. In this paper, we have modified and applied the non-equilibrium chemistry to RSG winds using the chemical model NECSA (non-equilibrium chemistry solver algorithm), previously used for supernovae of various types and AGN winds \citep{aman_2026a, sarangi_2025a, sarangi_2022a, sarangi_2019, sar15, sar13}. 

We follow the chemical evolution of each parcel of gas. When we assume a uniform mass loss rate, the evolution is identical, and at a snapshot just before the CC, the molecule and dust distribution is determined by how far in radius the wind has diffused. For variable mass loss, which reflects the reality of the last decades of an RSG, the distribution of molecules and dust in the CSM depends on the time of the snapshot and the time taken by each parcel to form dust after it was launched from the stellar surface. We have assumed mass loss to be prevalent for the last one million years leading to CC. 

The density, described by Equations \ref{eq:betalaw} and \ref{eq_CSMdensity} presents the motion of a gas parcel in the radial direction and the resulting CSM density. Based on the wind velocity profile, defined by the acceleration coefficient \bet, the density close to the stellar surface alters by a couple of orders of magnitude (see Figure \ref{fig:csm_density_with_velocity_profile}). This largely impacts the formation of molecules and dust, as explained in the following sections. The temperature profiles given in Figure \ref{fig:CSM_Temperature_profile} are also used as input to the chemistry code NECSA. 

The elemental abundances in the wind remain unchanged in our model throughout the evolutionary phase of an RSG. We consider about 600 gas-phase reactions simultaneously to account for the formation of molecules and molecular clusters. As Table \ref{tab:abundance_summary} suggests, the wind is O-rich in nature with a C/O ratio being less than 1. Such composition is expected to form predominantly O-rich diatomic molecules (CO, SiO, H$_2$O, AlO, SO, etc.) and dust grains such as Mg-silicate, alumina predominantly, along with other oxides in trace amounts \citep{cherchneff2013b, sarangi2018book}. The H atoms go on to form H$_2$ molecules along with H$_2$O, OH, NH$_3$, HCN, etc. 

The reaction network leads to the formation of Mg-silicates, of the chemical form [Mg$_2$SiO$_4$]$_n$, otherwise known as Forsterite. The silicate formation network, based on \cite{gou12, sar13, gob16}, proceeds through the oxidation of small clusters of Mg--Si--O molecules by O$_2$, H$_2$O, and SO. The dimer, [Mg$_2$SiO$_4$]$_2$ is considered the stable dust precursor, which grows to form silicate dust via coagulation. Alumina, of chemical form [Al$_2$O$_3$]$_n$, is formed through the series of AlO, O$_2$, and H$_2$O reactions. We consider the tetramer, [Al$_2$O$_3$]$_4$ as a dust precursor. 

The dust precursors are stable molecular clusters, which grow via coagulation at a very efficient rate to form dust grains of sizes between 0.01 to 1 \mic. As \cite{sar15} found, the efficiency of the dust precursors to condense to dust grains is close to 99\%. In this study, we have used the synthesis of dust precursors as a measure of dust formation, without deriving the grain-size distribution.  

We have separately discussed the two scenarios of wind acceleration, one with constant velocity \bet\ = 0 and one with an acceleration coefficient of \bet\ = 1.2. In both of these cases, the wind propagates identically beyond 100 stellar radii. The following sections show that the slower velocity, close to the stellar surface when we use \bet\ = 1.2, largely favours molecule and dust formation. 

\begin{figure*}
\centering

\begin{subfigure}{0.48\textwidth}
    \centering
    \includegraphics[width=\linewidth]
    {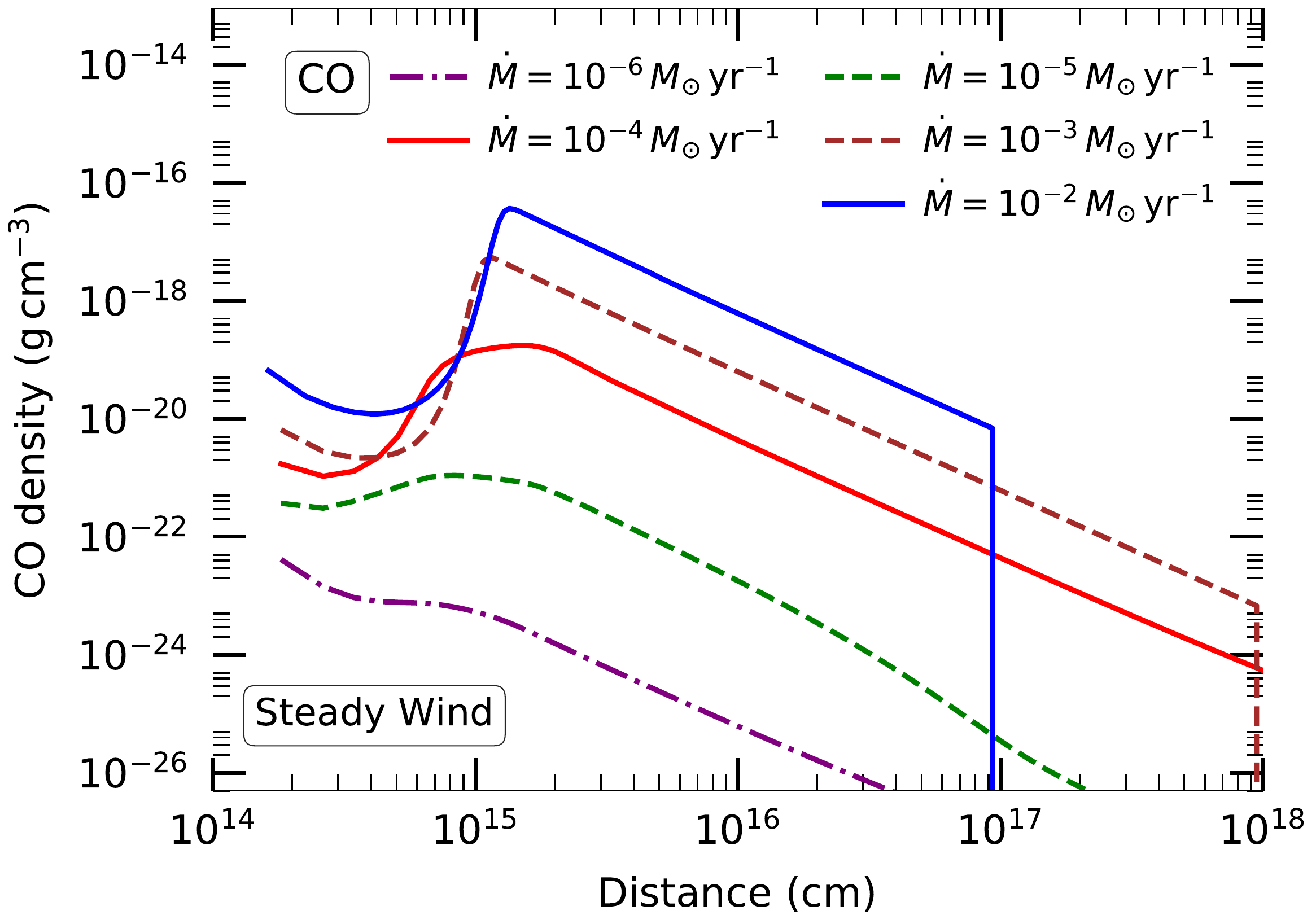}
\end{subfigure}
\hfill
\begin{subfigure}{0.48\textwidth}
    \centering
    \includegraphics[width=\linewidth]
    {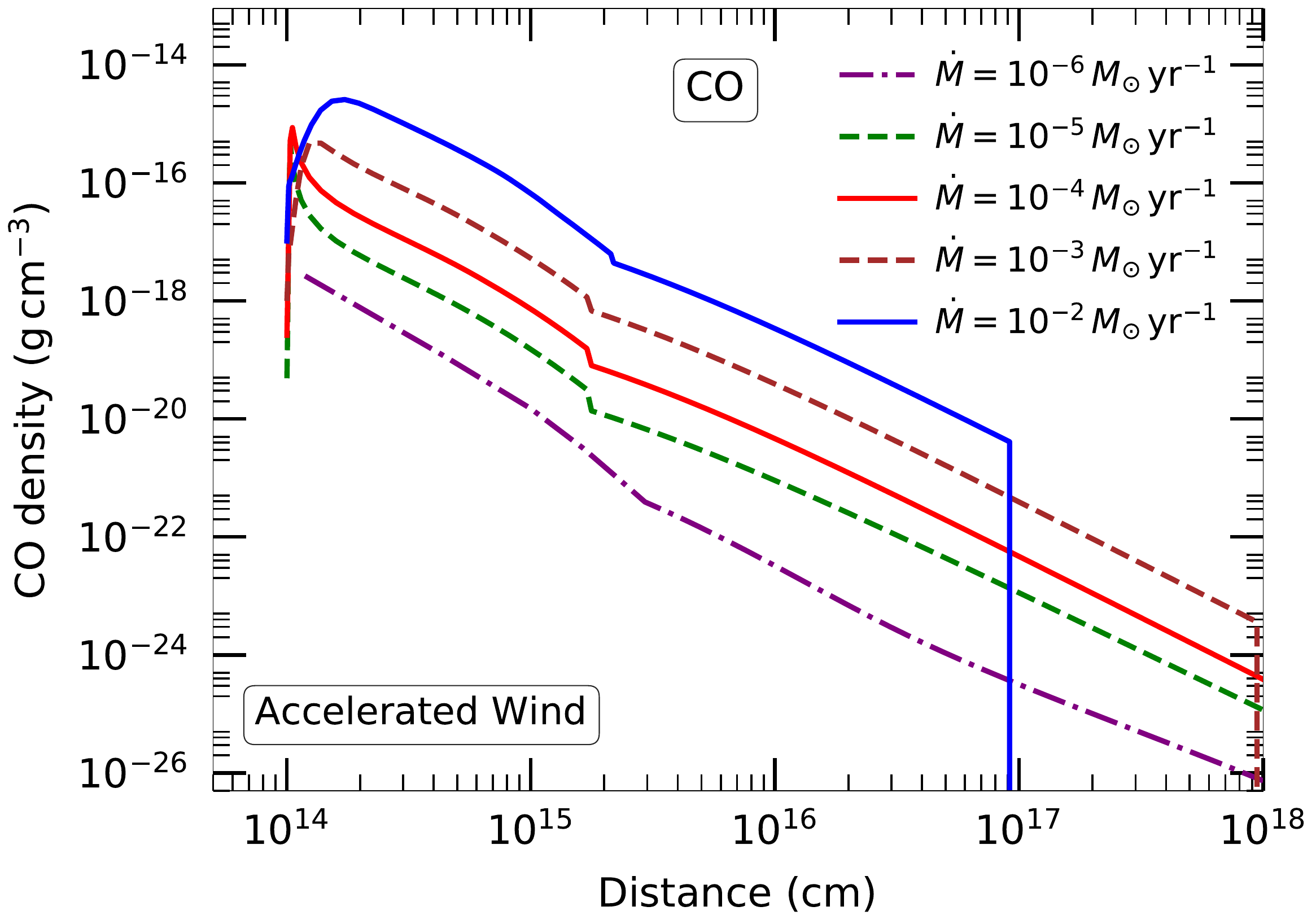}
\end{subfigure}

\vspace{0.25cm}

\begin{subfigure}{0.48\textwidth}
    \centering
    \includegraphics[width=\linewidth]
    {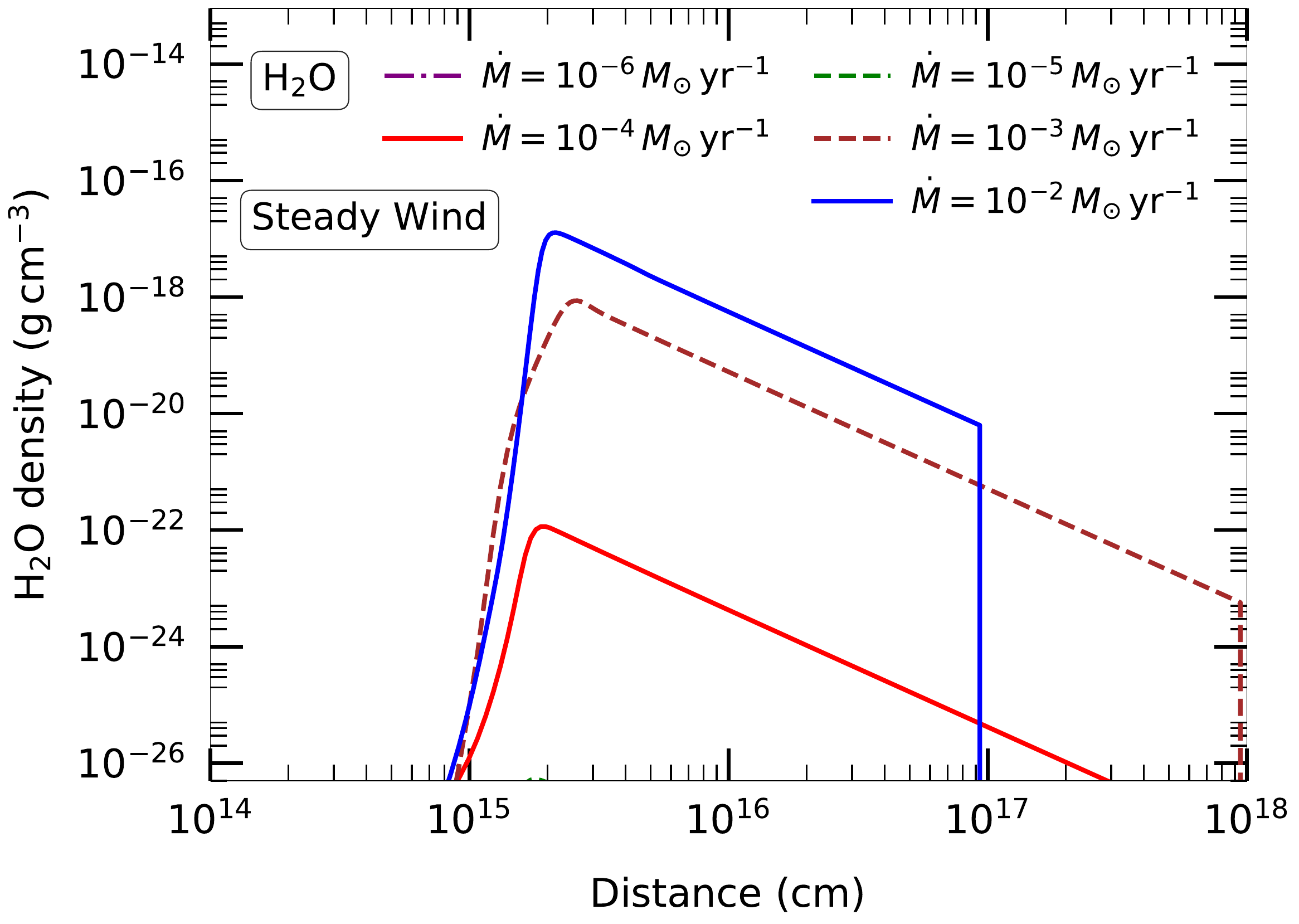}
\end{subfigure}
\hfill
\begin{subfigure}{0.48\textwidth}
    \centering
    \includegraphics[width=\linewidth]
    {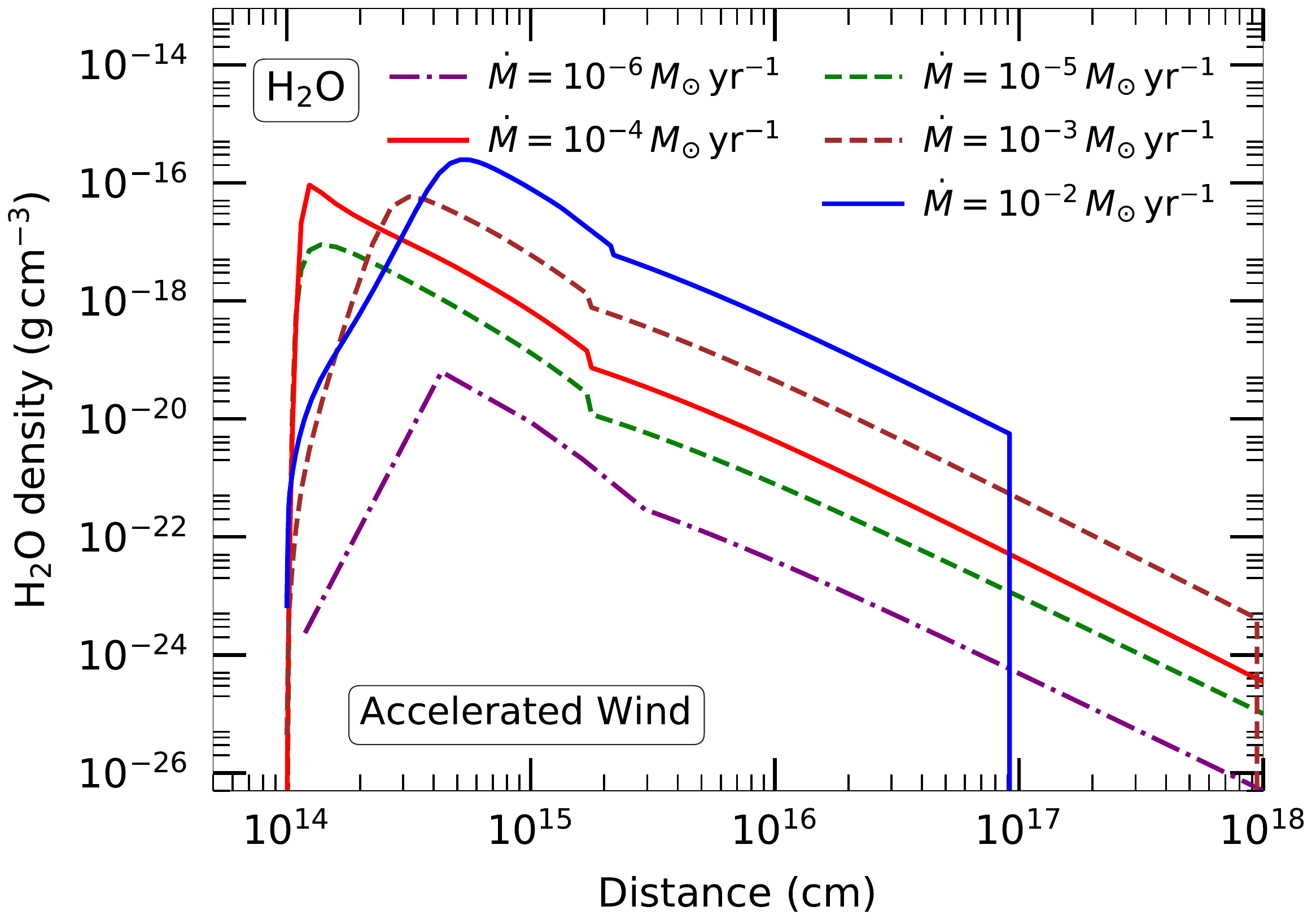}
\end{subfigure}

\vspace{0.25cm}

\begin{subfigure}{0.48\textwidth}
    \centering
    \includegraphics[width=\linewidth]
    {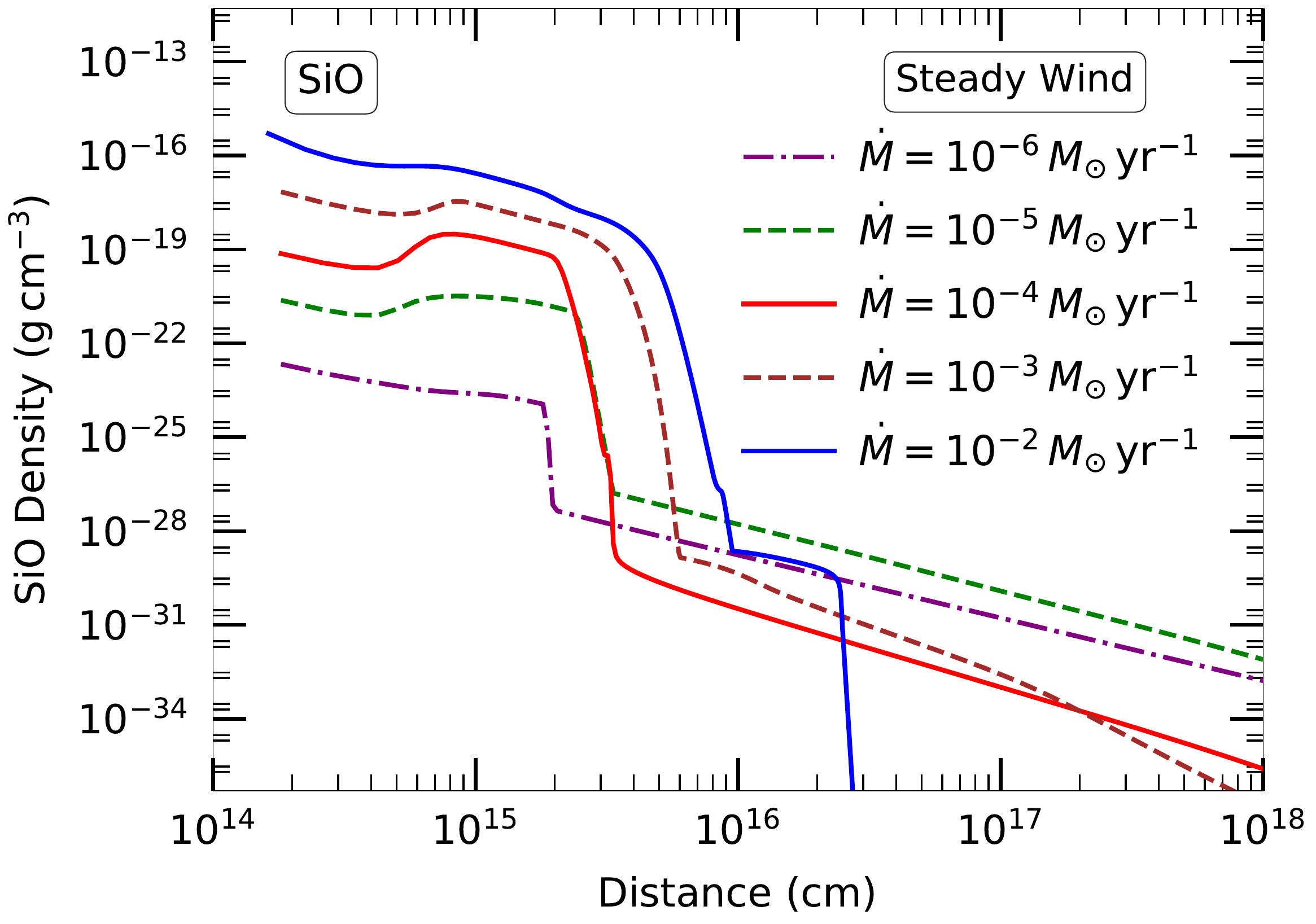}
\end{subfigure}
\hfill
\begin{subfigure}{0.48\textwidth}
    \centering
    \includegraphics[width=\linewidth]
    {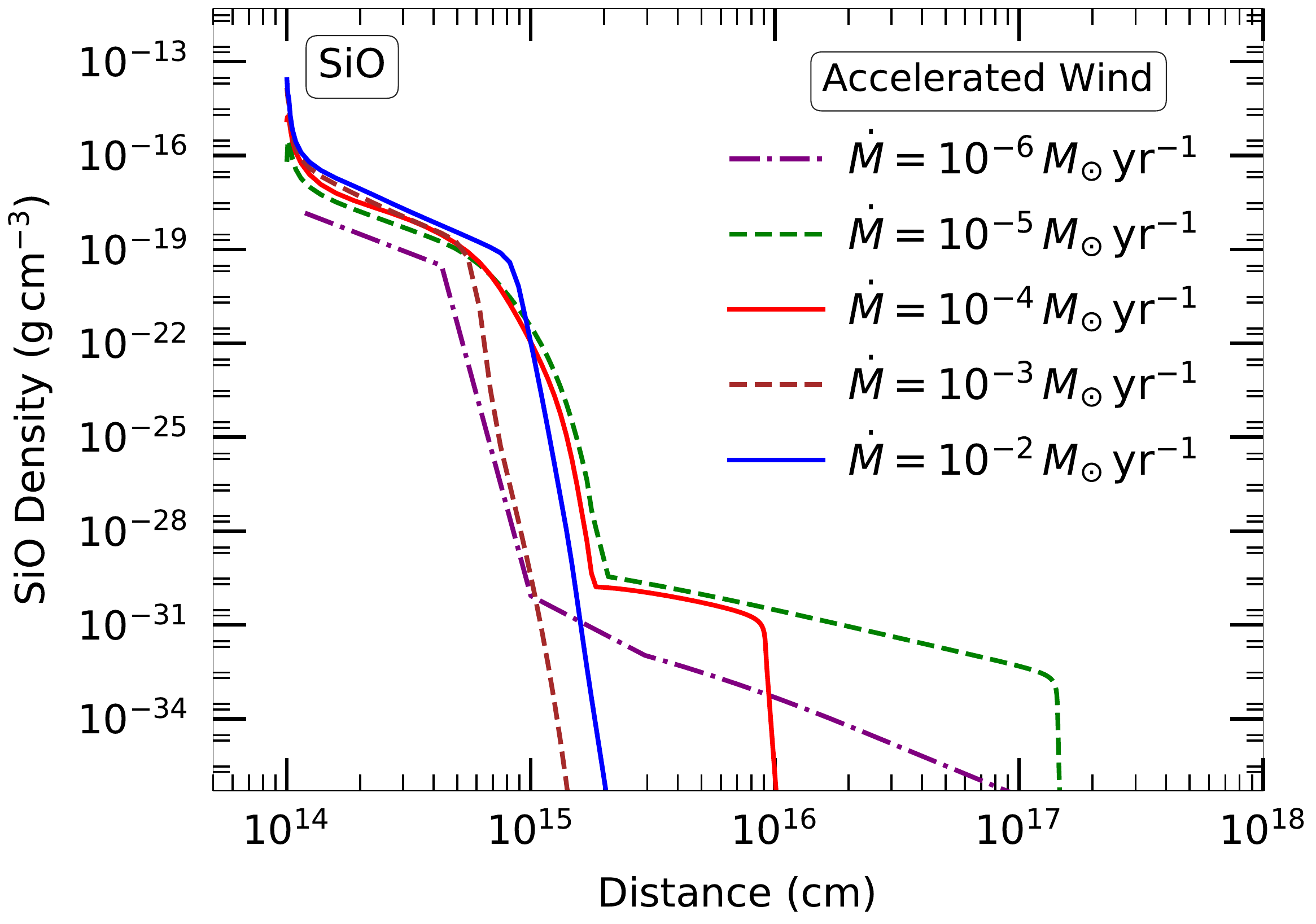}
\end{subfigure}

\caption{
Radial density profiles of molecules CO, H$_2$O, and SiO in constant (\textit{left panel}) and accelerating (\textit{right panel})
winds for \mdot\ spanning between
$10^{-6}$--$10^{-2}\,M_{\odot}\,{\rm yr^{-1}}$.
From top to bottom, the rows show the density profiles of CO, H$_2$O,
and SiO, respectively. 
}
\label{fig:molecule_co_h2o_sio}

\end{figure*}

\begin{table*}[!t]
\centering

\caption{
Column densities of molecules in the steady and accelerated RSG wind
models for constant mass-loss rates between
$10^{-6}$ and $10^{-2}\,M_{\odot}\,\mathrm{yr}^{-1}$.
All column densities are given in units of $\mathrm{cm}^{-2}$.
}
\label{tab:molecule_column_densities}

\vspace{0.2cm}

\setlength{\tabcolsep}{2.0pt}
\renewcommand{\arraystretch}{1.18}

\makebox[\textwidth][r]{%
\resizebox{1.13\textwidth}{!}{%
\begin{tabular}{lcccccccccc}

\hline\hline

\multirow{2}{*}{\raisebox{-2.0ex}{\textbf{Species}}}
&
\multicolumn{2}{c}{$\dot{M}=10^{-6}\,M_{\odot}\,\mathrm{yr}^{-1}$}
&
\multicolumn{2}{c}{$\dot{M}=10^{-5}\,M_{\odot}\,\mathrm{yr}^{-1}$}
&
\multicolumn{2}{c}{$\dot{M}=10^{-4}\,M_{\odot}\,\mathrm{yr}^{-1}$}
&
\multicolumn{2}{c}{$\dot{M}=10^{-3}\,M_{\odot}\,\mathrm{yr}^{-1}$}
&
\multicolumn{2}{c}{$\dot{M}=10^{-2}\,M_{\odot}\,\mathrm{yr}^{-1}$}
\\[1.5mm]

\cline{2-3}
\cline{4-5}
\cline{6-7}
\cline{8-9}
\cline{10-11}

&
Steady
&
Accelerated
&
Steady
&
Accelerated
&
Steady
&
Accelerated
&
Steady
&
Accelerated
&
Steady
&
Accelerated
\\[1.5mm]

\hline

$\mathrm{CO}$
&
$3.1\times10^{14}$
&
$1.0\times10^{19}$
&
$5.3\times10^{16}$
&
$9.2\times10^{19}$
&
$9.5\times10^{18}$
&
$3.5\times10^{20}$
&
$1.4\times10^{20}$
&
$1.3\times10^{21}$
&
$1.2\times10^{21}$
&
$1.2\times10^{22}$
\\[1.5mm]

$\mathrm{SiO}$
&
$8.1\times10^{13}$
&
$3.3\times10^{18}$
&
$6.3\times10^{16}$
&
$2.7\times10^{19}$
&
$3.6\times10^{18}$
&
$1.1\times10^{20}$
&
$5.2\times10^{19}$
&
$3.2\times10^{20}$
&
$1.0\times10^{21}$
&
$5.0\times10^{20}$
\\[1.5mm]

$\mathrm{H_2O}$
&
$6.1\times10^{6}$
&
$1.2\times10^{18}$
&
$3.8\times10^{11}$
&
$5.1\times10^{19}$
&
$8.6\times10^{15}$
&
$2.8\times10^{20}$
&
$8.2\times10^{19}$
&
$7.7\times10^{20}$
&
$1.0\times10^{21}$
&
$5.2\times10^{21}$
\\[1.5mm]

$\mathrm{CS}$
&
$1.4\times10^{7}$
&
$6.1\times10^{13}$
&
$1.9\times10^{9}$
&
$1.3\times10^{15}$
&
$6.6\times10^{11}$
&
$6.2\times10^{16}$
&
$1.5\times10^{15}$
&
$6.0\times10^{18}$
&
$2.2\times10^{17}$
&
$2.8\times10^{19}$
\\[1.5mm]

$\mathrm{HCN}$
&
$1.7\times10^{2}$
&
$3.0\times10^{11}$
&
$1.0\times10^{6}$
&
$3.2\times10^{13}$
&
$1.3\times10^{9}$
&
$5.4\times10^{14}$
&
$3.0\times10^{14}$
&
$5.2\times10^{18}$
&
$2.1\times10^{15}$
&
$1.4\times10^{20}$
\\[1.5mm]

$\mathrm{NH_3}$
&
$7.1\times10^{-8}$
&
$1.5\times10^{3}$
&
$2.5\times10^{-5}$
&
$9.2\times10^{6}$
&
$3.5\times10^{-1}$
&
$1.1\times10^{10}$
&
$5.9\times10^{5}$
&
$1.1\times10^{14}$
&
$9.2\times10^{9}$
&
$4.0\times10^{18}$
\\[1.5mm]

$\mathrm{OH}$
&
$3.8\times10^{10}$
&
$9.0\times10^{15}$
&
$1.7\times10^{13}$
&
$4.6\times10^{18}$
&
$9.6\times10^{14}$
&
$1.7\times10^{19}$
&
$2.1\times10^{18}$
&
$9.9\times10^{19}$
&
$4.4\times10^{19}$
&
$1.2\times10^{21}$
\\[1.5mm]

$\mathrm{H_2}$
&
$1.1\times10^{15}$
&
$6.3\times10^{19}$
&
$9.5\times10^{17}$
&
$6.2\times10^{22}$
&
$1.5\times10^{20}$
&
$6.6\times10^{23}$
&
$5.4\times10^{22}$
&
$1.5\times10^{24}$
&
$1.9\times10^{24}$
&
$9.7\times10^{24}$
\\[1.5mm]

$\mathrm{SO}$
&
$1.1\times10^{8}$
&
$7.1\times10^{12}$
&
$2.3\times10^{11}$
&
$1.1\times10^{17}$
&
$1.6\times10^{14}$
&
$3.4\times10^{17}$
&
$1.2\times10^{17}$
&
$1.6\times10^{17}$
&
$1.8\times10^{18}$
&
$9.0\times10^{17}$
\\[1.5mm]

\hline\hline

\end{tabular}%
}%
}

\end{table*}

\begin{figure*}[htbp]
    \centering

    \begin{subfigure}[t]{0.48\textwidth}
        \centering
        \includegraphics[width=\linewidth]
        {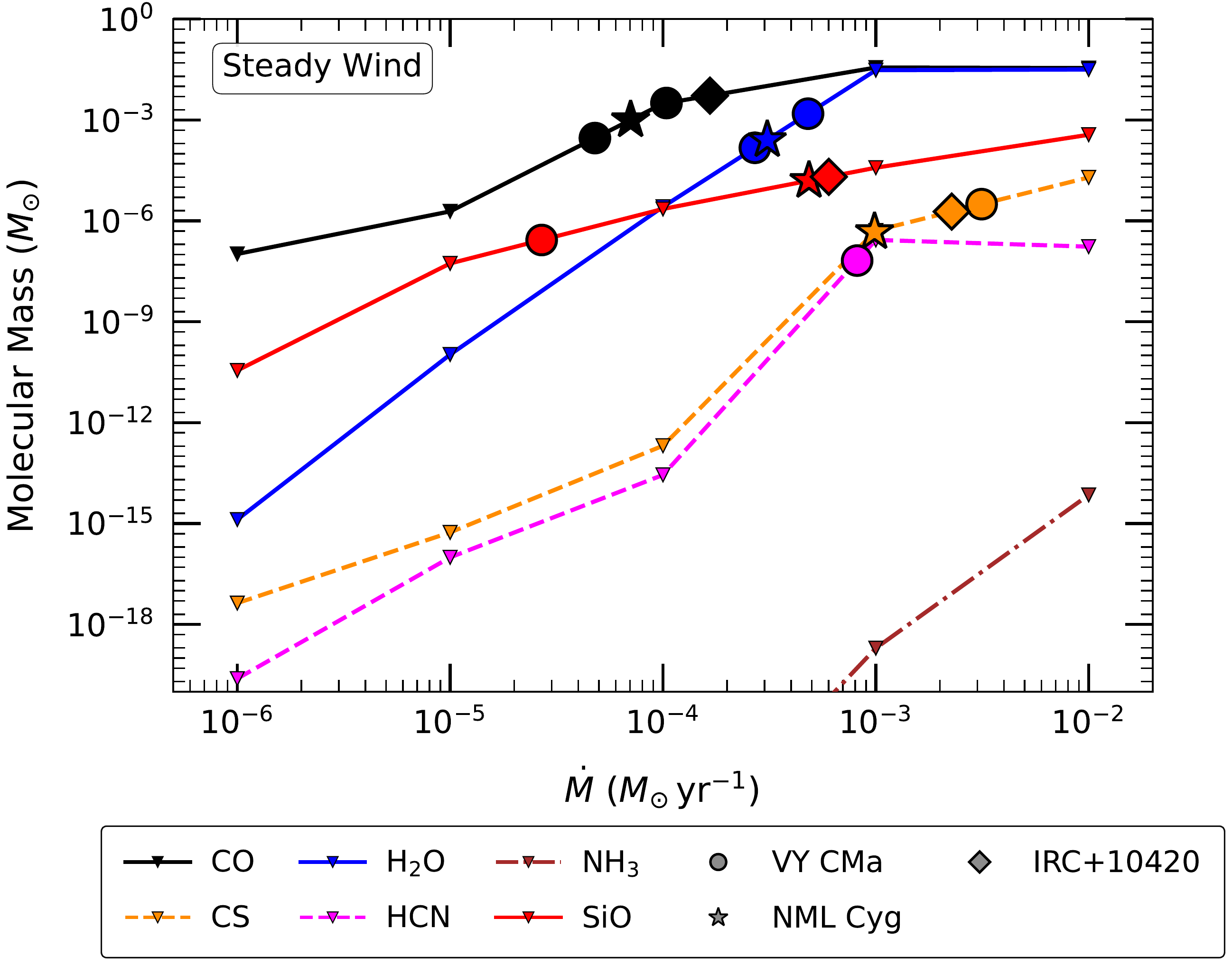}
    \end{subfigure}
    \hfill
    \begin{subfigure}[t]{0.48\textwidth}
        \centering
        \includegraphics[width=\linewidth]
        {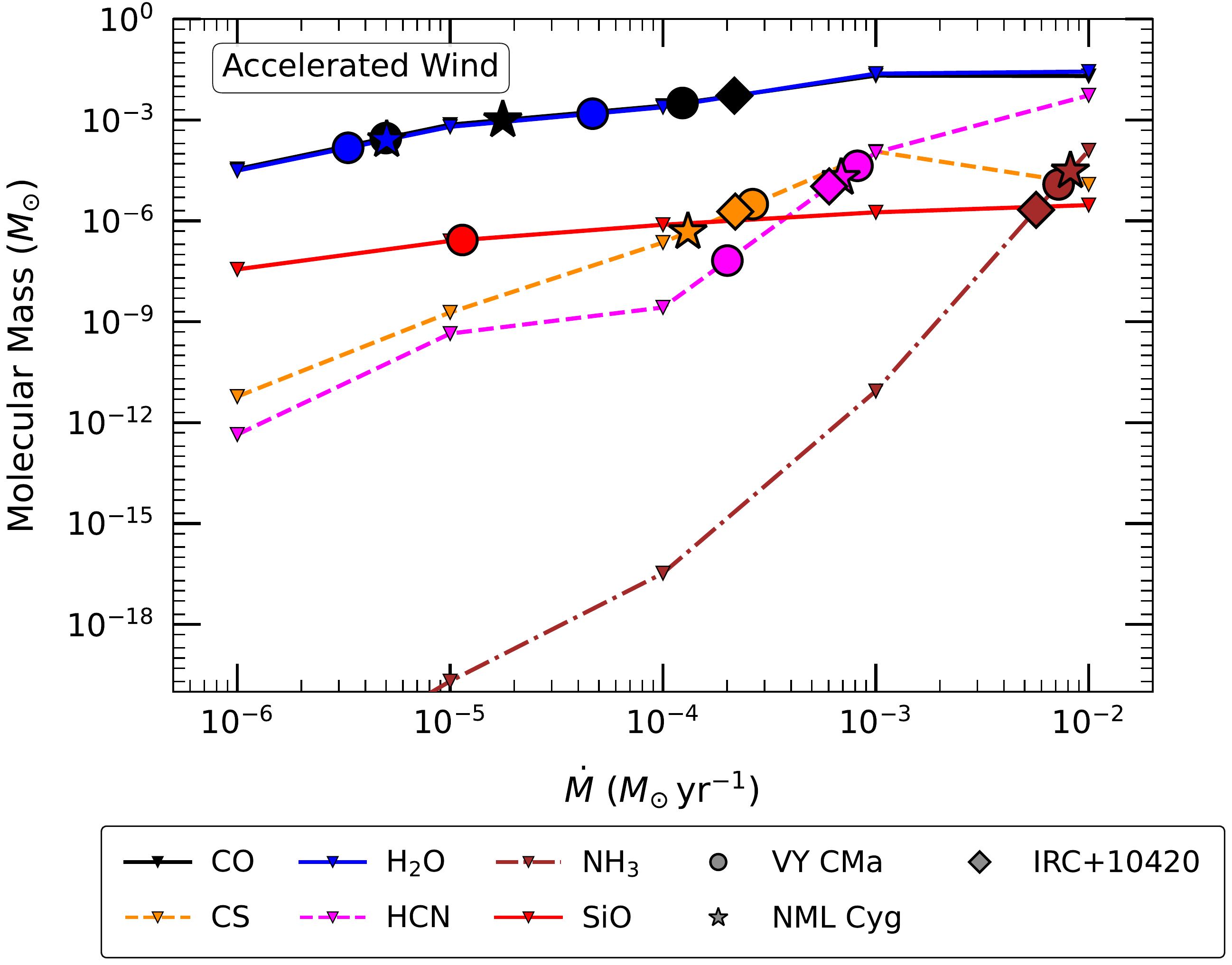}
    \end{subfigure}

    \caption{
    \textit{Left panel}: Total molecular mass as a function of the \mdot\ for
    constant wind velocity. The curves show the integrated masses
    of CO, CS, H$_2$O, HCN, NH$_3$, OH, and SiO for \mdot\ between
    $10^{-6}$--$10^{-2}\,M_{\odot}\,\mathrm{yr}^{-1}$.
    \textit{Right panel}: Same as the left panel, but for an accelerating wind described by
    the adopted \bet\ = 1.2. In both panels, the molecular
    masses are obtained by integrating the corresponding radial molecular
    density profiles in the CSM upto $\sim 1$~parsec. The observational
    estimates of molecular masses in VY CMa, NML Cyg, and IRC+10420  are adopted from
    \citet{matsuura_2014, ziurys_2025}.  The observed masses of each molecule are placed at the \mdot\ where the masses matches our model estimates. Markers: Circles, Stars, and Diamonds represent VY CMa, NML Cyg, and IRC+10420, respectively, while the colours identify the molecular species.
    }
    \label{fig:molecular_mass_steady_accelerated}

\end{figure*}

\section{Molecular yields}
\label{sec:molecules}

Molecules are formed in the wind depending on the abundances of elements and the physical conditions. Some molecules, such as SiO, are direct indicators of dust formation in the ejecta. CO and H$_2$, if formed abundantly, act as efficient coolants in the gas. We account for the preferred chemical pathways, molecular concentrations, their distribution in the CSM, and the timescales of formation. The results are summarized in Figure \ref{fig:molecule_co_h2o_sio} and \ref{fig:molecular_mass_steady_accelerated}; along with Table \ref{tab:molecule_column_densities}. 

\subsection{Molecules formed in steady wind}

We find efficient formation of CO, SiO, H$_2$O, and OH molecules, along with trace amounts of CH, SO, NH$_3$, and HCN. Being the most abundant element in the gas, most H atoms rapidly form H$_2$ molecules, while O atoms form O$_2$ molecules. Figure \ref{fig:molecule_co_h2o_sio} (left panel) presents the densities of CO, SiO and H$_2$O molecules as a function of radial distance from the centre of the star, considering a steady wind (\bet\ = 0, v$_w$ = 20 km s$^{-1}$). The mass-loss rate, \mdot, is varied from 10$^{-6}$ to 10$^{-2}$ \Mdot. 

In this case, molecule formation begins close to the stellar surface, with SiO being among the first species to form, followed by CO, AlO, OH, and H$_2$O. Around 10 stellar radii (\about\ 10$^{15}$ cm), the molecular abundances reach its maximum, as indicated by the figures. Once saturation is reached, the CO density declines with radius, following the overall gas density. For all mass-loss rates, the SiO density shows a clear depletion between 10$^{15}$ and 10$^{16}$ cm, which is attributed to the formation of silicate dust within the same region. H$_2$O forms rapidly around 10$^{15}$ cm, while OH molecules begin to deplete around the same radius for all values of \mdot. Although H$_2$O is involved in the oxidation reactions leading to silicate dust formation, a noticeable depletion in its abundance at larger radii is not seen, primarily because of its high abundance.

The molecular abundances increase with the assumed mass-loss rate. For the lower mass-loss rates of \mdot\ = 10$^{-6}$, 10$^{-5}$, and 10$^{-4}$ \Mdot, the maximum molecular abundance scales approximately as the square of the mass-loss rate. At higher values of \mdot, the scaling becomes shallower, with a power-law index between 1 and 2. The depletion of SiO into silicate dust also depends on \mdot, with higher mass-loss rates leading to more efficient depletion. Thus, at higher \mdot, SiO initially forms in greater abundance but is also depleted more efficiently into silicate dust, resulting in lower SiO abundances at larger radii. 

\subsection{Molecules formed in accelerating wind}

In the alternative scenario, the ejected mass gradually accelerates until it reaches a terminal velocity of 20 km s$^{-1}$. We adopt \bet\ = 1.2 in Equation \ref{eq:betalaw}, which best reproduces the CSM structures of several Type IIn SNe \citep{Sengupta_Sujit_Sarangi2026}. This results in an enhancement of the gas density out to 10$^{15}$ cm (see Figure \ref{fig:csm_density_with_velocity_profile}). Individual parcels of gas remain at high densities for a considerably longer time, until they reach more than 10 stellar radii ($> 10^{15}$ cm). As a result, molecules form rapidly and closer to the star, where the gas is hotter and denser. Compared to the steady-wind scenario, the molecular abundances are also higher due to the enhanced gas densities. The effect of the higher gas density is more pronounced at lower values of \mdot. Figure \ref{fig:molecule_co_h2o_sio} (right panel) presents the density distributions of CO, SiO, and H$_2$O molecules as a function of distance from the centre of the star for the different mass-loss rates.

We find that CO reaches its maximum abundance within 3 stellar radii (3$\times$10$^{14}$ cm). Compared to the steady-wind scenario, the CO abundance is approximately one order of magnitude higher for \mdot\ = 10$^{-2}$ \Mdot, while it is nearly six orders of magnitude higher for \mdot\ = 10$^{-6}$ \Mdot. As expected, CO does not show any significant depletion, and hence its radial profile closely follows the gas density. In contrast, SiO is efficiently depleted into silicate dust within 10 stellar radii, resulting in SiO molecules being largely confined within 10$^{15}$ cm. H$_2$O molecules also form close to the star and gradually increase in abundance, reaching their maximum between 10$^{14}$ and 10$^{15}$ cm. For the highest \mdot, we find a relatively modest formation of H$_2$O molecules, accompanied by efficient formation of O-rich dust grains. Interestingly, considering the balance between the enhanced gas density and molecular depletion into dust for all \mdot, the case with \mdot\ = 10$^{-4}$ \Mdot\ shows the most rapid formation of H$_2$O.

In Figure \ref{fig:molecular_mass_steady_accelerated}, we present our model predictions for the masses of CO, SiO, CS, H$_2$O, HCN, and NH$_3$ formed within one parsec around the star, alongside the derived molecular masses of the supergiants VY CMa, NML Cyg, and IRC+10420 \citep{ziurys_2025, kaminski_2013a, kaminski_2013b, cherchneff2013b, matsuura_2014}. The steady wind and accelerated wind scenarios are presented in separate panels. As expected, we find an increasing trend in molecular masses with \mdot. However, in the steady wind scenario, the molecular masses are much lower, particularly at lower \mdot. In the figure \ref{fig:molecular_mass_steady_accelerated}, we have placed the observed masses of the respective molecules at the \mdot\ values where they coincide with our model predictions. Table \ref{tab:molecule_column_densities} summarizes the column densities of all the important molecules (CO, SiO, H$_2$O, CS, HCN, NH$_3$, OH, H$_2$, SO) formed in the CSM. 

The estimated mass loss rates for VY CMa, NML Cyg, and IRC+10420 lie between 10$^{-4}$ and 10$^{-3}$ \Mdot\ \citep{singh_2022, ziurys_2025, ziurys_2009}. We find a similar range of \mdot\ for CO, H$_2$O, HCN, and CS. The observed SiO mass of VY CMa, when compared with our models, corresponds to a lower \mdot. This mainly reflects the efficiency of silicate dust condensation rather than the amount of SiO molecules formed. In the case of the accelerated wind, the SiO masses found in NML Cyg and IRC+10420 are larger than those predicted by our models for any \mdot. Comparing the SiO and H$_2$O column densities in the atmosphere of Betelgeuse \citep{perrin_2007} with our models, we find a good match for H$_2$O at \mdot\ = 10$^{-5}$ \Mdot. In contrast, the observed SiO column density corresponds to \mdot\ = 10$^{-4}$ \Mdot\ or higher. These inferred mass loss rates are significantly larger than the values of $\sim$ 10$^{-6}$ \Mdot\ derived from CO rotational lines \citep{debeck_2010}.  The predicted NH$_3$ mass matches the observations only at high \mdot\ values and only for the accelerated wind scenario. It is important to note that the molecular lines are affected by extinction due to dust formed in the CSM, which may influence the molecular masses estimated for the different supergiants. In addition, the mass loss in any typical RSG goes through several phases of high and low \mdot\ over its lifetime of about a million years, due to changes in luminosity, remaining envelope mass, pulsations, and shocks \citep{Sengupta_Sujit_Sarangi2026}, and hence the molecular concentration in the CSM will carry that signature.

\begin{figure*}[htbp]
    \centering


    \begin{subfigure}[t]{0.45\textwidth}
        \centering
        \includegraphics[width=\linewidth]
        {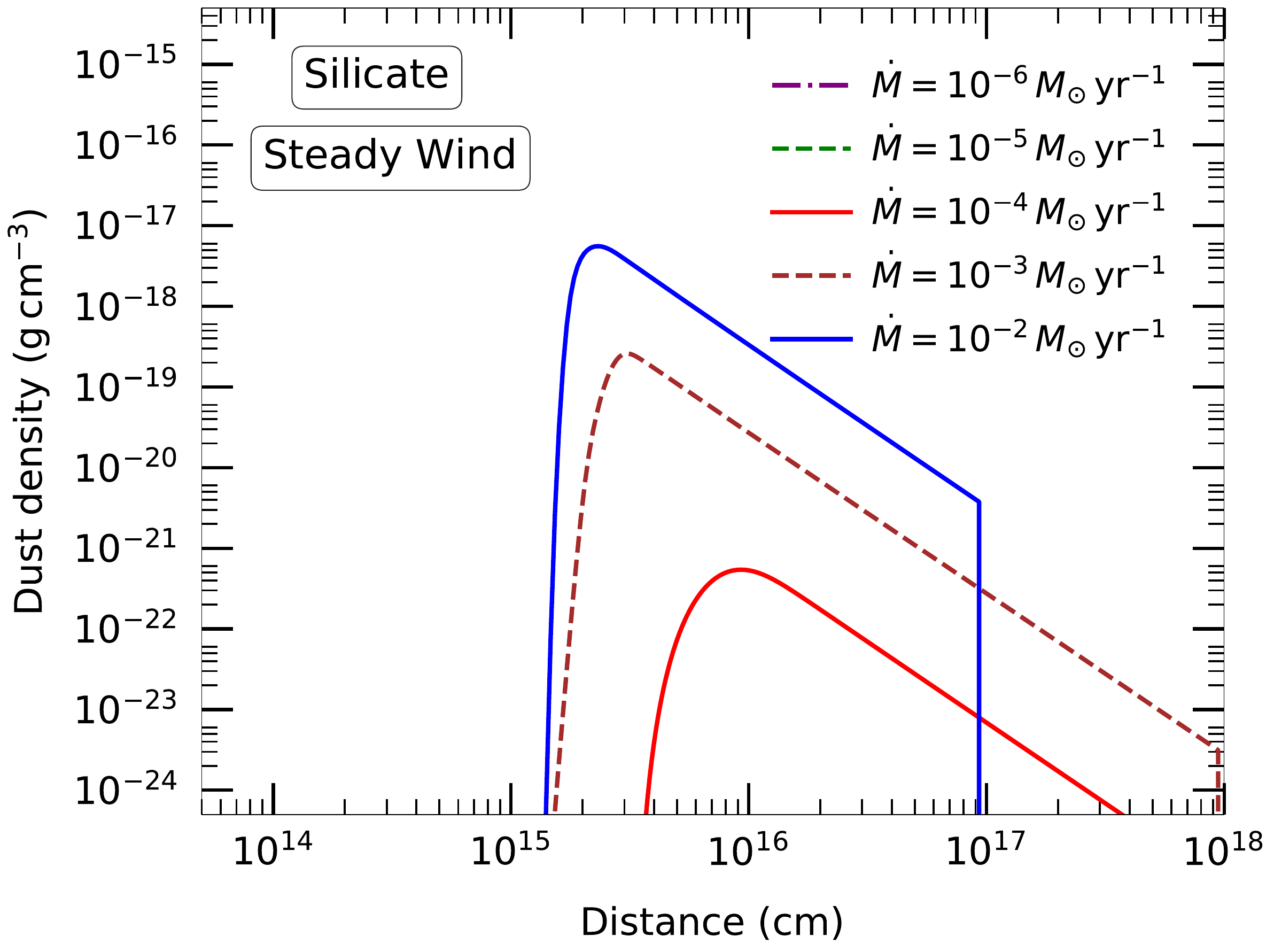}
        \label{fig:silicate_density_steady}
    \end{subfigure}
    \hfill
    \begin{subfigure}[t]{0.48\textwidth}
        \centering
        \includegraphics[width=\linewidth]
        {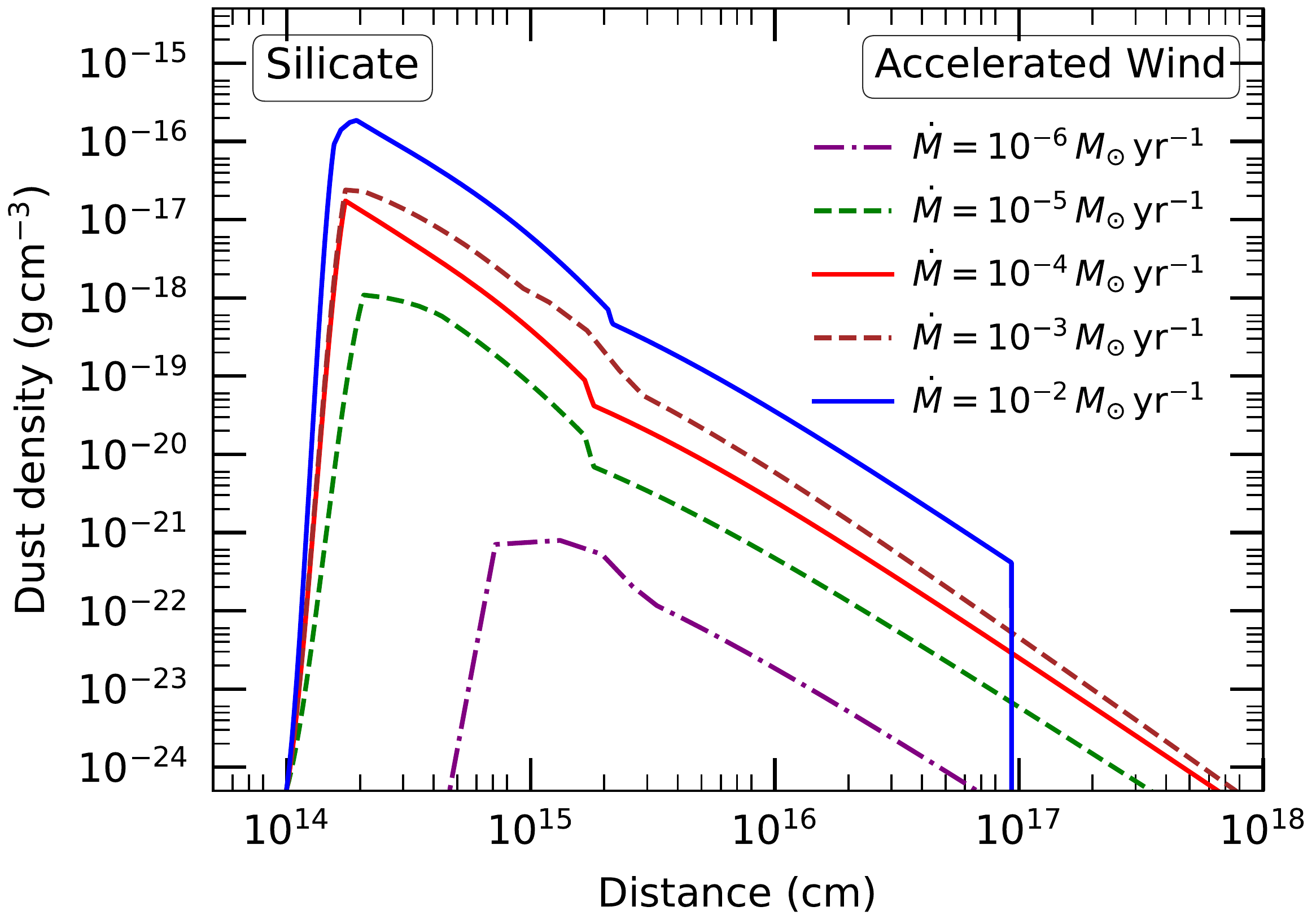}
        
        \label{fig:silicate_density_accelerated}
    \end{subfigure}

    \vspace{0.6em}


    \begin{subfigure}[t]{0.48\textwidth}
        \centering
        \includegraphics[width=\linewidth]
        {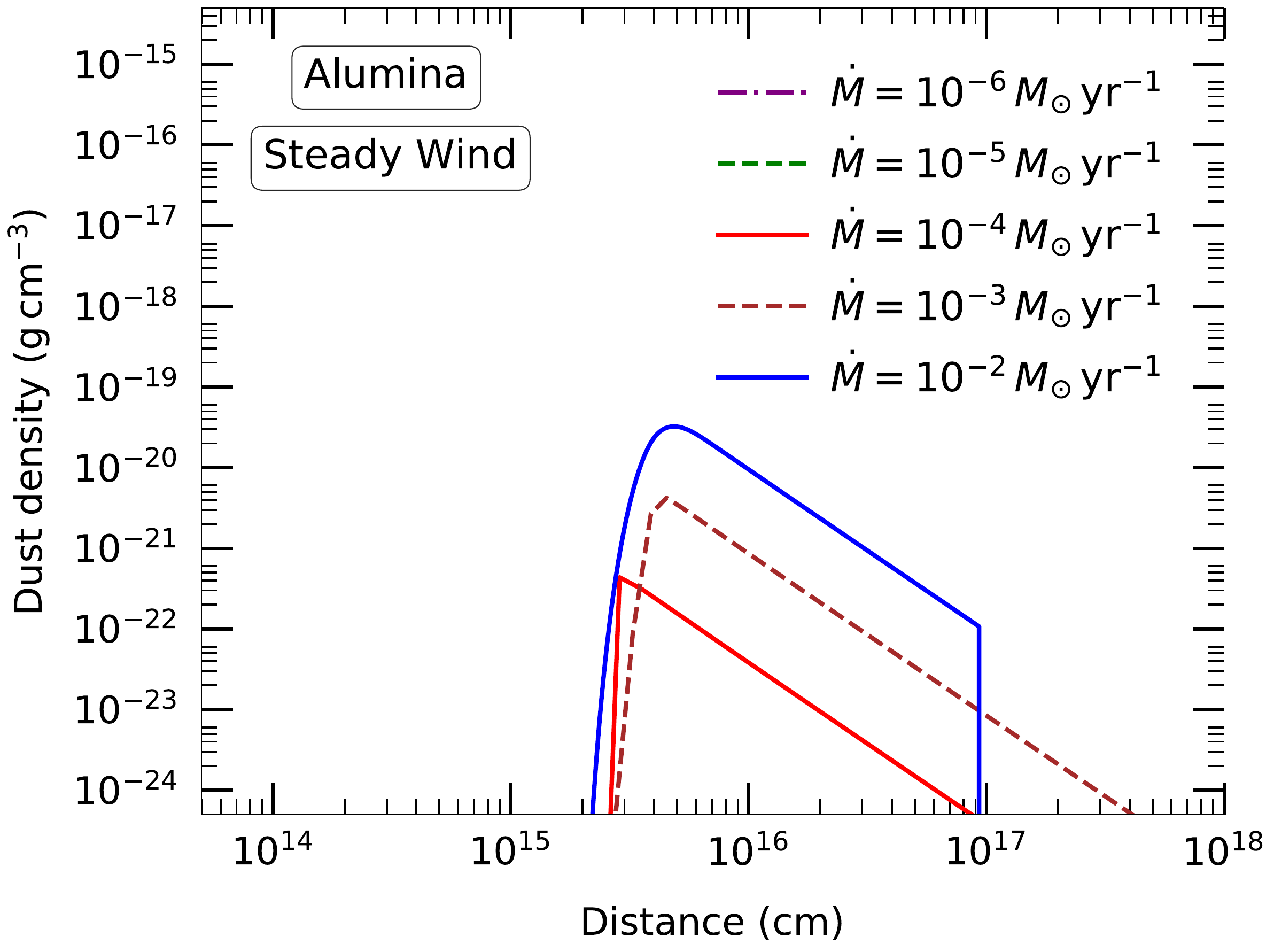}
        
        \label{fig:alumina_density_steady}
    \end{subfigure}
    \hfill
    \begin{subfigure}[t]{0.48\textwidth}
        \centering
        \includegraphics[width=\linewidth]
        {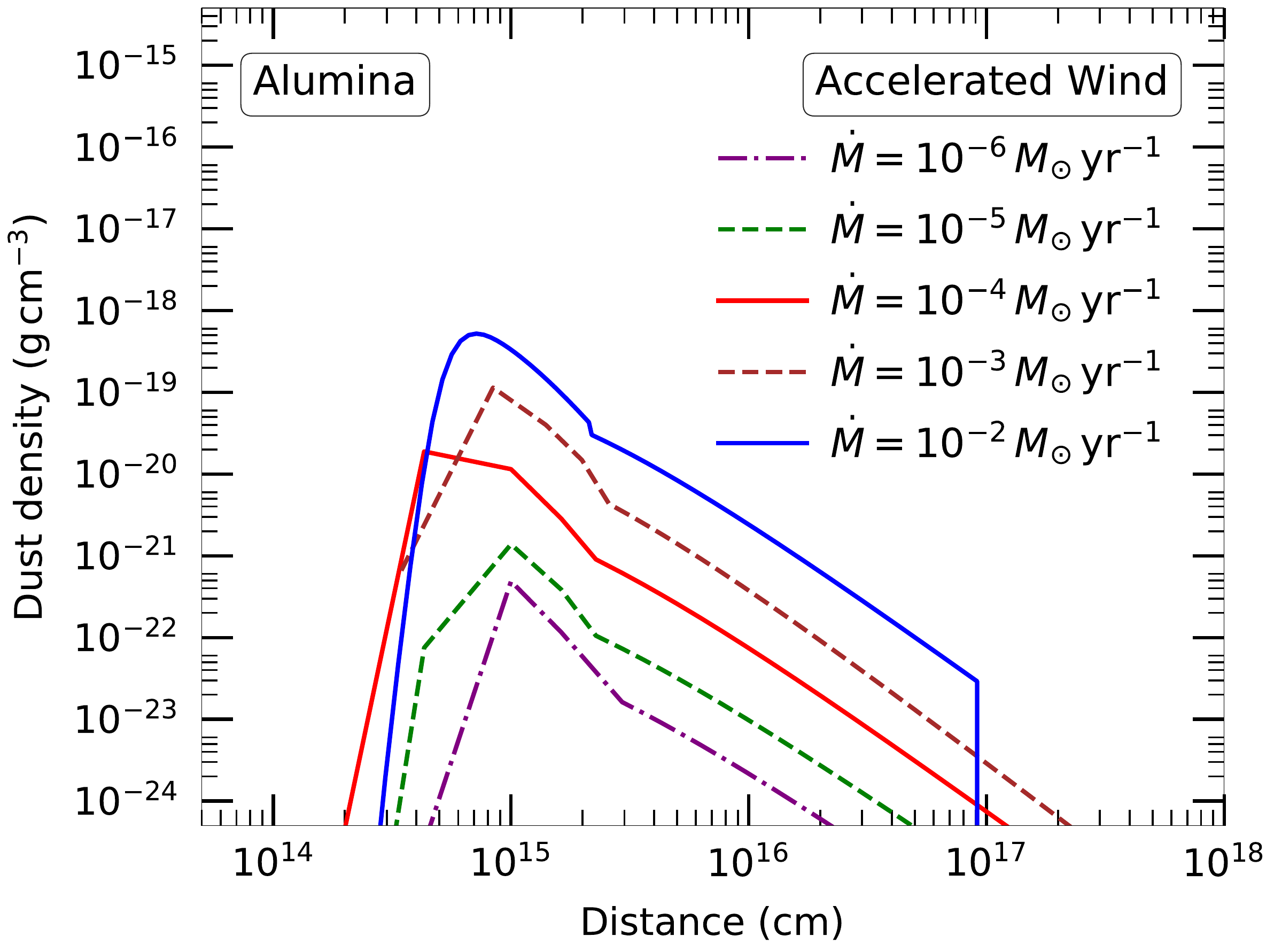}
       
        \label{fig:alumina_density_accelerated}
    \end{subfigure}

    \caption{
    Radial dust density profiles for all \mdot\ ranging between
    $10^{-6}$--$10^{-2}$ \Mdot. 
    The upper panels show the silicate dust, while the lower panels show the alumina.
    The left panels correspond to a constant wind velocity (\bet\ = 0, v = 20 km s$^{-1}$), whereas the right
    panels correspond to accelerated wind (\bet\ = 1.2, Equation~\ref{eq:betalaw}). See Section \ref{sec:dust_general} for details. 
    }
    \label{fig:dust_density_silicate_alumina} 

\end{figure*}



\section{The dusty CSM}
\label{sec:dust_general}

We considered the formation of Mg-silicates, aluminum oxide or alumina, amorphous carbon, and silicon carbide dust. The elemental abundances and molecular chemistry define the pathways leading to dust formation. As shown in Section \ref{sec:molecules}, molecules such as CO, SiO, H$_2$O, H$_2$, O$_2$, SO, and HCN dominate the molecular yields. Almost all C atoms rapidly form CO molecules, along with a trace amount of CS. We find that amorphous carbon and silicon carbide dust do not form when such O-rich chemistry prevails, consistent with previous studies \citep{cherchneff2013b, sarangi2018book}.
The primary dust components formed in the winds are Mg-silicates of the chemical type [Mg$_2$SiO$_4$]$_n$ and alumina of the chemical type [Al$_2$O$_3$]$_n$. We also find some Si--O molecular clusters, a fraction of which may eventually form [SiO$_2$]$_n$-type dust. We have not considered this dust component here because of the uncertainty in the efficiency with which these clusters condense into dust grains.

\begin{figure*}[htbp]
    \centering

    \begin{subfigure}{0.48\textwidth}
        \centering
        \includegraphics[width=\linewidth]{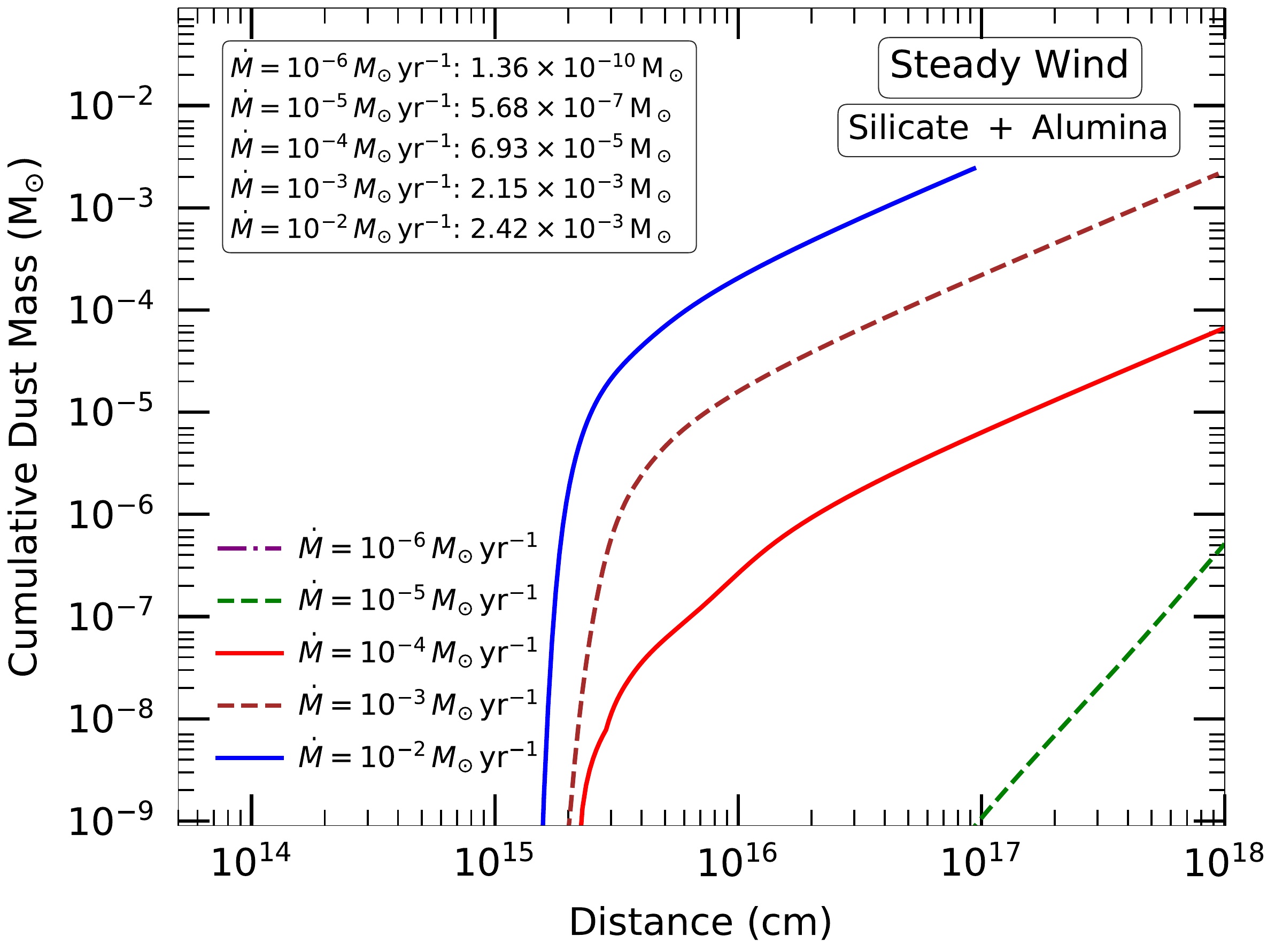}
    \end{subfigure}
    \hfill
    \begin{subfigure}{0.48\textwidth}
        \centering
        \includegraphics[width=\linewidth]{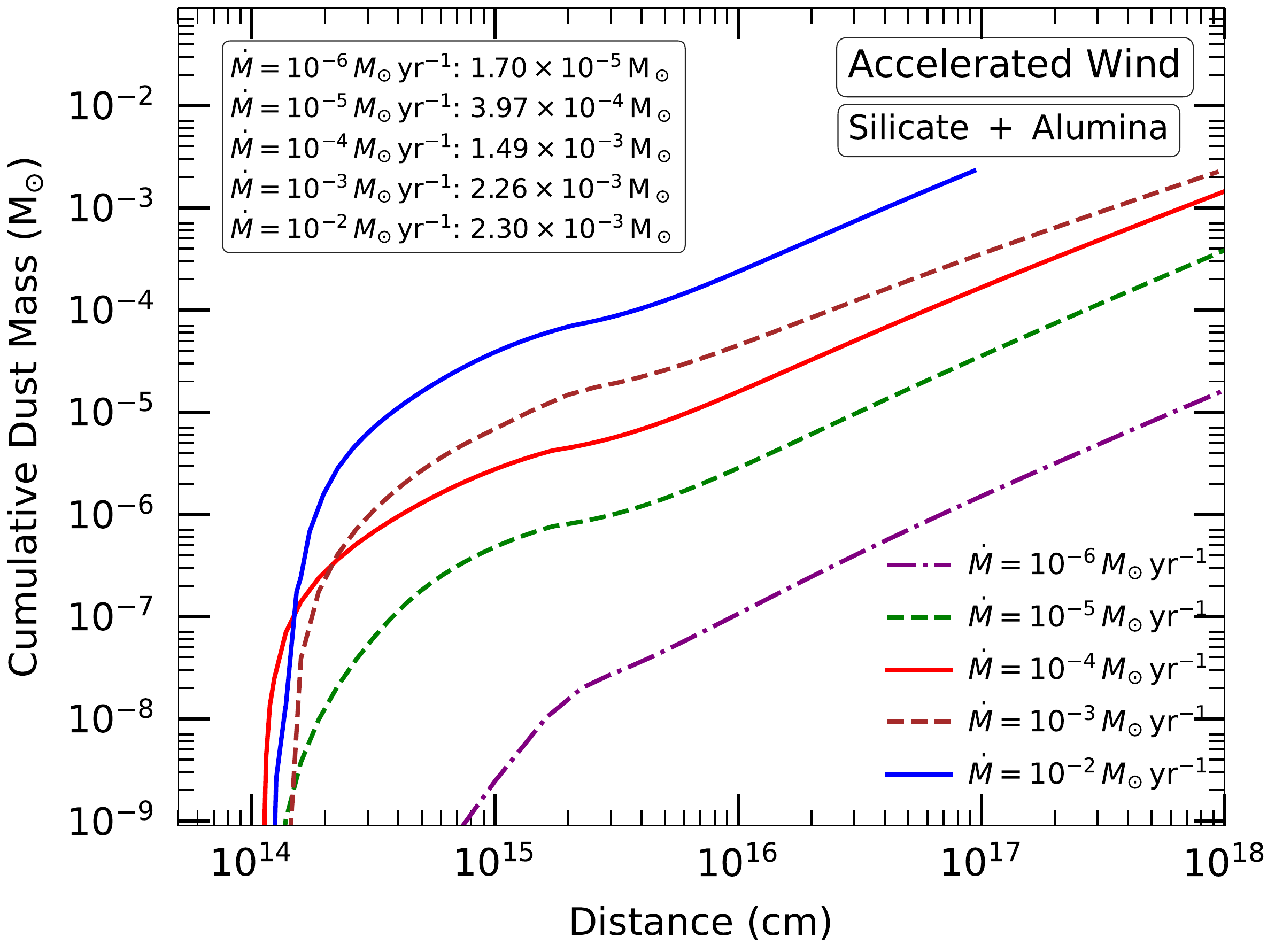}
    \end{subfigure}

    \caption{
    The cumulative dust mass, obtained by summing the masses of silicate and alumina over the radial bins, is shown for all \mdot\ for the constant wind velocity (\textit{left panel}) and accelerating wind velocity (\textit{right panel}) cases. The total dust mass enclosed within one parsec around the star just before CC is indicated in the figure and discussed in Section \ref{sec:dust_general}. The dust masses appear to continue increasing with radius for all \mdot, which is an artifact of assuming a time-independent \mdot\ in the model. In a realistic scenario, gas located at distances approaching one parsec at the time of CC would have been ejected much earlier, when the mass loss rates were likely much lower. Consequently, little dust would be expected to form in those winds. 
    }
    \label{fig:cumulative_mass_constant_beta}

\end{figure*}

\begin{figure*}[htbp]
    \centering

    \begin{subfigure}{0.48\textwidth}
        \centering
        \includegraphics[width=\linewidth]{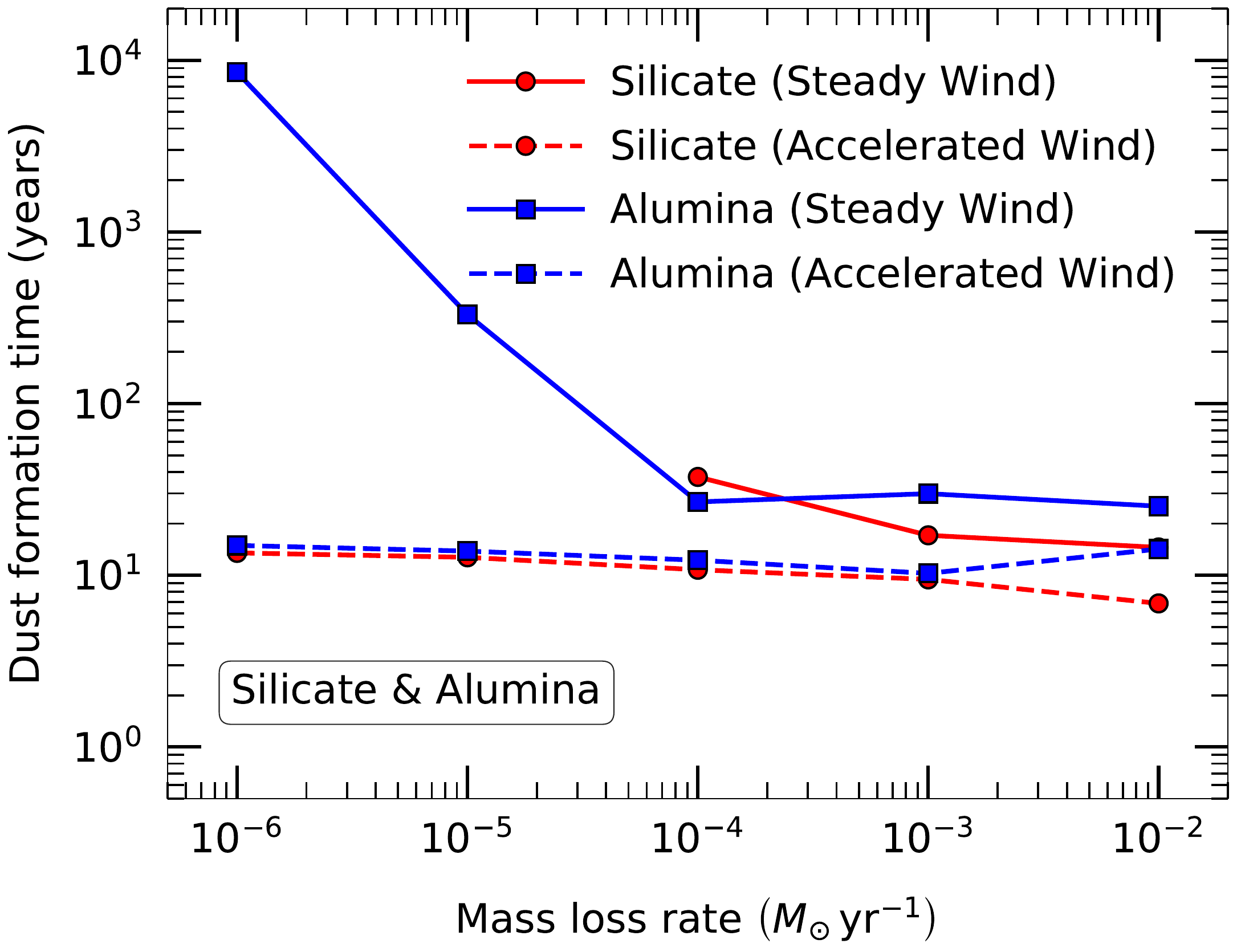}
    \end{subfigure}
    \hfill
    \begin{subfigure}{0.48\textwidth}
        \centering
        \includegraphics[width=\linewidth]{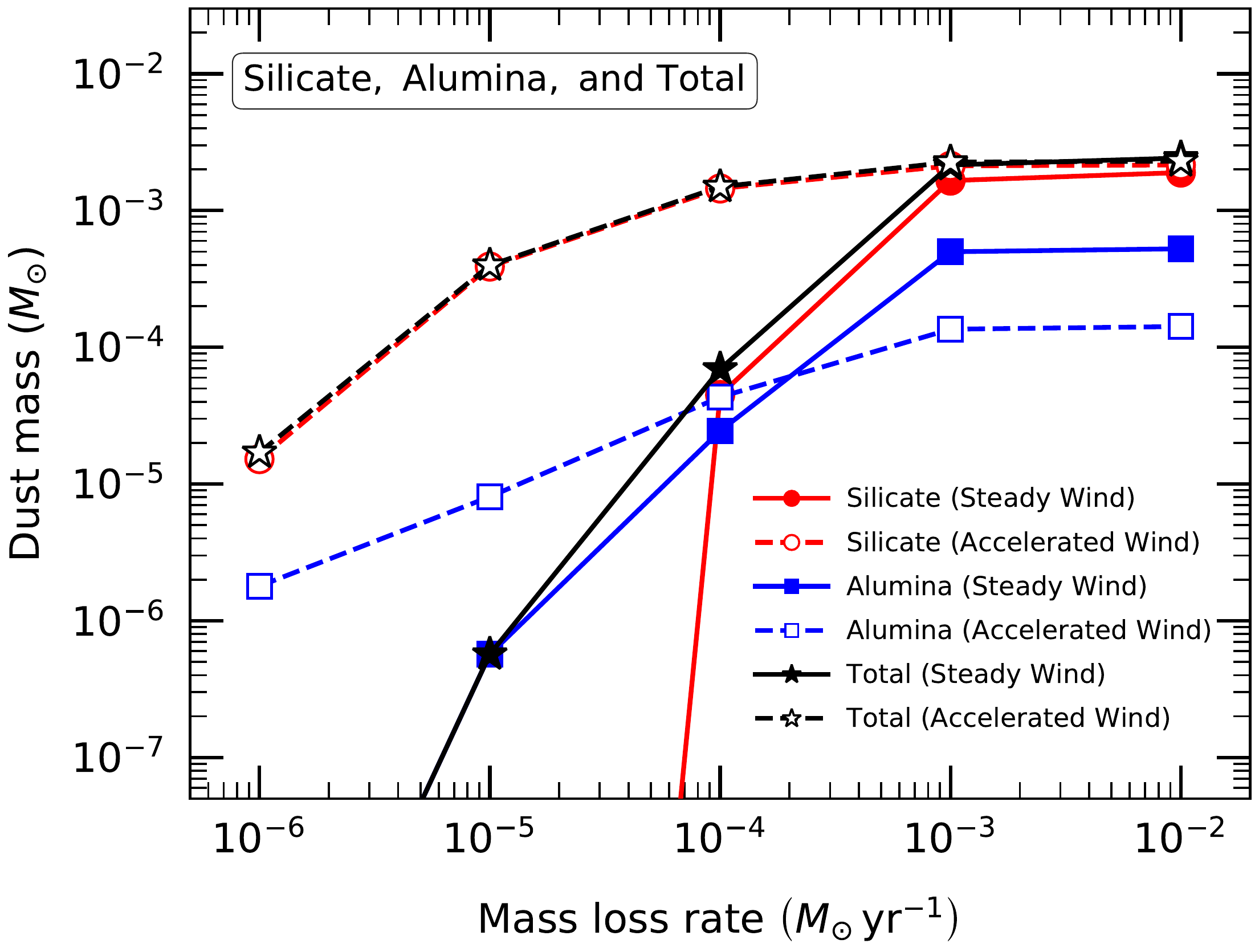}
    \end{subfigure}

    \caption{
    \textit{Left panel}: Dust formation time for any given parcel of gas (time taken to form dust after it is launched from the stellar surface) is shown as a function of the various \mdot\ for silicate and alumina dust, in steady and accelerated wind models. 
    \textit{Right panel}: The total mass (enclosed within one parsec) of silicates, alumina, and their sum is shown as a function of various \mdot, for steady and accelerated wind models. 
    }
    \label{fig:dust_formation_time_and_mass}

\end{figure*}

\subsection{Dust formation in steady wind}
\label{sec:dust_steady}

Assuming a constant wind speed of 20 km s$^{-1}$, Figure~\ref{fig:dust_density_silicate_alumina} (left panel) presents the dust density as a function of radial distance from the centre of the star for various mass loss rates. The radius of the RSG is 10$^{14}$ cm. We find that the formation of silicates and alumina is preceded by the formation of SiO and AlO molecules, respectively. The formation of these dust grains is therefore accompanied by the depletion of the corresponding molecules from the gas. For \mdot\ less than 10$^{-4}$ \Mdot, we find that dust formation is negligible because the densities are too low. Even at such low \mdot, SiO molecules form small molecular clusters, but these clusters do not grow efficiently through nucleation to form the dust precursors that eventually condense into silicate dust. At higher \mdot, silicate dust and alumina are found to form beyond 10 stellar radii ($>$ 10$^{15}$ cm). The condensation efficiency, as reflected in the dust density, increases with \mdot\ at these mass loss rates. The densities of silicates are about two orders of magnitude higher than those of alumina. However, trace amounts of alumina are formed even for \mdot\ = 10$^{-5}$ and 10$^{-6}$ \Mdot; these are not visible in the figure because their densities are sufficiently low.

The cumulative dust mass as a function of distance, expected to be present within one parsec around a typical RSG, for various \mdot, is shown in Figure~\ref{fig:cumulative_mass_constant_beta} (left panel). As the figure suggests, for \mdot\ = 10$^{-2}$ \Mdot, the mass of dust increases quickly from 20 stellar radii, and reaches values up to 10$^{-4}$ \Ms, within a spherical volume of 50 stellar radii. The mass confined within a parsec is about 2 $\times$ 10$^{-3}$ \Ms. On the contrary, it is only about 10$^{-7}$ \Ms\ for the lower \mdot\ of 10$^{-5}$ \Mdot. The dust formation zone also shifts outwards with decreasing \mdot, as the lower gas density increases the timescale required for dust formation. As seen in Figure~\ref{fig:cumulative_mass_constant_beta}, in the steady-wind scenario, the characteristic location of dust formation shifts from tens to thousands of stellar radii as the mass loss rate decreases.

\subsection{Dust formation in accelerated wind}
\label{sec:dust_accelerated}

When we assume an accelerating wind profile with \bet\ = 1.2, the resulting densities of silicates and alumina within one parsec of the star are shown for various mass-loss rates in Figure~\ref{fig:dust_density_silicate_alumina} (right panel). Close to the star, the low velocities of each parcel of gas result in enhanced densities (see Figure \ref{fig:csm_density_with_velocity_profile}). More time spent at higher densities facilitates for the formation of molecules and dust grains. As a result, silicates form as close as two stellar radii (\about 2$\times$10$^{14}$ cm), and reach the maximum efficiency well before ten stellar radii, thereafter following the gas density. The density of silicates is two orders of magnitude more than that of alumina, as expected, due to the larger abundances of Si and Mg, compared to Al. In the case of accelerating wind, we find that dust forms for all \mdot\ from 10$^{-6}$ to 10$^{-2}$ \Mdot.

The cumulative dust mass within one parsec around the star, combining silicates and alumina, is presented in Figure~\ref{fig:cumulative_mass_constant_beta} (right panel). We find that, for all values of \mdot\ except 10$^{-6}$ \Mdot, the dust mass begins to build up at around 2 $\times$ 10$^{14}$ cm. For \mdot\ = 10$^{-2}$ \Mdot, the enclosed dust mass already reaches about 10$^{-4}$ \Ms\ within ten stellar radii and increases to 2 $\times$ 10$^{-3}$ \Ms\ within a thousand stellar radii. For the lowest \mdot\ of 10$^{-6}$ \Mdot, the total enclosed dust mass within one parsec is about 10$^{-5}$ \Ms.

In Figure~\ref{fig:dust_formation_time_and_mass}, we compare the dust mass yields and dust formation times for the steady and accelerated wind scenarios as a function of mass loss rate. For \mdot\ varying from 10$^{-6}$ to 10$^{-2}$ \Mdot, the time required for dust to form in a given parcel of gas in the steady wind varies from about 10$^4$ years to 15 years. At low \mdot, only alumina forms, with formation timescales of thousands of years. In the accelerated wind scenario, over the same range of \mdot, the dust formation timescales are considerably shorter, varying between 15 and 5 years. Silicates and alumina form almost simultaneously in this scenario.

The enclosed dust masses within one parsec for the two wind profiles are compared in the right panel of Figure~\ref{fig:dust_formation_time_and_mass}. The total dust masses for the steady and accelerated winds are nearly the same for \mdot\ = 10$^{-3}$ \Mdot\ or higher. Below this mass loss rate, however, clear differences emerge in both the total dust masses and the relative abundances of silicates and alumina. For the steady wind, the dust mass decreases rapidly with decreasing \mdot, reaching about 10$^{-10}$ \Ms\ at the lowest \mdot, whereas it remains as high as 10$^{-5}$ \Ms\ in the accelerated wind scenario.

For both the steady and accelerated wind models, the cumulative dust mass continues to rise with radius and does not reach a saturation even at 10$^{18}$ cm. This behavior results from our assumption of a constant time-independent \mdot. In reality, material reaching distances of order a parsec at the time of CC would have been expelled much earlier, when the mass loss rate was likely substantially lower, and efficient dust formation would therefore be unlikely. Consequently, the cumulative dust mass is expected to level off beyond a certain radius. This emphasizes the necessity of a time-dependent \mdot, which we explore in the following section using the progenitor of SN~2023ixf as a specific case.

\begin{figure*}[htbp]
    \centering

    \begin{subfigure}{0.48\textwidth}
        \centering
        \includegraphics[width=\linewidth]{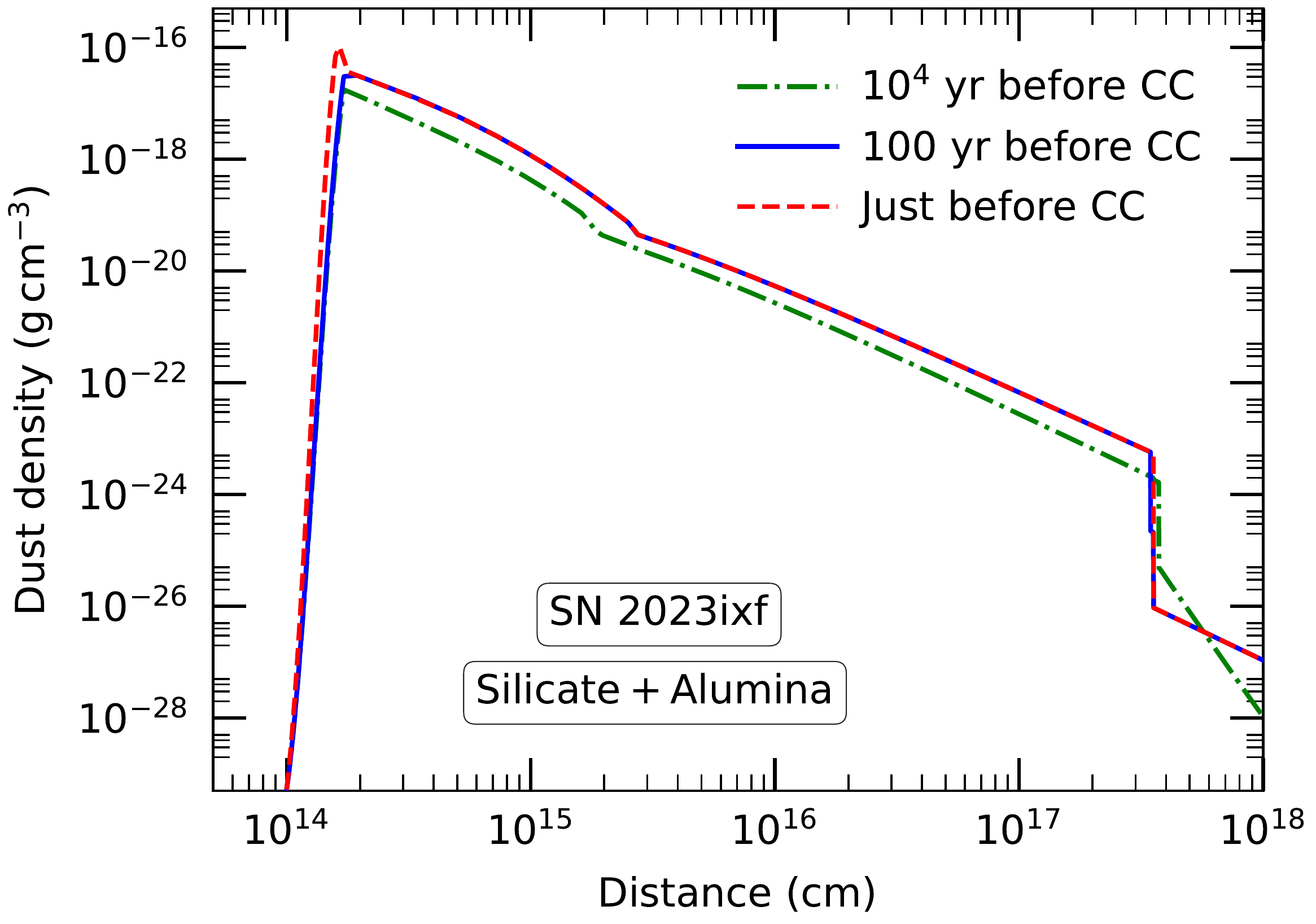}
    \end{subfigure}
    \hfill
    \begin{subfigure}{0.48\textwidth}
        \centering
        \includegraphics[width=\linewidth]{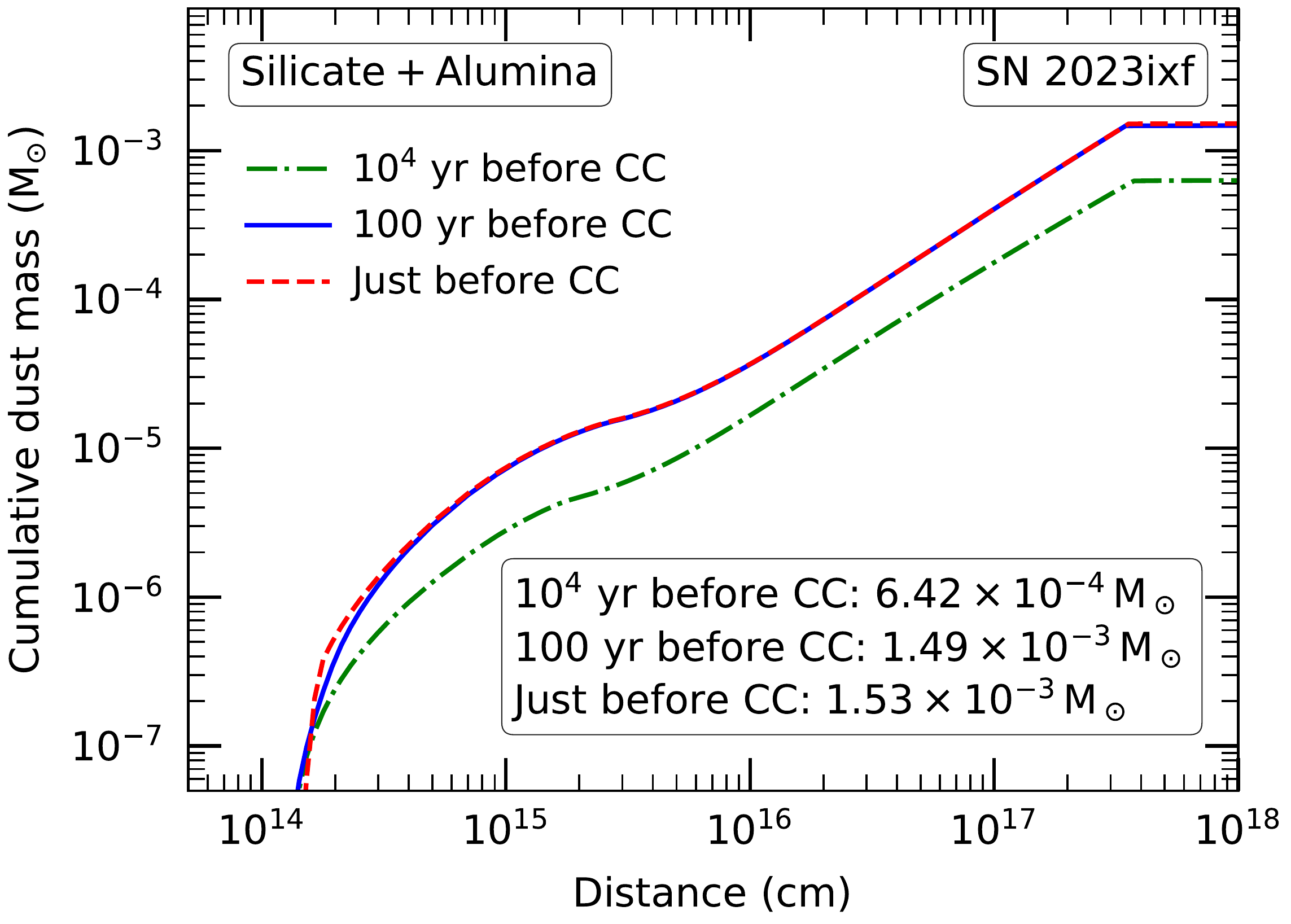}
    \end{subfigure}

    \caption{
    \textit{Left panel}: Radial dust density profile (silicates $+$ alumina)
    in the CSM of SN~2023ixf progenitor at $10^{4}$ yr before CC,
    100 yr before CC, and just before CC. 
    \textit{Right panel}: Cumulative total dust mass (silicates $+$ alumina),
    as a function of radial distance for the same three periods. The total dust masses enclosed within one parsec are $6.42\times10^{-4}$, $1.49\times10^{-3}$, and
    $1.53\times10^{-3}\,M_{\odot}$, respectively. See Section \ref{sec:dust_SN2023ixf} for details. 
    }
    \label{fig:sn2023ixf_total_dust_density_cumulative_mass}

\end{figure*} 
\begin{figure*}[htbp]
    \centering

    \begin{subfigure}{0.47\textwidth}
        \centering
        \includegraphics[width=\linewidth]
        {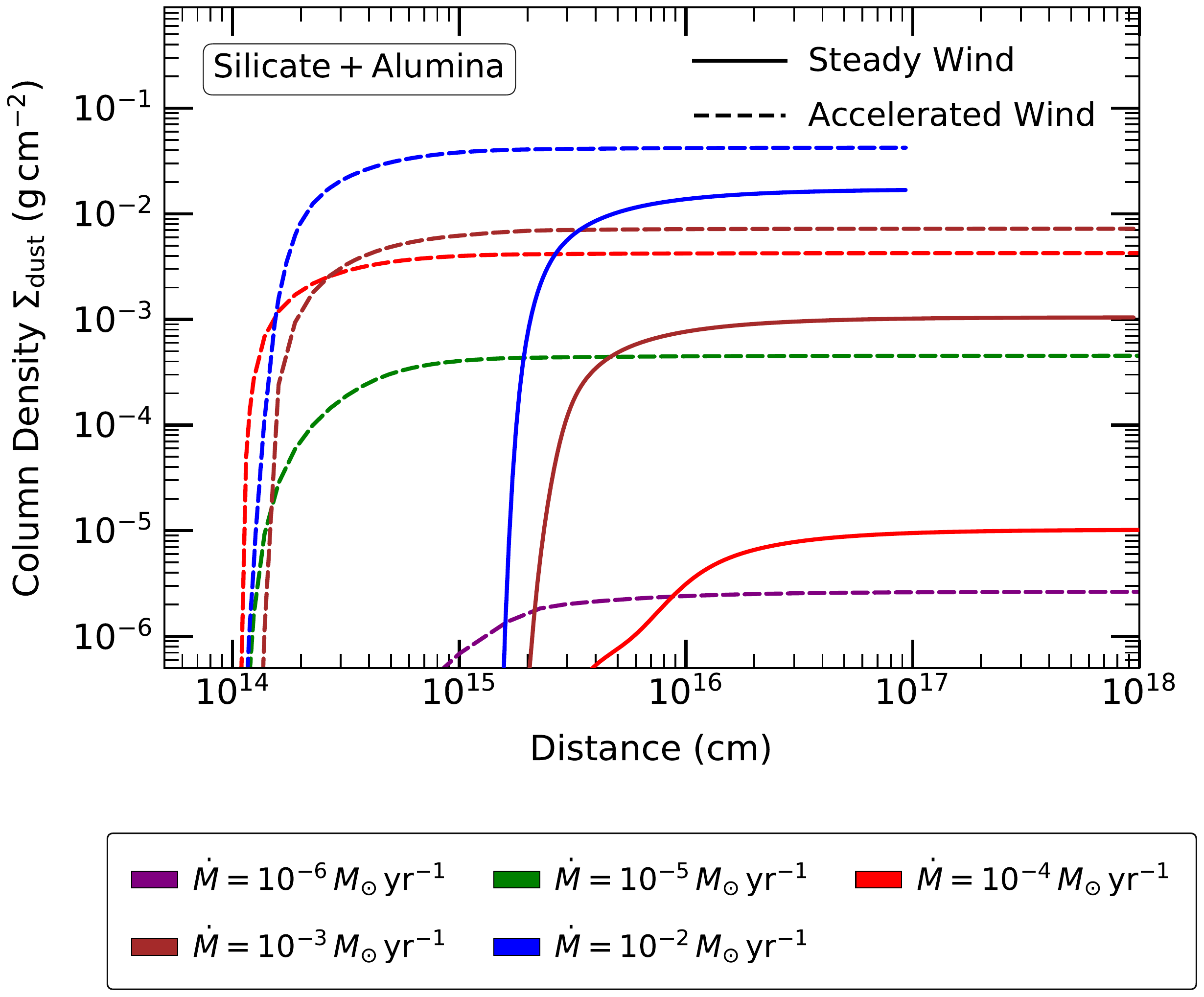}
    \end{subfigure}
    \hfill
    \begin{subfigure}{0.49\textwidth}
        \centering
        \includegraphics[width=\linewidth]
        {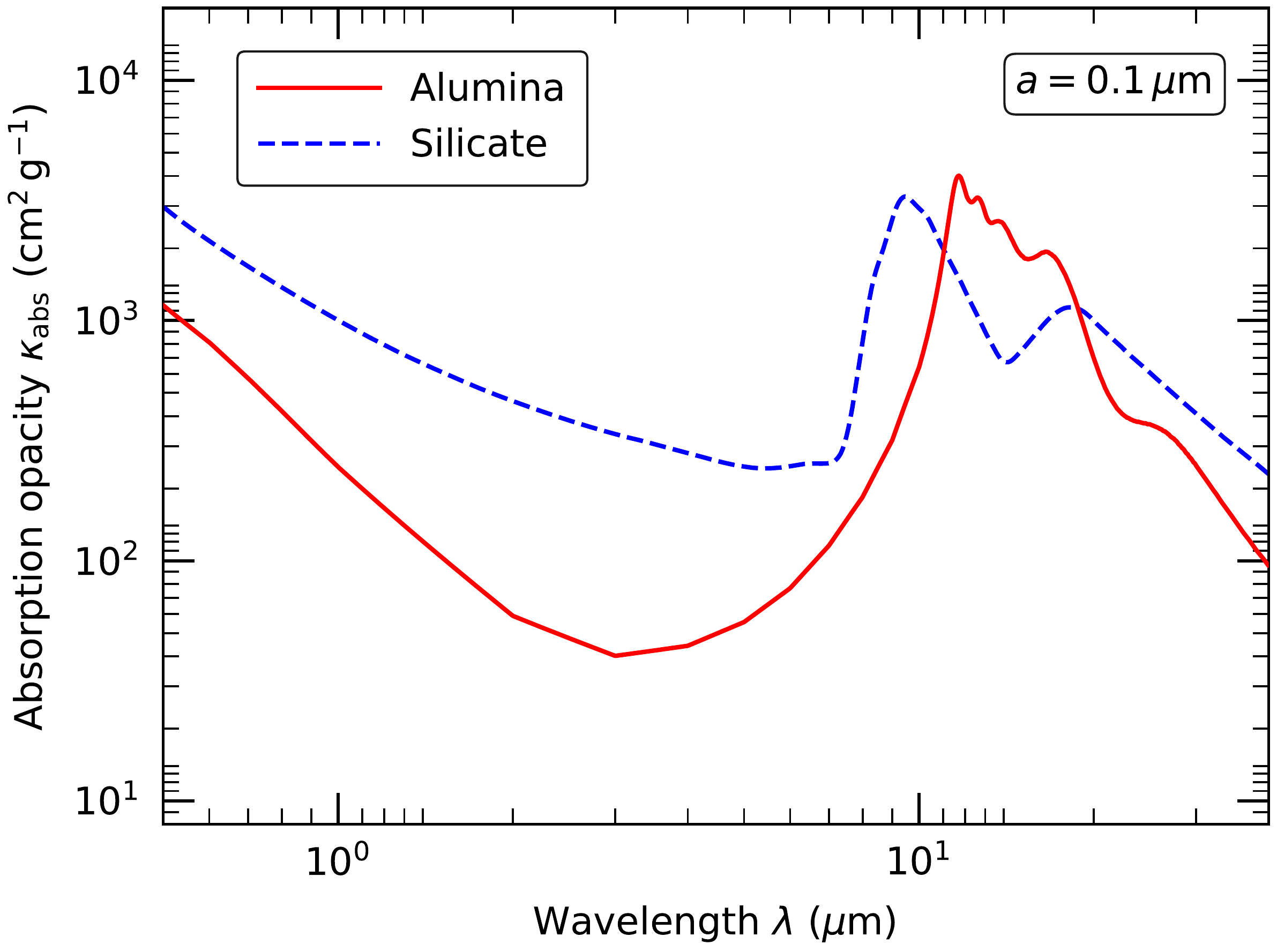}
    \end{subfigure}

    \caption{
    \textit{Left panel}: Cumulative radial column densities of the total dust,
    consisting of silicate and alumina, as a function of distance for all \mdot\  within the range of 
    $10^{-6}$--$10^{-2}\,M_{\odot}\,\mathrm{yr}^{-1}$, for both steady as well as accelerated wind. 
    \textit{Right panel}: Absorption coefficients, $\kappa_{\mathrm{abs}}(\lambda,a)$, of alumina and
    silicate dust as a function of wavelength for a grain radius of
    $a=0.1$ \mic. The silicate dust opacities are taken from \citet{dra07}.
    The alumina opacities were calculated using the package \texttt{OPTOOL}
    \citep{Dominik_2021a}, matching with the optical constants from \citet{koike_1995}.
    }
    \label{fig:opacity_and_dust_column_density}

\end{figure*}

\subsection{Dust formation in progenitor of SN~2023ixf}
\label{sec:dust_SN2023ixf}


In the previous sections, we assumed a constant mass loss rate, which does not reflect the realistic time-dependent mass loss variations that characterize the late stages of RSG evolution prior to a CCSN explosion.
To emulate the pre-explosion mass loss history of the progenitor of the nearby supernova SN~2023ixf, which exhibited clear signatures of a dusty CSM \citep{Kilpatrick2023, vandyk_2024}, we compile the dust and molecule formation in wind parcels with varying mass loss rates that collectively constitute the CSM of SN~2023ixf \citep{Sengupta_Sujit_Sarangi2026}. As seen in Figure~\ref{fig:mass-loss_rate_vs_time}, the mass loss rates during the last million years vary between 10$^{-6}$ and 10$^{-2}$ \Mdot, with \mdot\ reaching 10$^{-2}$ \Mdot\ only during two brief phases in the last decade before the explosion. In Sections~\ref{sec:dust_steady} and \ref{sec:dust_accelerated}, we described how dust forms over this range of \mdot; however, there we assumed \mdot\ to be constant. Our choice of \bet\ = 1.2 for the accelerated wind is motivated by the CSM profile of SN~2023ixf, and therefore, when modeling the progenitor of SN~2023ixf, we consider only the accelerated wind scenario.

Following individual parcels of gas, we construct the dust density in the CSM of SN~2023ixf progenitor, and compute the cumulative mass of dust enclosed within one parsec around the star.  In Figure~\ref{fig:sn2023ixf_total_dust_density_cumulative_mass}, we show the dust density and the cumulative dust mass in the CSM at 10$^4$, 10$^2$, and 4 yrs prior to CC. We find that dust formation occurs within two stellar radii, and therefore creates a dense dusty shell around the central RSG. The dust density keeps increasing gradually as the RSG reaches closer to CC; however, in the last 10$^4$ yrs we find the density to increase only by two times, while the geometry of the dusty shell remains almost consistent. The dust mass that could have been present in the CSM, before the explosion of SN~2023ixf is found to grow from 6$\times$10$^{-4}$ \Ms, 10,000 years before the explosion, to 1.5$\times$10$^{-3}$ \Ms\ before CC. However, not much change in dust density or dust masses in the CSM is found in the last 100 years. The two short phases of high mass loss (\about\ 10$^{-2}$ \Mdot) due to dynamical ejection of mass in the last 10 years (see Figure~\ref{fig:mass-loss_rate_vs_time}) did not contribute much to the dust in the CSM, since it takes about 6--10 years to form dust in the wind, even for high \mdot. 

The density of silicate dust in the inner CSM of SN~2023ixf progenitor, used by \cite{vandyk_2024}, to fits the UV-optical spectrum is between 10$^{-18}$ and 10$^{-17}$ g cm$^{-3}$. From our model, we find the dust density in the inner CSM (10 stellar radii) is between 10$^{-18}$ and 10$^{-16}$ g cm$^{-3}$, which is composed of both silicates and alumina, where silicates constitute more than 95\% of the total dust. The dust mass estimate from a fit depends on the geometry and assumed dust opacities. In the following section, we will derive synthetic spectra of an RSG using the dust yields from our models.

\section{The synthetic spectra}
\label{sec_synthetic_spectra}

The continuum part of a stellar spectrum is the combination of the emerging RSG spectrum after extinction by the circumstellar dust, and the re-radiated emission from dust that emerges after self-extinction. We have assumed a spherically symmetric geometry for the CSM. The necessary quantities to calculate the fluxes are (a) the unattenuated stellar spectrum, (b) densities for all individual dust compositions, (c) the corresponding column densities, (d) the optical constants for each dust type, (e) the distribution of dust temperature, and (f) the grain size distribution. 

We used the same unattenuated stellar spectrum that is used for modeling the gas kinetics. The RSG model is based on SN~2023ixf progenitor \citep{Sengupta_Sujit_Sarangi2026}, with a luminosity $L_{\star}$ of $\sim$ 10$^{5}$~\Ls, an effective temperature $T_{\star}$ of $\sim$ $~3067$~K, and stellar radius R$_{\star}$ of \about\ 10$^{14}$~cm. The dust densities for silicates ($\rho_{sil}$) and alumina ($\rho_{alu}$) are calculated in our study, as explained in Section \ref{sec:dust_general}. Integrating the dust densities over the radial bins at any given time, we get the dust column density. In Figure \ref{fig:opacity_and_dust_column_density}, the cumulative dust column densities are shown for various mass loss rates, for the steady and accelerated wind. We use the absorption coefficients $\kappa(\lambda)$ (shown in Figure \ref{fig:opacity_and_dust_column_density}) for Mg-silicates from \cite{dra07} and calculate the absorption coefficients for alumina using the code \texttt{OPTOOL} \citep{Dominik_2021a}, motivated by the optical constants derived by \cite{koike_1995}. The optical depth, $\tau_d(\lambda)$, as a function of radial distance and wavelength is given in Equation \ref{eq:opticaldepth}. 

\begin{equation}
\begin{split}
\label{eq:opticaldepth}
&\tau_d(\lambda, R) = \int^{R}_{R_{\star}} {\rm d}\tau_d(\lambda, r) \\
& = \int^{R}_{R_{\star}} \Big[\rho_{sil}(r)k_{sil}(\lambda) + \rho_{alu}(r)k_{alu}(\lambda)\Big] \mathrm{d}r  
\end{split}
\end{equation}

We have assumed a uniform grain size of radius $a = 0.1$~\mic, since in the Rayleigh limit ($\lambda_{IR} \gg a $) the grain size distribution does not significantly alter the IR luminosities, as also argued in previous studies \citep{fox_2010, shahbandeh_2023, sarangi_2025a}. 

\begin{figure*}[htbp]
    \centering

    \begin{subfigure}{0.48\textwidth}
        \centering
        \includegraphics[width=\linewidth]
        {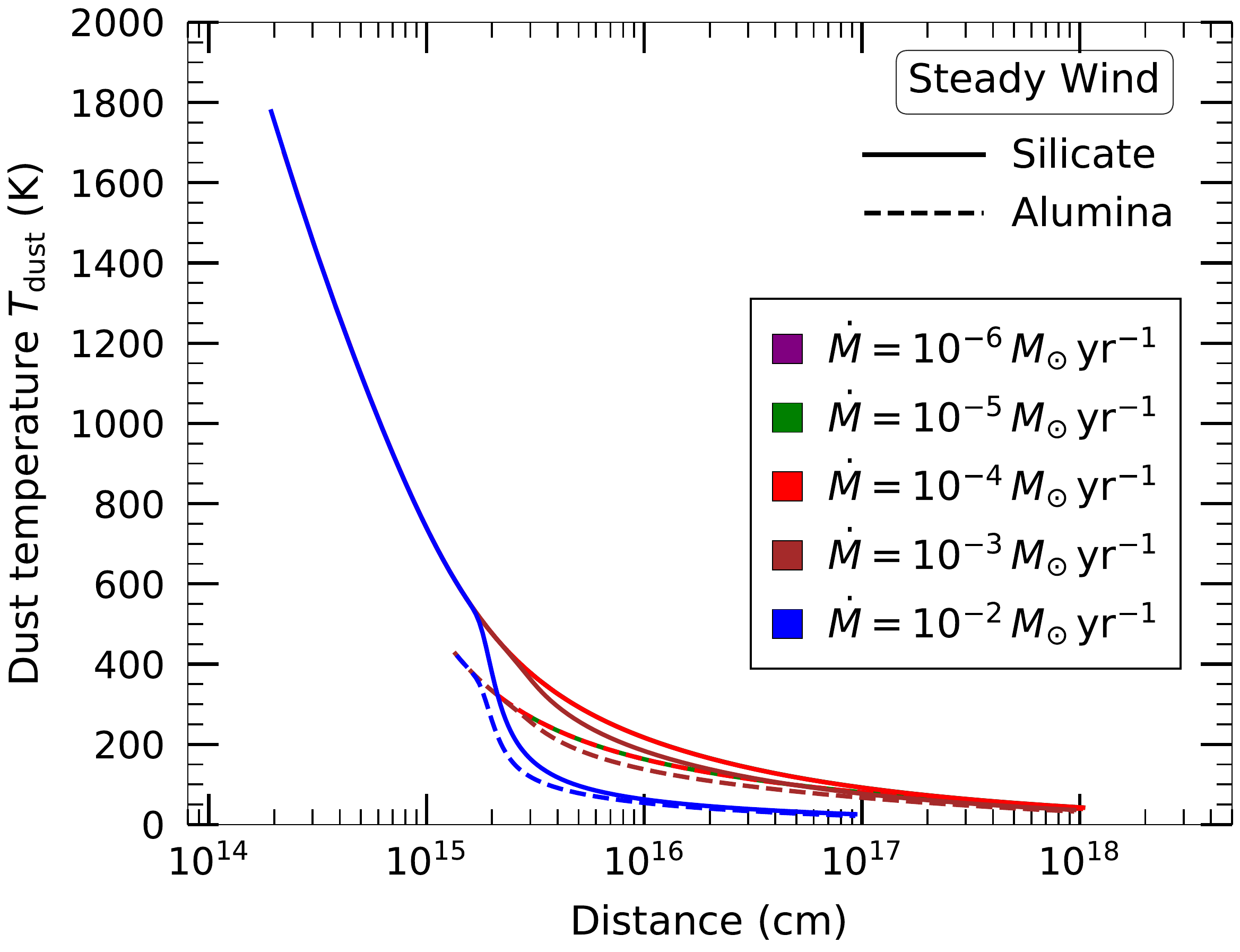}
    \end{subfigure}
    \hfill
    \begin{subfigure}{0.48\textwidth}
        \centering
        \includegraphics[width=\linewidth]
        {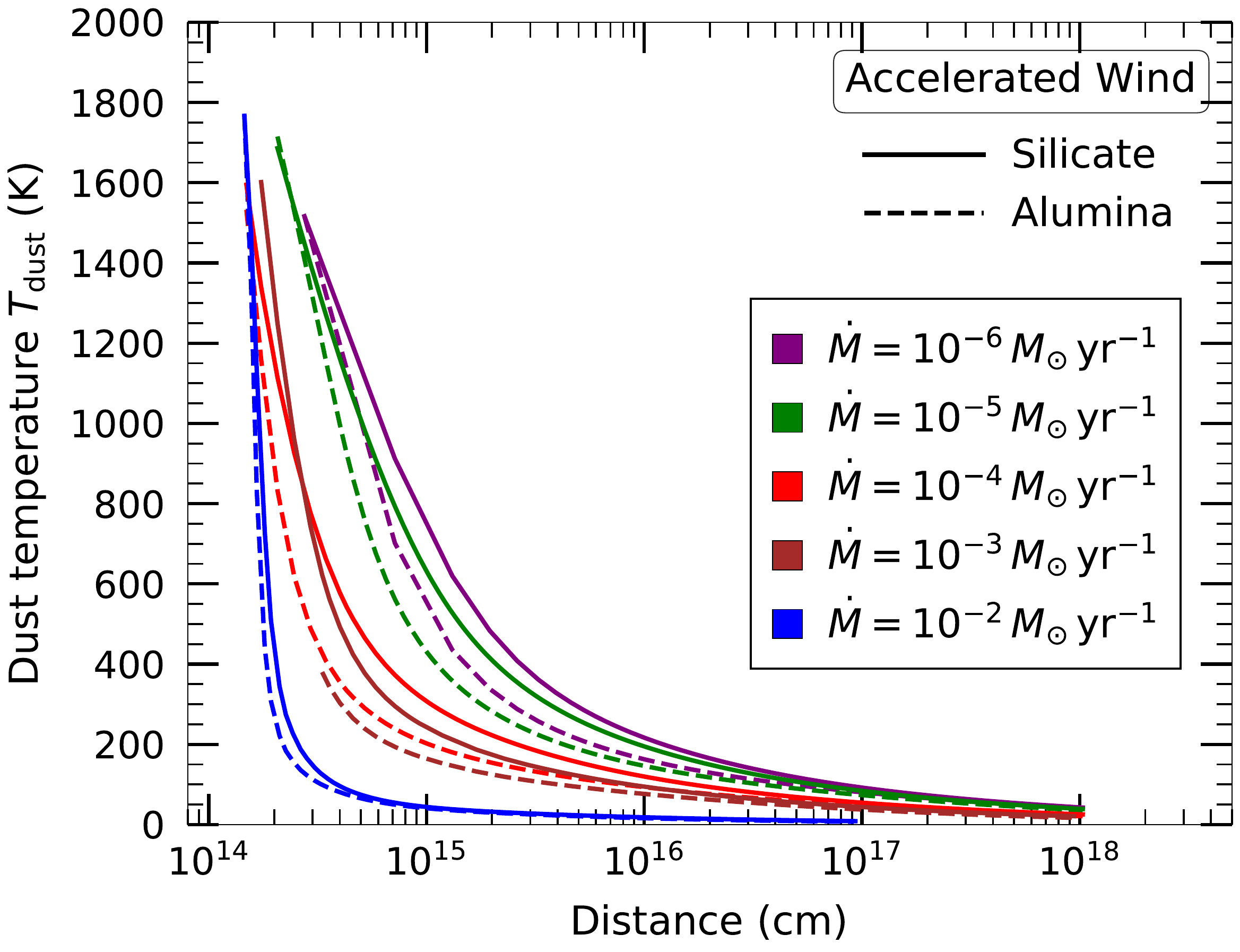}
    \end{subfigure}

    \caption{
    Radial temperature profiles of silicate and alumina dust for \mdot\ spanning
    $10^{-6}$--$10^{-2}\,M_{\odot}\,\mathrm{yr}^{-1}$, for constant (\textit{left panel}) and accelerated wind (\textit{right panel}) scenarios. See Equations~\ref{eq:dust_T} and \ref{eq:coolingrate} for referenece. 
    }
    \label{fig:dust_temperature_steady_accelerated_mdot}

\end{figure*}

To estimate the emerging fluxes, we calculate the dust temperature in each radial bin at each snapshot. The temperature of a silicate grain, $T_{sil,R}(R)$, at a given radius is calculated by the equilibrium temperature that matches the rate of absorption (heating) and the rate of emission (cooling). In a similar way, the temperature of an alumina grain, $T_{alu,R}(R)$, is also calculated. The rate of absorption of a silicate grain at distance $R$, when heated by the RSG star, depends on the flux of the star that is incident at radius $R$ after attenuation by the dust grains present between $R$ and $R_{\star}$. The attenuation is measured by the escape probability, $P_{esc}(\lambda)$, of any photon of wavelength $\lambda$ from a shell to the next. For a thin, concentric spherical shell, $P_{esc}(\lambda)$ is estimated as $e^{-\tau(\lambda)}$ \citep{inoue_2020, sarangi_2022b}. A numerical solution for $P_{esc}(\lambda)$ in a thin homogeneous shell is provided by \cite{dwek_2024}, however, we found that the numerical solution provides values identical to $e^{-\tau(\lambda)}$. The rate of absorption also depends on the geometric cross section of the grain of radius $a$, and the absorption coefficient $\kappa_{sil}(\lambda, a)$. The CSM at any given time $t$ in our model is constructed by a series of concentric spherical shells with unique dust densities $\rho_{sil}$ and $\rho_{alu}$. The resulting heating rate, $H_{sil}(R,t)$, for a silicate grain at distance $R$ is given by,

\begin{equation}
\label{eq:dust_T}
\begin{split}
H_{sil}(R,t) &= \int_{\lambda} H_{sil,\lambda}(R,t)\,\mathrm{d}\lambda \\
     &= \int_{\lambda}
        \frac{R_{\star}^2}{R^2}
        \pi \mathrm{B}_{\lambda}(\lambda,T_{\star})
        \exp\left[
        -\int_{R_{\star}}^{R}\mathrm{d}\tau_d(\lambda,r)
        \right] \\
     &\quad \times \pi a^2
        \frac{4a\Omega_{\rm sil}}{3}\,
        \kappa_{sil}(\lambda,a)\,\mathrm{d}\lambda,
\end{split}
\end{equation}

where $\mathrm{B}_{\lambda}(\lambda,T)$ is the Planck function, and $\Omega$ is the material density, taken as 3.3 and 3.5 g cm$^{-3}$ for silicate and alumina, respectively. The grain radius $a$ was taken as 0.1 \mic\ in all cases. The rate of cooling, $L_{sil}(R,t)$, of a single silicate dust grain at $R$ is given by, 

\begin{equation}
\label{eq:coolingrate}
\int L_{sil,\lambda}(R,t)\mathrm{d}\lambda  = \int 4 m_d \kappa_{sil}(\lambda,a) \pi \mathrm{B}_{\lambda}(\lambda,T_{sil}(R)) \mathrm{d}\lambda,
\end{equation}

 where $T_{sil}(R)$ is the dust temperature at $R$. Equating $H_{sil}(R,t)$ and $L_{sil}(R,t)$, we find the temperature at each $R$ for silicate, and similarly for alumina dust. In Figure~\ref{fig:dust_temperature_steady_accelerated_mdot}, dust temperatures are presented as a function of $R$ for all \mdot values in the scenarios of steady (\bet\ = 0) and accelerated (\bet\ = 1.2) winds. 

 We set the outer radius of the CSM $R_{out} = 1$ parsec. The emerging spectrum at $R_{out}$ at any given time is the combination of the unattenuated stellar spectra that is not absorbed by dust and the re-radiated emission by dust in the CSM. Integrating over all the radial shells with unique dust temperature, dust density, and optical depths, we get the emerging SED for silicate ($F_{\lambda}^{sil}$) and alumina ($F_{\lambda}^{alu}$) at any given time. The attenuated stellar SED ($F_{\lambda}^{S}$) is a result of absorption by the dusty CSM, measured by the total optical depth of the CSM. 
 The resulting SED, $F_{\lambda}$, at a distance of $D$ from the star, is calculated by Equation \ref{eq:emflux} below. 

\begin{equation}
\begin{split}
\label{eq:emflux}
F_{\lambda}^{S}(\lambda,t)  = & \frac{R_{star}^2}{D^2} \pi \mathrm{B}_{\lambda}(\lambda,T_{star}) \exp\left[-\int_{R_{star}}^{R_{out}}\mathrm{d}\tau_d(\lambda,r) \right] \\
F_{\lambda}^{sil}(\lambda,t) & = \frac{1}{D^2}\int^{R^{out}}_{R_{\star}} 4 r^2 \rho_{sil}(r) \kappa_{sil}(\lambda) {\rm d}r \\
& \times \pi \mathrm{B}_{\lambda}(\lambda,T_{sil}(r)) \exp\left[-\int_{r}^{R_{out}}\mathrm{d}\tau_d(\lambda,r) \right] \\
F_{\lambda}^{alu}(\lambda,t) & = \frac{1}{D^2}\int^{R^{out}}_{R_{\star}} 4 r^2 \rho_{alu}(r) \kappa_{alu}(\lambda) {\rm d}r \\
& \times \pi \mathrm{B}_{\lambda}(\lambda,T_{alu}(r)) \exp\left[-\int_{r}^{R_{out}}\mathrm{d}\tau_d(\lambda,r) \right] \\
F_{\lambda}(\lambda,t) & = F_{\lambda}^{S}(\lambda,t) + F_{\lambda}^{sil}(\lambda,t) + F_{\lambda}^{alu}(\lambda,t)
\end{split}
\end{equation}

The time dependence of the SED reflects the time-dependent dust density and stellar properties such as luminosity, temperature, and mass loss rate. We have not accounted for the effects of scattering; given the red-dominated spectrum of the RSG, scattering is expected to be minimal.   

\begin{figure*}[htbp]
    \centering

    \begin{subfigure}{0.48\textwidth}
        \centering
        \includegraphics[width=\linewidth]{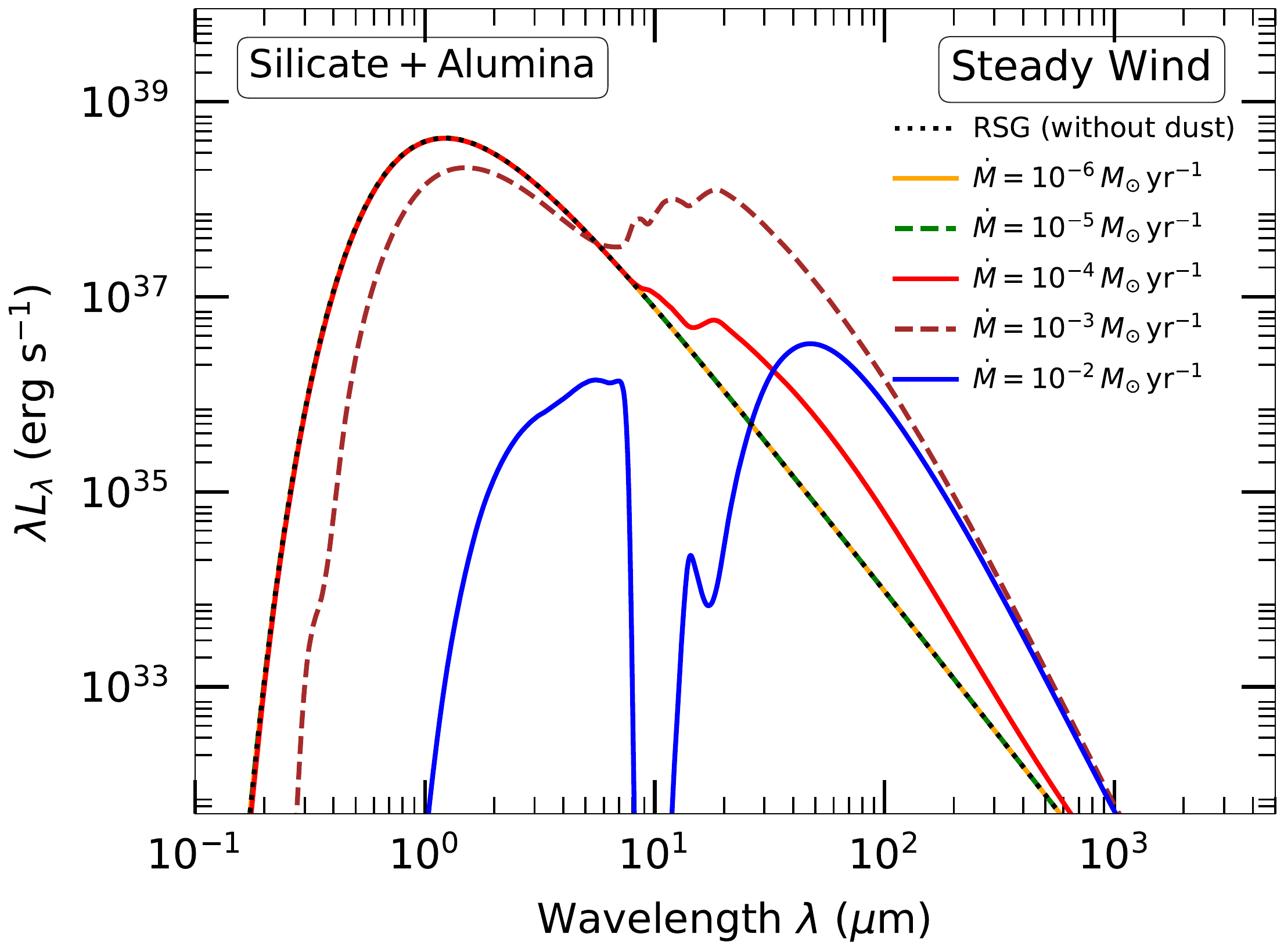}
    \end{subfigure}
    \hfill
    \begin{subfigure}{0.48\textwidth}
        \centering
        \includegraphics[width=\linewidth]{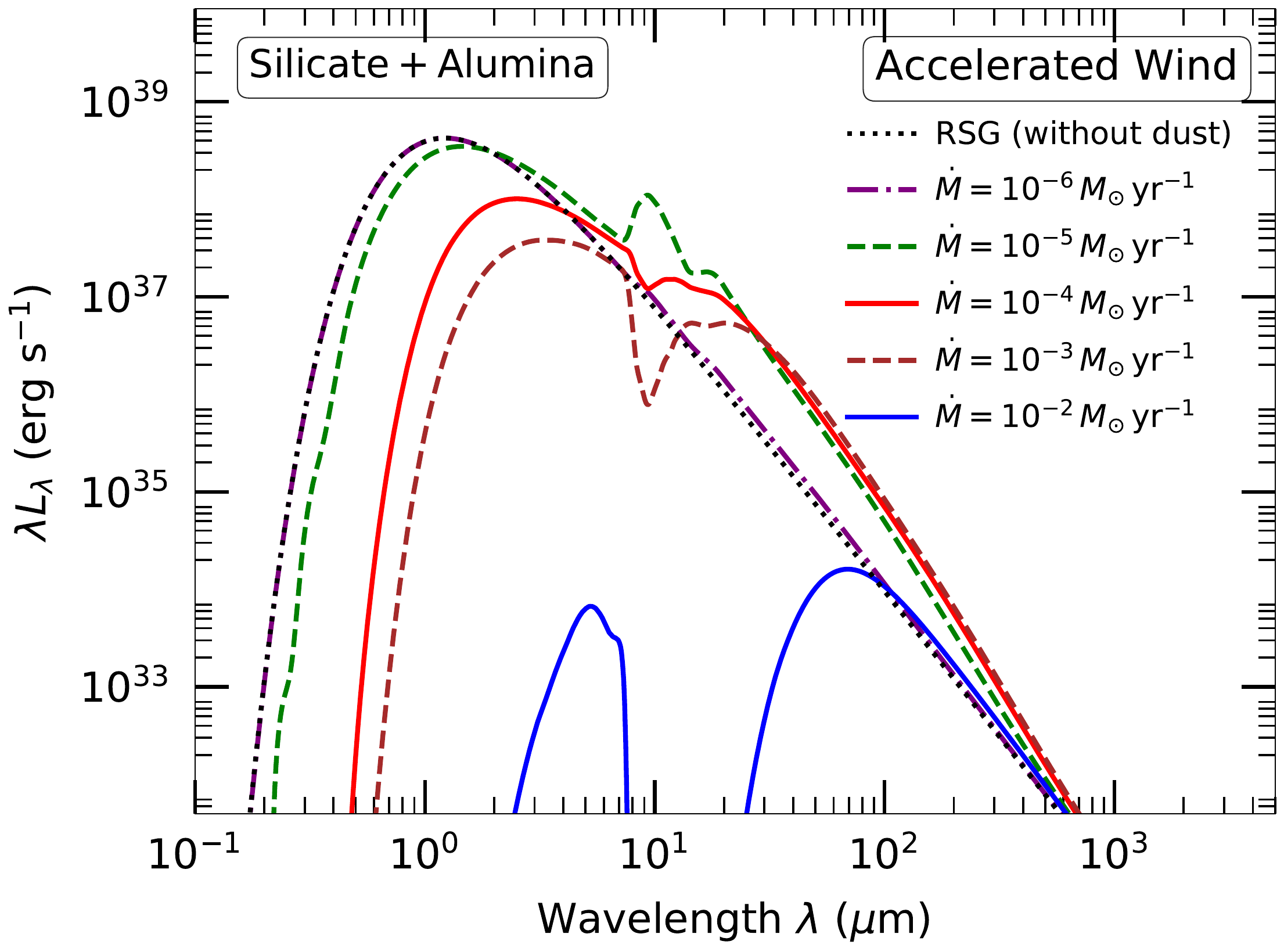}
    \end{subfigure}

    \caption{
    Emergent SEDs for constant (\textit{left panel}) and accelerated (\textit{right panel}) wind cases for all \mdot\ in the range of 10$^{-6}$ to 10$^{-2}$ \Mdot. The corresponding dust densities are shown in Figure~\ref{fig:dust_density_silicate_alumina} and dust temperatures in Figure~\ref{fig:dust_temperature_steady_accelerated_mdot}, and the formulation is explained in Section \ref{sec_synthetic_spectra}. 
    }
    \label{fig:sed_constant_beta}

\end{figure*}



\subsection{Emerging SED for steady wind}
\label{sec:SED_steadywind}

Figure~\ref{fig:dust_temperature_steady_accelerated_mdot} (left panel) shows the dust temperatures for \mdot\ = 10$^{-6}$ -- 10$^{-2}$ \Mdot\ in a radial distance from the centre of the star. As expected, the dust temperature drops with distance from the star, for both silicates and alumina. In addition, dust temperature in a given parcel of gas also reflects the optical depths due to layers of dust present between that parcel and the stellar surface. When we assume \mdot\ = 10$^{-2}$ \Mdot, the dust densities are largest near 2$\times$10$^{15}$ cm (see Figure \ref{fig:dust_density_silicate_alumina}), and that is reflected in the sudden drop in dust temperature beyond this radius due to large optical depths. We find the dust temperatures to vary between 600 and 50~K for all \mdot. The larger the \mdot, the faster the temperature drops along the radial direction. This is because a larger \mdot\ leads to a larger dust mass (see Figure~\ref{fig:dust_formation_time_and_mass}), and so larger optical depths. 

The resulting SED, given by Equation \ref{eq:emflux}, is shown in Figure~\ref{fig:sed_constant_beta} (left panel). For \mdot\ = 10$^{-6}$,10$^{-5}$ \Mdot, the dust mass is negligible, so the SED resembles the RSG. For \mdot\ = 10$^{-4}$ \Mdot, the SED shows a minor excess in the IR. The IR becomes prominent at \mdot\ = 10$^{-3}$ \Mdot, with considerable attenuation of the original stellar spectrum. A combination of silicate and alumina dust features is noticeable in the spectrum. Finally, when \mdot\ = 10$^{-2}$ \Mdot, the stellar spectrum is attenuated by more than three orders of magnitude. The SED has a sharp drop at 9.7 \mic\ representing the silicate absorption feature. The self-attenuation of the IR emission within the dusty CSM leads to an overall lower luminosity.   

\subsection{Emerging SED for accelerated wind}
\label{sec:SED_acceleratedwind}

Dust forms close to the star when we consider the accelerated wind. Dust masses are also larger than the steady wind case (see Figure~\ref{fig:dust_formation_time_and_mass}). In the innermost layers of the CSM, the dust temperatures are large and reach close to the sublimation temperature, as presented in Figure~\ref{fig:dust_temperature_steady_accelerated_mdot} (right panel). Due to larger dust densities, the optical depths are also large, hence the temperature drops fast as a function of radial distance. The dust temperatures vary between 1800 and 50~K for all \mdot. The drop is steepest for \mdot\ = 10$^{-2}$ \Mdot\ and gradually becomes shallower with decreasing \mdot.

Figure~\ref{fig:sed_constant_beta} (right panel) presents the model SEDs for all \mdot\ values ranging between 10$^{-6}$ and 10$^{-2}$ \Mdot. For \mdot\ = 10$^{-6}$ \Mdot, the original RSG spectrum mostly retains its shape, while the IR emission from dust remains minimal. At \mdot\ = 10$^{-5}$ \Mdot, we find moderate attenuation of the original stellar spectrum, along with a clear silicate emission from the CSM. When \mdot\ is increased to 10$^{-4}$ \Mdot, the attenuation of the RSG spectrum reaches about one order of magnitude, and the overall spectrum develops a strong IR excess. However, the silicate feature at 9.7 \mic\ becomes flatter due to self-absorption in the dusty CSM. For \mdot\ = 10$^{-3}$ \Mdot, in addition to the attenuation of the original RSG spectrum, the model SED shows strong silicate absorption due to self-absorption. At the highest \mdot\ of 10$^{-2}$ \Mdot, the spectrum shows significant attenuation of more than three orders of magnitude in both the optical and mid-IR, with only the far-IR emission remaining prominent. The strong silicate absorption at 9.7 \mic\ produces a pronounced dip in the emergent model SED.

\begin{figure}[h]
    \centering
    \includegraphics[width=0.48\textwidth]
    {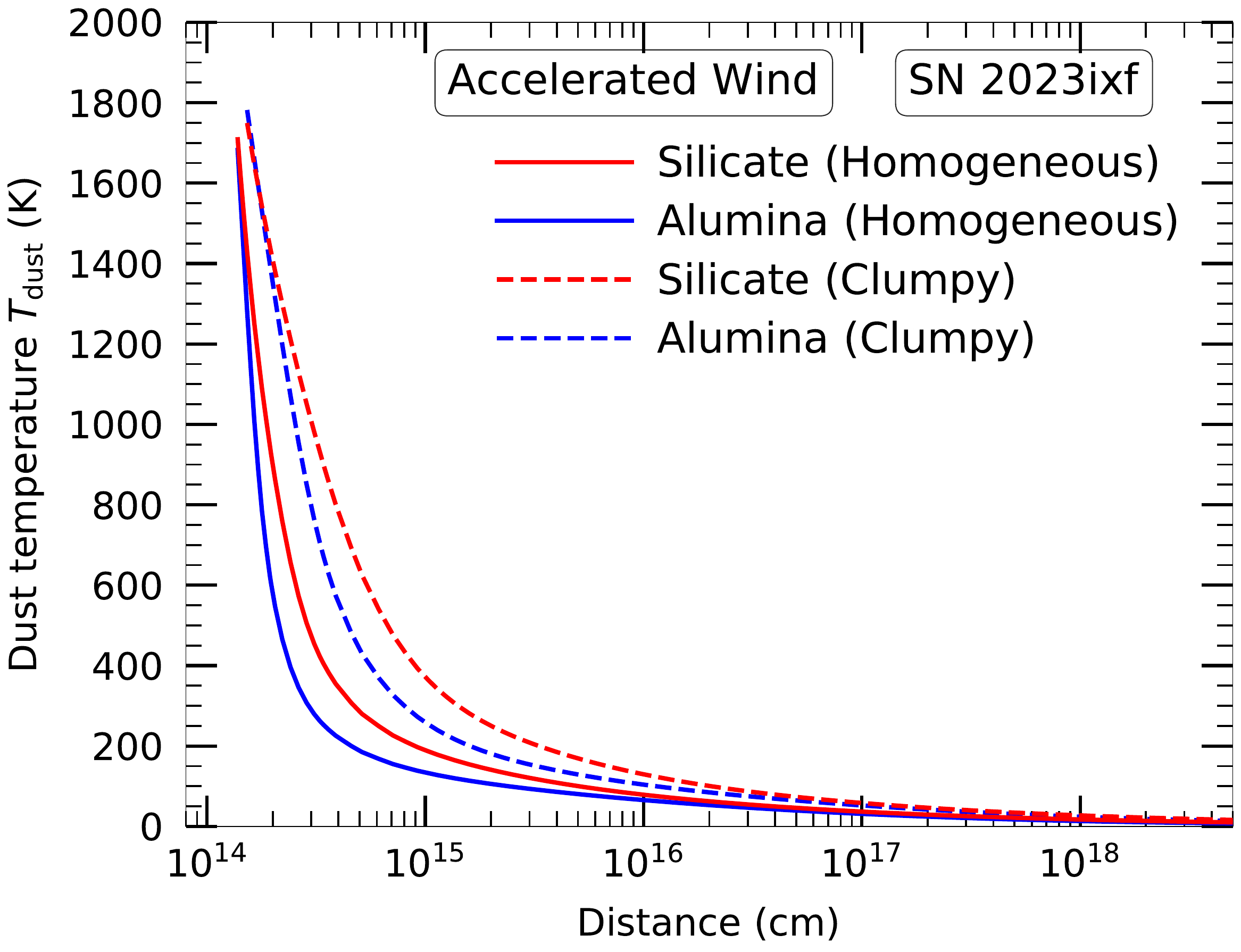}

    \caption{
    Radial temperature profiles of silicate and alumina dust in the CSM of 
    SN~2023ixf progenitor, at 4 years before CC. We show the cases of a homogeneous as well as a clumpy CSM 
    (using filling factor $f_{\rm cl}=0.1$ and clump radius $r_{\rm cl}=0.1r$). 
    }
    \label{fig:sn2023ixf_dust_temperature_homogeneous_clumpy}

\end{figure}

\begin{figure*}[htbp]
    \centering

    \begin{subfigure}{0.48\textwidth}
        \centering
        \includegraphics[width=\linewidth]
        {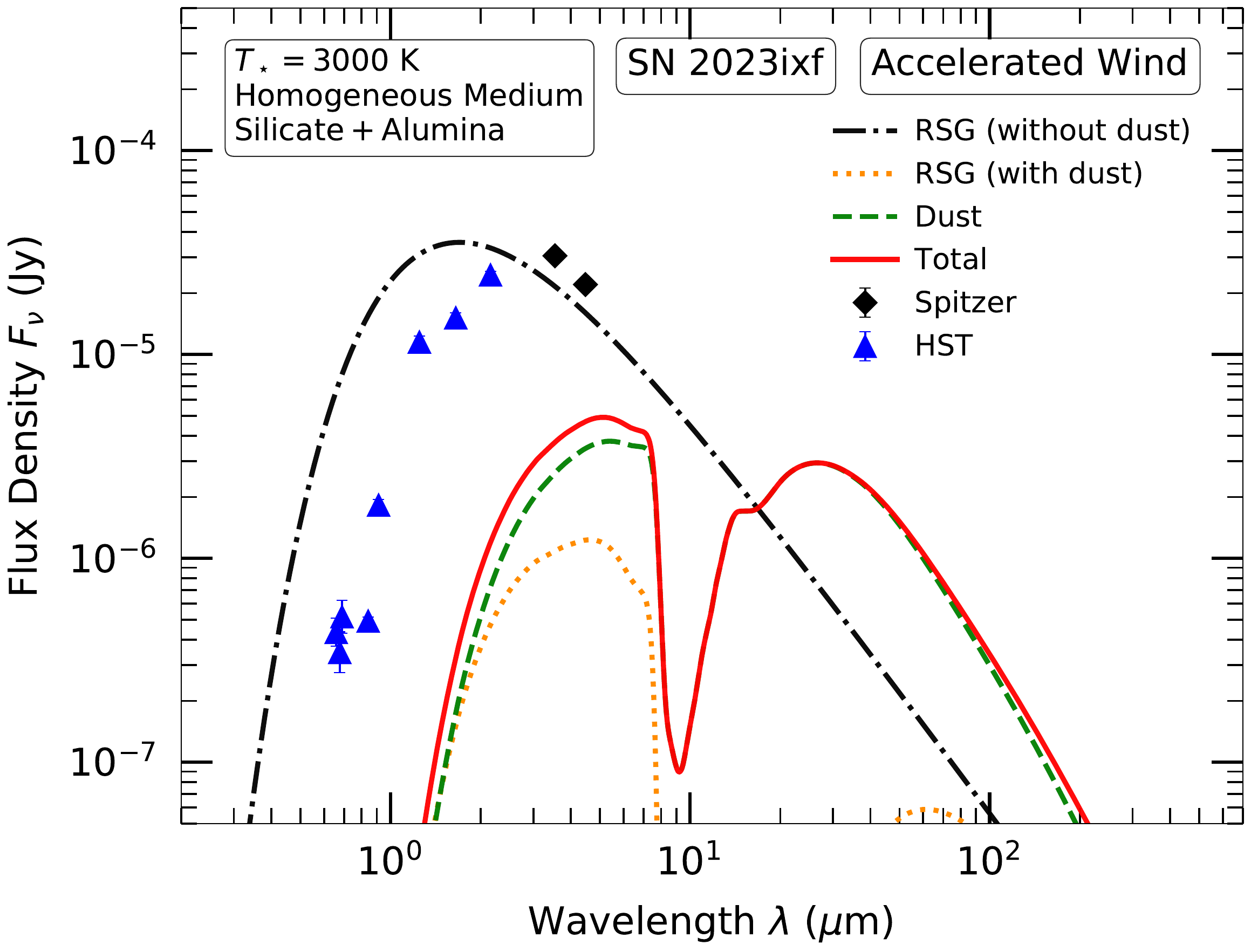}
    \end{subfigure}
    \hfill
    \begin{subfigure}{0.48\textwidth}
        \centering
        \includegraphics[width=\linewidth]
        {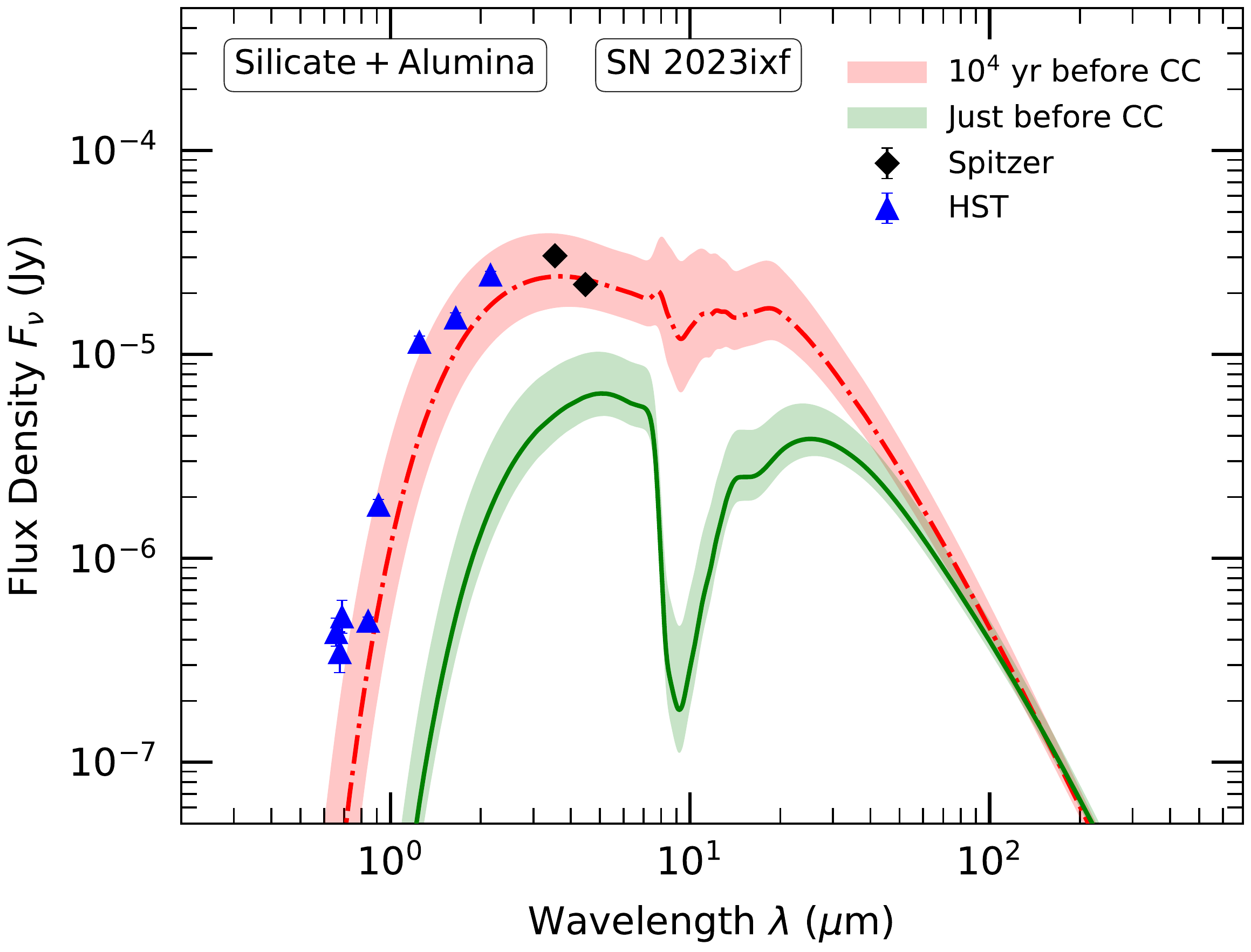}
    \end{subfigure}

    \caption{
    Synthetic SED of the SN~2023ixf progenitor is compared to the \texttt{HST} optical/NIR photometric data and the \texttt{Spitzer}
    3.6 and 4.5~$\mu$m fluxes taken from \citet{vandyk_2024}. The corresponding dust densities and temperatures are given in Figure~\ref{fig:sn2023ixf_total_dust_density_cumulative_mass} and \ref{fig:sn2023ixf_dust_temperature_homogeneous_clumpy} respectively. \textit{Left panel}: The original RSG spectrum, the attenuated RSG spectrum, and the reprocessed radiation from dust are shown alongside the expected SED that emerges from the CSM. \textit{Right panel}: The expected range of SED is shown as bands when we vary the $T_{\star}$ from 2800~K to 3300~K, for times 4 years before CC, and 10$^4$ years before CC. The solid line corresponds to $T_{\star}$ = 3000~K. See Section \ref{sec_sed_2023ixf} for details. The dust temperatures are calculated uniquely for each $T_{\star}$. 
}
    \label{fig:sn2023ixf_homogeneous_sed_silicate_alumina}

\end{figure*}

\subsection{The spectra of the progenitor of SN~2023ixf}
\label{sec_sed_2023ixf}

To compare our results with the progenitor of SN~2023ixf, we use a distance of $D=6.9$~Mpc \citep{JacobsonGalan2025}. 
The optical/NIR and IR photometry data provided by \cite{vandyk_2024}, obtained by \texttt{Hubble Space Telescope (HST)} and \texttt{Spitzer} 3.6 and 4.5~$\mu$m filters, corresponding to a time about 4 years before the explosion, were used as a reference.   
The dust densities in the CSM of the SN~2023ixf progenitor, shown in Figure~\ref{fig:sn2023ixf_total_dust_density_cumulative_mass}, is used in Equation \ref{eq:emflux} to derive the model SED at various epochs pre-explosion. 

In Figure~\ref{fig:sn2023ixf_dust_temperature_homogeneous_clumpy}, the synthetic dust temperatures are shown for silicates and alumina as a function of the radial distance. The temperature is found to drop rapidly from about 1800~K to 200~K within ten stellar radii. This indicates that most of the IR emission will originate from the region close to the star. Figure \ref{fig:sn2023ixf_homogeneous_sed_silicate_alumina} (left panel) presents our model spectrum, alongside the observed fluxes of SN~2023ixf progenitor \citep{vandyk_2024}. The original, unattenuated stellar spectrum is also shown, reflecting the substantial attenuation expected in our model. There is also a very prominent silicate absorption feature predicted in the SED; however, observations in the \texttt{Spitzer} bands cannot be used to verify this feature. We find that, due to self-attenuation in the dusty CSM, our model SED at 4~years before the explosion does not match well with the observed optical/IR fluxes. In Figure \ref{fig:sn2023ixf_homogeneous_sed_silicate_alumina} (right panel), the model SEDs for 10$^4$~years pre-explosion and 4~years pre-explosion are presented, where we have varied the stellar surface temperature $T_{\star}$ between 2800 and 3300~K. Coincidentally, the observed fluxes match our model SED generated by a CSM dust density about 10$^4$~years pre-explosion. The total dust mass in the CSM for such epochs is estimated as 7$\times$10$^{-4}$ \Ms\ in our model. As the dust mass and density in the CSM increased in the last thousand years, the self-absorption became more prominent, and the model fluxes in the optical and mid-IR systematically decreased.

We explored possible changes in the SED by assuming a clumpy CSM \citep{hillel_2025}. There is growing observational evidence that the CSM around RSGs is clumpy in nature, as also confirmed for SN progenitors and SN remnants, e.g., VY CMa \citep{kaminski_2019}, the progenitor of SN~2024qiw \citep{nagao_2026}, and Cassiopeia~A \citep{weil_2020b}. 

Based on the treatment of opacities in a clumpy medium by \cite{VarosiDwek1999}, we construct the CSM assuming a volume filling factor $f_{\rm cl}$ and the clump radius at $r$ written as $r_{\rm cl}=qr$ ($q$ being a factor less than 1). For one clump located at distance $r$, the optical depth, $\tau_{\rm cl}(\lambda,r)$, is given as,

\begin{equation}
\tau_{\rm cl}(\lambda,r) = r_{\rm cl} f_{\rm cl}^{-1} \times \big(\rho_{sil}(r) \kappa_{sil}(\lambda) + \rho_{alu}(r) \kappa_{alu}(\lambda)\big), 
\end{equation}
considering that the density inside the clump has been enhanced by a factor of $f_{\rm cl}^{-1}$. We use the mega-grain approximation \citep{VarosiDwek1999, inoue_2020}, assuming an individual clump acts like an isolated dust grain. In this approximation, the effective optical depth of the clumpy medium, $\tau_{eff}(\lambda, R)$, from the surface of the star to any given radius $R$ is given by, 

\begin{equation}
\tau_{eff}(\lambda,R) = \int_{R_{\star}}^{R} n_{\rm cl} \pi r_{\rm cl}^2 Q_{\rm cl}(\lambda) {\rm d}r = \int\frac{3f_{\rm cl}}{4r_{\rm cl}} P_{\rm int}(\tau) {\rm d}r
\end{equation}
where the number density of clumps is expressed as $n_{\rm cl} = f/v_{\rm cl}$, $v_{\rm cl} = (4/3) \pi r_{\rm cl}^3$, and $r_{\rm cl}=qr$. The probability of interaction of a photon of wavelength $\lambda$ with the clump is given by $Q_{\rm cl}(\lambda)$, which is derived in \cite{VarosiDwek1999} as $P_{\rm int}$, 

\begin{equation}
P_{\rm int}(\tau_{\rm cl}) = 1-\frac{1}{2\tau_{\rm cl}^2} + \Big(\frac{1}{\tau_{\rm cl}} + \frac{1}{2\tau_{\rm cl}^2}\Big)e^{-2\tau_{\rm cl}}
\end{equation}

\begin{figure*}[htbp]
\centering

\begin{subfigure}{0.48\textwidth}
  \centering
  \includegraphics[width=\linewidth]
  {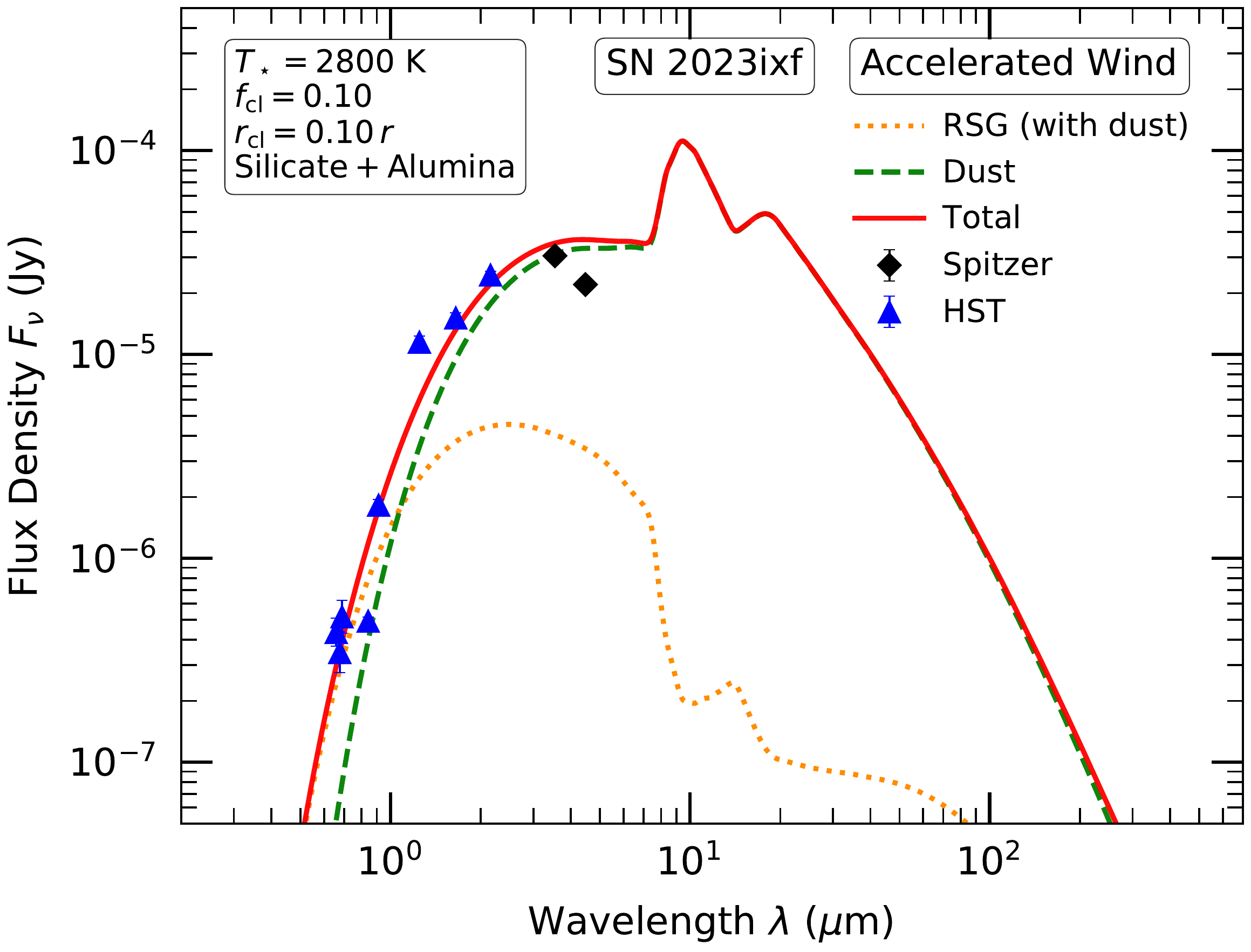}
\end{subfigure}
\hfill
\begin{subfigure}{0.48\textwidth}
  \centering
  \includegraphics[width=\linewidth]
  {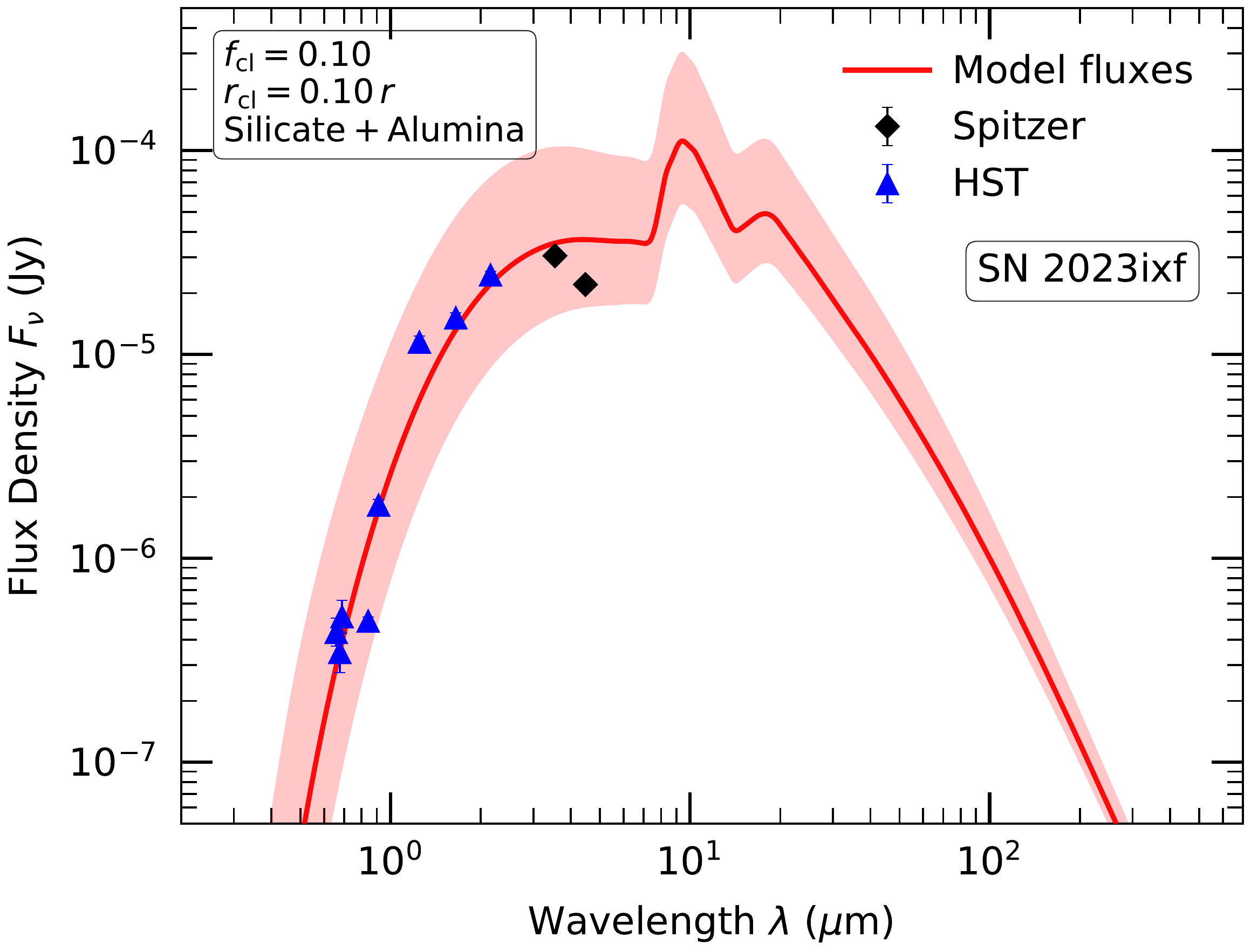}
\end{subfigure}

\caption{
Model SED of the SN~2023ixf progenitor
computed for a clumpy CSM containing silicate and alumina
dust, using a volume filling
factor $f_{\rm cl}=0.1$ and characteristic clump radius
$r_{\rm cl}=0.1r$. The synthetic spectrum is compared to the \texttt{HST} optical/NIR photometric data and the \texttt{Spitzer}
    3.6 and 4.5~$\mu$m fluxes taken from \citet{vandyk_2024}. The formalism is explained in Section \ref{sec_sed_2023ixf}. 
\textit{Left panel}: Model SED for clumpy CSM and $T_{\star}=2800$~K, showing the attenuated stellar
spectrum, thermal dust emission, and total emergent SED. 
\textit{Right panel}: Total model SED for $T_{\star}$ in range between
2500--3300~K. The shaded band shows the range of
SEDs over this temperature interval, while the solid curve shows
the $T_{\star}=2800$~K.
}
\label{fig:sn2023ixf_clumpy_sed_silicate_alumina}

\end{figure*}

For the clumpy medium, we use the effective optical depth in Equation \ref{eq:emflux} and estimate the emerging SED. When we assume a clumpy nature for the CSM, the optical depth is reduced compared to the homogeneous medium. As a result, dust temperatures drop at a slower rate with distance, as seen in Figure~\ref{fig:sn2023ixf_dust_temperature_homogeneous_clumpy}. 

In Figure~\ref{fig:sn2023ixf_clumpy_sed_silicate_alumina} (left panel), our estimated SED for a time 4 years before the explosion is compared with the observed UV/optical and mid-IR fluxes from a similar time. The attenuation of the original stellar spectrum is also shown in the same figure, along with the emission reradiated by the dust. In this case, the model SED from the dusty CSM matches the observed fluxes very well. We explored various filling factors and clump radii and found that the best match is obtained for $f_{\rm cl}$ = 0.1 and $r_{\rm cl}$ = 0.1$r$. To be clear, there is some degeneracy between $f_{\rm cl}$ and $r_{\rm cl}$ when evaluating the emergent SED. However, in this work, we do not aim to determine the exact filling factor or clump size, but rather to qualitatively explain the origin of the observed fluxes in a dusty CSM. Our resulting fluxes do not match the 4.5 \mic\ flux from the SN~2023ixf progenitor particularly well. The comparison in this band indicates that the dust may be somewhat cooler than estimated by our model. The silicate features (9.7 \mic\ and 18 \mic) in the clumpy CSM appear in emission, in contrast to the homogeneous case, where the silicate features appear in absorption. 

In Figure~\ref{fig:sn2023ixf_clumpy_sed_silicate_alumina} (right panel), we show the variation in the SED with stellar temperature, ranging from 2500~K to 3300~K. We find that the best match to the observed SED is obtained at 2800~K, while our stellar model for the SN~2023ixf progenitor has a $T_{\star}$ of 3067~K. When deriving the X-ray column densities \citep{Sengupta_Sujit_Sarangi2026}, our models also suggested that a clumpy CSM provides a better fit to the observed column densities.

\section{Summary}
\label{sec_summary}


\begin{itemize}
    \item[] In this work, we have investigated the formation of molecules and dust in the winds of RSG stars of solar metallicity and demonstrated how late-stage mass loss can naturally lead to the formation of a dense, dust-rich CSM before CC. We couple time-dependent mass loss with the chemistry of the stellar wind and use a non-equilibrium kinetic approach to model the simultaneous formation of molecules, molecular clusters, and dust precursors. We explore mass loss rates ranging from 10$^{-6}$ to 10$^{-2}$ \Mdot\ and two wind acceleration laws, \bet\ = 0 (constant wind velocity) and 1.2 (accelerated wind velocity). Their effects on the molecular and dust yields, dust temperatures, and emergent SEDs are the primary focus of this paper. A schematic overview of our model is shown in Figure \ref{fig:conceptual_overview}. 

    \item[] We find that molecules such as CO, H$_2$O, SiO, HCN, CS, SO, and NH$_3$ can form in the CSM, along with H$_2$ and O$_2$. SiO is primarily confined to the region close to the star and becomes depleted as it is incorporated into dust at larger radii. The resulting molecular masses span a wide range, from 10$^{-15}$ to 10$^{-2}$ \Ms. The total dust mass in the CSM varies from 10$^{-8}$ to 3$\times$10$^{-3}$ \Ms\ over the explored range of \mdot. Silicates and alumina are the primary dust components, with trace amounts of other O-rich dust species also forming. However, we do not find viable pathways for the formation of C-rich dust in this O-rich CSM.

    \item[] The wind acceleration strongly influences the location and efficiency of dust formation. For \bet\ = 0, the wind velocity is constant, whereas for \bet\ = 1.2, the wind gradually accelerates from near-zero velocity close to the stellar surface to a terminal velocity of 20 km s$^{-1}$. In the accelerated wind, dust can form as close as two stellar radii, resulting in larger dust masses than in the constant-wind case. For \bet\ = 0, dust formation occurs predominantly at $\sim$10 stellar radii and beyond, where the lower gas densities make dust formation less efficient. In particular, for \mdot\ = 10$^{-6}$ \Mdot\ with a constant wind, the amount of dust formed is almost negligible.

    \item[] The dust temperatures of both silicates and alumina decrease steeply with radial distance because of the large optical depths of the dusty shells between a given location and the stellar surface. The dust temperatures range from about 1800~K to as low as 50~K. The average temperature is lower in the constant-wind case because dust forms farther from the star. The predicted SEDs for both wind profiles show a clear mid-IR excess produced by dust in the CSM. At lower \mdot, dust extinction is weak, and the emergent spectrum retains a substantial contribution from the original RSG emission. At higher \mdot, the SED becomes increasingly dominated by radiation reprocessed by the CSM.

    \item[] We find that the well-known silicate features at 9.7 and 18 \mic\ can appear either in emission or absorption, with their strength depending primarily on the optical depth of the CSM. At high \mdot\ of 10$^{-2}$ and 10$^{-3}$ \Mdot, strong self-absorption of IR photons results in silicate features appearing predominantly in absorption. At lower \mdot\ of 10$^{-4}$ and 10$^{-5}$ \Mdot, the lower optical depths allow the silicate features to appear in emission. These features are more pronounced in the accelerated-wind case because dust forms closer to the star, where the temperatures are favorable for producing strong silicate emission.

    \item[] In the accelerated-wind models, the optical depth of the CSM can become large because dust forms close to the stellar surface, where the wind density is high. This produces substantial self-extinction in the emergent SED and leads to a gradual decline in the near- and mid-IR luminosities with increasing \mdot, despite the increasing dust mass. The energy absorbed by the dusty CSM is instead expected to emerge predominantly at longer wavelengths (submm).

    \item[] Finally, we model a more realistic time-dependent mass loss history corresponding to the progenitor of SN~2023ixf \citep{Sengupta_Sujit_Sarangi2026}. Based on the observed column densities in the CSM, an accelerated wind profile with \bet\ = 1.2 provides the most appropriate description of the wind structure. We find a gradual increase in the CSM dust mass from 6$\times$10$^{-4}$ \Ms\ at 10$^4$ yr before the explosion to 1.5$\times$10$^{-3}$ \Ms\ immediately before the explosion. Dust forms close to the stellar surface, and the dust density decreases steadily from approximately two stellar radii to about 0.1 parsec.

    \item[] We compare our model-predicted SED with the observed optical and IR fluxes of the SN~2023ixf progenitor \citep{vandyk_2024}. The model SED for the dense dusty CSM does not reproduce the observed fluxes particularly well at 4 yr before the explosion, whereas the model corresponding to a time of 10$^4$ yr before the explosion provides a somewhat better match. We show that a clumpy CSM can reproduce the observed fluxes much more closely. A filling factor of approximately 0.1 and a stellar surface temperature of about 2800~K provide a suitable match to the observed fluxes.

\end{itemize}

\section{Discussion}
\label{sec_discussion}

The mass-loss rate, $\dot M$, of a RSG directly influences the formation of its dusty CSM. The expected mass-loss rates for RSGs are in the range of 10$^{-9}$ to 10$^{-5}$ \Mdot, with an average of about 10$^{-6}$ \Mdot \citep{antoniadis_2024, fuller_2024, Moriya2018}. Dust production rates are estimated to be of the order of 10$^{-9}$ \Mdot. In our models, for a \mdot\ of 10$^{-6}$ \Mdot, dust formation in the constant wind velocity scenario is negligible. Even for an accelerating wind, the resulting dust masses are only modest, of the order of 10$^{-5}$ \Ms, and produce little or no detectable IR excess in the SED. This suggests that significant dust formation in the CSM around RSGs is likely to occur during phases of enhanced mass loss, when the mass-loss rates aresubstantially higher than the typical quiescent values for RSGs. Such enhanced mass loss is characteristic of the very late stages of RSG evolution, typically during the last 10$^4$--10$^5$ years before explosion, as evidenced by the dense CSM observed around Type II CCSNe \citep{Sengupta_Sujit_Sarangi2026, Nayana2025ApJ...985...51N, baer-way_2025, chandra_2018, smith_2017}. Furthermore, as shown in our previous work \citep{Sengupta_Sujit_Sarangi2026}, episodic mass loss events such as ejections and/or eruptions, are also expected to be more prominent in relatively massive RSGs, with initial masses of about 16 \Ms\ or higher. Connecting this expectation with our results, we propose that stars with initial masses of $\gtrsim$16 \Ms\ may typically develop a dense, dusty CSM before exploding as CCSNe. The mass and time dependence of dust formation in RSG winds may therefore be an important factor in understanding the so-called `RSG progenitor problem' \citep{walmswell_2012}. The galactic surveys looking at stellar populations in nearby galaxies are crucial to understanding how common dusty RSGs are, and if there is a systematic reddening of the typical RSG spectrum \citep{massey_2021, sarbadhicary_2025, dai_2026}. 

\cite{davies_2022} suggest that enhanced mass loss due to a `superwind' \citep{YoonCantiello2010} can cause significant extinction (10-100 times) to the original RSG spectrum, and so it is unlikely to be a general characteristic of RSGs, and instead proposes mass loss by eruptions prior to explosion. To explain the column densities, we find that an overall enhanced mass loss is necessary in the last few thousand years \citep{Sengupta_Sujit_Sarangi2026}. However, we also agree to the time dependent nature of RSG mass loss, and the formation of dust in short phases instead of continuously. 

Importantly, if dust formation occurs only during the very late evolutionary phases and is associated with an accelerating wind, the dust will be concentrated closer to the stellar surface and will remain particularly susceptible to destruction following the explosion. For relatively high initial masses, typically $\gtrsim$19 \Ms, enhanced mass loss may begin much earlier in the RSG phase \citep{YoonCantiello2010}, allowing dust to travel farther from the star and potentially survive the explosion. The destruction of CSM dust by the SN flash and the subsequent forward shock will be investigated in detail in our future study. 

Our models of Type II SN progenitors \citep{Sengupta_Sujit_Sarangi2026}, applied to SN~2005ip, SN~2017hcc, SN~2023ixf, SN~2020ywx, found that the observed CSM column densities are best reproduced when the RSG wind is characterized by a \bet-law in which the wind is launched from the stellar surface with a near-zero velocity and gradually accelerates outward until it reaches a constant terminal speed \citep{Moriya2018}. Our choice of \bet\ = 1.2 is based on observed wind profiles of RSGs (refer Appenndix A in \citealt{Sengupta_Sujit_Sarangi2026}).
The physical mechanism responsible for launching and accelerating winds from the surfaces of cool stars such as RSGs remains poorly understood \citep{vanLoon_2025}. Radiation pressure on dust grains is often proposed as a mechanism for driving winds in cool stars, including RSGs and AGB stars \citep{hofner_2018, kee_2021}. Because dust grains have a large absorption cross-section in the red part of the spectrum compared with the gas, radiation pressure on dust can provide a significant driving force. However, as demonstrated in this study, dust formation in a gas parcel requires time and suitable density and temperature conditions. Before dust forms, radiation pressure on dust cannot efficiently drive the wind. If dust formation occurs farther from the star, the radiative flux is already reduced. Moreover, increasing dust densities lead to higher optical depths, which can further reduce the radiative flux and hence the radiation pressure through attenuation. We are currently investigating the physical mechanism responsible for wind acceleration to deduce a more appropriate characterization of RSG wind dynamics than what is done emperically in terms of a simple \bet-law, originally derived for radiation-driven winds of hot stars. 

In our models, we have explored all possible chemical reactions leading to dust precursors forming in the gas. We found that most of the C-atoms will go on to form CO molecules, with a trace amount of HCN. Large masses of CO are indeed expected, as found in the CSM around many supergiants such as VY CMa, IRC+10420, etc. \citep{matsuura_2014, ziurys_2025}. The elemental abundances correspond to the outer envelope of massive stars, where C/O $<$ 1, and the chemical processes lead to only O-rich dust (silicates and alumina) forming in the wind. There are no suitable pathways in such conditions to produce C-rich dust such as amorphous carbon, graphite, or silicon carbide. 
One of the primary ways to distinguish between silicate and carbon-type dust is through the presence of 9.7 and 18 \mic\ features of silicates. In our model SEDs, we found that the silicates are either in absorption or emission, based on the optical depth of the CSM, at any given time. There can be scenarios of mass loss (around 10$^{-4}$ \Mdot) which show the transition from emission to absorption (see Figure~\ref{fig:sed_constant_beta}), and the silicate feature may not be prominent over the continuum. 
Incidentally, observed spectra of supergiant stars in the mid-IR, such as [W60] B90 \citep{Munoz-Sanchez2024} or WOH G64 \citep{Ohnaka2024}, or progenitors of CCSNe such as SN~2025pht \citep{Kilpatrick2025} or SN~2023ixf \citep{niu_2023} often do not exhibit the silicate features, and graphite is taken as suitable dust type that may fit the SED. While our models suggest that, due to opacities, the silicate features could be damped, it is indeed difficult to explain the absence of any notable spectral character around those bands for any supergiants in general. 

When comparing the UV/optical/IR fluxes of SN~2023ixf \citep{vandyk_2024} with our time-dependent SED of the SN~2023ixf progenitor, we found that a clumpy CSM is necessary to reproduce the observations. However, our treatment of the clumpy model is incomplete, as we did not evaluate the full chemical budget using a clumpy wind structure. The enhanced density within clumps may accelerate dust formation and could also increase the total dust mass under certain conditions. Moreover, accounting for the density contrast between the clumps and the interclump medium would be necessary for a more realistic treatment of the CSM.

In our simulations, we predict the formation of numerous molecular clusters with chemical compositions of the form [SiO$_2$]$_n$, which do not directly condense into dust. However, these clusters could potentially contribute to the O-rich dust mass through subsequent accretion or coagulation with silicate grains. When comparing the molecular masses of VY CMa, IRC+10420, and NML Cyg with our models (see Figure~\ref{fig:molecular_mass_steady_accelerated}), we assumed \mdot\ to remain constant for several thousand years. The model \mdot\ values ranging from 10$^{-6}$ to 10$^{-2}$ \Mdot\ are associated with the observed molecular masses. This assumption introduces an inaccuracy because, at higher \mdot, maintaining such mass-loss rates over several thousand years would result in unrealistically large CSM masses of 1--10 \Ms. A more realistic comparison would therefore require a time-dependent mass-loss history, similar to that adopted for the SN~2023ixf progenitor.


\section{Software}

\software{
\textsc{MESA} \citep{Paxton2011,Paxton2013,Paxton2015,Paxton2018,Paxton2019,Jermyn2023},
\textsc{CLOUDY} \citep{Ferland2013},
\textsc{OPTOOL} \citep{Dominik_2021a}
}


\section{Acknowledgments}

The authors acknowledge the support of the Department of Science and
Technology (DST), Government of India. Part of this research made use of the
High Performance Computing (HPC)
resources\footnote{\url{https://www.iiap.res.in/centers/main-campus/computing-and-i-t/}}
maintained by Mr. Anish Parwage, Engineer D and Head of the Computer \& IT
Infrastructure Division at the Indian Institute of Astrophysics (IIA), Bengaluru,
whose continuous support and assistance with HPC-related issues are gratefully
acknowledged. We also thank Dr. Pritam Banerjee and Ms. Khushbu Kathpalia
for helpful discussions.

\bibliographystyle{aasjournalv7}
\bibliography{Bibliography_sarangi,bibliography}

\begin{thebibliography}{}
\expandafter\ifx\csname natexlab\endcsname\relax\def\natexlab#1{#1}\fi
\providecommand{\url}[1]{\href{#1}{#1}}
\providecommand{\dodoi}[1]{doi:~\href{http://doi.org/#1}{\nolinkurl{#1}}}
\providecommand{\doeprint}[1]{\href{http://ascl.net/#1}{\nolinkurl{http://ascl.net/#1}}}
\providecommand{\doarXiv}[1]{\href{https://arxiv.org/abs/#1}{\nolinkurl{https://arxiv.org/abs/#1}}}

\bibitem[{K. {Antoniadis} {et~al.}(2024){Antoniadis}, {Bonanos}, {de Wit},
  {Zapartas}, {Munoz-Sanchez}, \& {Maravelias}}]{antoniadis_2024}
{Antoniadis}, K., {Bonanos}, A.~Z., {de Wit}, S., {et~al.} 2024,
  \bibinfo{title}{{Establishing a mass-loss rate relation for red supergiants
  in the Large Magellanic Cloud},} \aap, 686, A88,
  \dodoi{10.1051/0004-6361/202449383}

\bibitem[{R. {Baer-Way} {et~al.}(2025){Baer-Way}, {Chandra}, {Modjaz}, {Kumar},
  {Pellegrino}, {Chevalier}, {Crawford}, {Sarangi}, {Smith}, {Maeda}, {Nayana},
  {Filippenko}, {Andrews}, {Arcavi}, {Bostroem}, {Brink}, {Dong}, {Dwarkadas},
  {Farah}, {Howell}, {Hiramatsu}, {Hosseinzadeh}, {McCully}, {Meza}, {Newsome},
  {Padilla Gonzalez}, {Pearson}, {Sand}, {Shrestha}, {Terreran}, {Valenti},
  {Wyatt}, {Yang}, \& {Zheng}}]{baer-way_2025}
{Baer-Way}, R., {Chandra}, P., {Modjaz}, M., {et~al.} 2025, \bibinfo{title}{{A
  Multiwavelength Autopsy of the Interacting Type IIn Supernova 2020ywx:
  Tracing Its Progenitor Mass-loss History for 100 Yr Before Death},} \apj,
  983, 101, \dodoi{10.3847/1538-4357/adc00a}

\bibitem[{E.~R. {Beasor} \& N. {Smith}(2022){Beasor} \& {Smith}}]{beasor_2022}
{Beasor}, E.~R., \& {Smith}, N. 2022, \bibinfo{title}{{The Extreme Scarcity of
  Dust-enshrouded Red Supergiants: Consequences for Producing Stripped Stars
  via Winds},} \apj, 933, 41, \dodoi{10.3847/1538-4357/ac6dcf}

\bibitem[{K.~A. {Bostroem} {et~al.}(2026){Bostroem}, {Valenti}, {Sand},
  {Pearson}, {Shrestha}, {Andrews}, {Dessart}, {Jacobson-Gal{\'a}n}, {Hsu},
  {Ravi}, {Andrews}, {Christy}, {Dong}, {Farah}, {Franz}, {Filippenko}, {Gill},
  {Hoang}, {Hosseinzadeh}, {Howell}, {Janzen}, {Jencson}, {Jha}, {Kwok},
  {Lundquist}, {Martas}, {McCully}, {Mehta}, {Newsome}, {Padilla-Gonzalez},
  {Ransome}, {Meza Retamal}, {Smith}, {Subrayan}, \& {Terreran}}]{Bostroem2026}
{Bostroem}, K.~A., {Valenti}, S., {Sand}, D.~J., {et~al.} 2026,
  \bibinfo{title}{{Late-time Hubble Space Telescope Ultraviolet Spectra of SN
  2023ixf and SN 2024ggi Show Ongoing Interaction with Circumstellar
  Material},} \apj, 1004, 23, \dodoi{10.3847/1538-4357/ae644e}

\bibitem[{E. {Cannon} {et~al.}(2021){Cannon}, {Montarg{\`e}s}, {de Koter},
  {Decin}, {Min}, {Lagadec}, {Kervella}, {Sundqvist}, \& {Sana}}]{cannon_2021}
{Cannon}, E., {Montarg{\`e}s}, M., {de Koter}, A., {et~al.} 2021,
  \bibinfo{title}{{The inner circumstellar dust of the red supergiant Antares
  as seen with VLT/SPHERE/ZIMPOL},} \mnras, 502, 369,
  \dodoi{10.1093/mnras/stab018}

\bibitem[{E. {Cannon} {et~al.}(2023){Cannon}, {Montarg{\`e}s}, {de Koter},
  {Matter}, {Sanchez-Bermudez}, {Norris}, {Paladini}, {Decin}, {Sana},
  {Sundqvist}, {Lagadec}, {Kervella}, {Chiavassa}, {Dupree}, {Perrin},
  {Scicluna}, {Stee}, {Kraus}, {Danchi}, {Lopez}, {Millour}, {Drevon},
  {Cruzal{\`e}bes}, {Berio}, {Robbe-Dubois}, \& {Rosales-Guzman}}]{cannon_2023}
{Cannon}, E., {Montarg{\`e}s}, M., {de Koter}, A., {et~al.} 2023,
  \bibinfo{title}{{The dusty circumstellar environment of Betelgeuse during the
  Great Dimming as seen by VLTI/MATISSE},} \aap, 675, A46,
  \dodoi{10.1051/0004-6361/202243611}

\bibitem[{P. {Chandra}(2018){Chandra}}]{chandra_2018}
{Chandra}, P. 2018, \bibinfo{title}{{Circumstellar Interaction in Supernovae in
  Dense Environments{\textemdash}An Observational Perspective},} \ssr, 214, 27,
  \dodoi{10.1007/s11214-017-0461-6}

\bibitem[{P. {Chandra} {et~al.}(2022){Chandra}, {Chevalier}, {James}, \&
  {Fox}}]{Chandra2022}
{Chandra}, P., {Chevalier}, R.~A., {James}, N. J.~H., \& {Fox}, O.~D. 2022,
  \bibinfo{title}{{The luminous type IIn supernova SN 2017hcc: Infrared bright,
  X-ray, and radio faint},} \mnras, 517, 4151, \dodoi{10.1093/mnras/stac2915}

\bibitem[{I. {Cherchneff}(2013){Cherchneff}}]{cherchneff2013b}
{Cherchneff}, I. 2013, in EAS Publications Series, Vol.~60, EAS Publications
  Series, ed. P.~{Kervella}, T.~{Le Bertre}, \& G.~{Perrin}, 175--184,
  \dodoi{10.1051/eas/1360020}

\bibitem[{G.~C. {Clayton} {et~al.}(2009){Clayton}, {Freeman}, {Bright},
  {Massey}, {Gordon}, {Levesque}, {Plez}, {Olsen}, \&
  {Nordhaus}}]{clayton_2009}
{Clayton}, G.~C., {Freeman}, W., {Bright}, S., {et~al.} 2009, in IAU Symposium,
  Vol. 256, The Magellanic System: Stars, Gas, and Galaxies, ed. J.~T. {Van
  Loon} \& J.~M. {Oliveira}, 411--414, \dodoi{10.1017/S1743921308028792}

\bibitem[{M. Clayton(2018)Clayton}]{clayton2018a}
Clayton, M. 2018, PhD thesis, University of Oxford.
\newblock
  \url{https://ora.ox.ac.uk/objects/uuid:3aa54f27-e13e-4b16-bafd-28b3e7059e8d}

\bibitem[{M. {Clayton} {et~al.}(2017){Clayton}, {Podsiadlowski}, {Ivanova}, \&
  {Justham}}]{Clayton2017}
{Clayton}, M., {Podsiadlowski}, P., {Ivanova}, N., \& {Justham}, S. 2017,
  \bibinfo{title}{{Episodic mass ejections from common-envelope objects},}
  \mnras, 470, 1788, \dodoi{10.1093/mnras/stx1290}

\bibitem[{M. {Dai} {et~al.}(2026){Dai}, {Wang}, \& {Jiang}}]{dai_2026}
{Dai}, M., {Wang}, S., \& {Jiang}, B. 2026, \bibinfo{title}{{JWST Reveals a
  Galaxy-Wide Association of Red Supergiants with OB Stars in NGC 5584},}
  \mnras, \dodoi{10.1093/mnras/stag1519}

\bibitem[{B. Davies \& E.~R. Beasor(2017)Davies \& Beasor}]{Davies2017}
Davies, B., \& Beasor, E.~R. 2017, \bibinfo{title}{The initial masses of the
  red supergiant progenitors to Type-II supernovae,} Monthly Notices of the
  Royal Astronomical Society, 474, 2116, \dodoi{10.1093/mnras/stx2734}

\bibitem[{B. {Davies} {et~al.}(2022){Davies}, {Plez}, \&
  {Petrault}}]{davies_2022}
{Davies}, B., {Plez}, B., \& {Petrault}, M. 2022, \bibinfo{title}{{Explosion
  imminent: the appearance of red supergiants at the point of core-collapse},}
  \mnras, 517, 1483, \dodoi{10.1093/mnras/stac2427}

\bibitem[{E. {De Beck} {et~al.}(2010){De Beck}, {Decin}, {de Koter},
  {Justtanont}, {Verhoelst}, {Kemper}, \& {Menten}}]{debeck_2010}
{De Beck}, E., {Decin}, L., {de Koter}, A., {et~al.} 2010,
  \bibinfo{title}{{Probing the mass-loss history of AGB and red supergiant
  stars from CO rotational line profiles. II. CO line survey of evolved stars:
  derivation of mass-loss rate formulae},} \aap, 523, A18,
  \dodoi{10.1051/0004-6361/200913771}

\bibitem[{ {Dessart, Luc}(2025){Dessart, Luc}}]{dessart2024}
{Dessart, Luc}. 2025, \bibinfo{title}{Probing red supergiant atmospheres and
  winds with early-time, high-cadence, high-resolution type II supernova
  spectra,} A\&A, 694, A132, \dodoi{10.1051/0004-6361/202452769}

\bibitem[{M. D\'{\i}az-Rodr\'{\i}guez {et~al.}(2021)D\'{\i}az-Rodr\'{\i}guez,
  Murphy, Williams, Dalcanton, \& Dolphin}]{DiazRodriguez2021}
D\'{\i}az-Rodr\'{\i}guez, M., Murphy, J.~W., Williams, B.~F., Dalcanton, J.~J.,
  \& Dolphin, A.~E. 2021, \bibinfo{title}{Progenitor mass distribution for 22
  historic core-collapse supernovae,} Monthly Notices of the Royal Astronomical
  Society, 506, 781, \dodoi{10.1093/mnras/stab1800}

\bibitem[{C. {Dominik} {et~al.}(2021){Dominik}, {Min}, \&
  {Tazaki}}]{Dominik_2021a}
{Dominik}, C., {Min}, M., \& {Tazaki}, R. 2021, \bibinfo{title}{{OpTool:
  Command-line driven tool for creating complex dust opacities},}, Astrophysics
  Source Code Library, record ascl:2104.010 \doeprint{2104.010}

\bibitem[{B.~T. {Draine} \& A. {Li}(2007){Draine} \& {Li}}]{dra07}
{Draine}, B.~T., \& {Li}, A. 2007, \bibinfo{title}{{Infrared Emission from
  Interstellar Dust. IV. The Silicate-Graphite-PAH Model in the Post-Spitzer
  Era},} \apj, 657, 810, \dodoi{10.1086/511055}

\bibitem[{E. {Dwek} \& R.~G. {Arendt}(2024){Dwek} \& {Arendt}}]{dwek_2024}
{Dwek}, E., \& {Arendt}, R.~G. 2024, \bibinfo{title}{{The Escape Probability of
  Photons Emitted in a Spherical Homogeneous Shell},} Research Notes of the
  American Astronomical Society, 8, 194, \dodoi{10.3847/2515-5172/ad68f9}

\bibitem[{A. {Fassia} {et~al.}(2001){Fassia}, {Meikle}, {Chugai}, {Geballe},
  {Lundqvist}, {Walton}, {Pollacco}, {Veilleux}, {Wright}, {Pettini}, {Kerr},
  {Puchnarewicz}, {Puxley}, {Irwin}, {Packham}, {Smartt}, \& {Harmer}}]{fas01}
{Fassia}, A., {Meikle}, W.~P.~S., {Chugai}, N., {et~al.} 2001,
  \bibinfo{title}{{Optical and infrared spectroscopy of the type IIn SN 1998S:
  days 3-127},} Monthly Notices of the Royal Astronomical Society, 325, 907,
  \dodoi{10.1046/j.1365-8711.2001.04282.x}

\bibitem[{G.~J. {Ferland} {et~al.}(2013){Ferland}, {Porter}, {van Hoof},
  {Williams}, {Abel}, {Lykins}, {Shaw}, {Henney}, \& {Stancil}}]{ferland_2013}
{Ferland}, G.~J., {Porter}, R.~L., {van Hoof}, P.~A.~M., {et~al.} 2013,
  \bibinfo{title}{{The 2013 Release of Cloudy},} \rmxaa, 49, 137.
\newblock \doarXiv{1302.4485}

\bibitem[{G.~J. {Ferland} {et~al.}(2017){Ferland}, {Chatzikos}, {Guzm{\'a}n},
  {Lykins}, {van Hoof}, {Williams}, {Abel}, {Badnell}, {Keenan}, {Porter}, \&
  {Stancil}}]{Ferland2013}
{Ferland}, G.~J., {Chatzikos}, M., {Guzm{\'a}n}, F., {et~al.} 2017,
  \bibinfo{title}{{The 2017 Release Cloudy},} \rmxaa, 53, 385,
  \dodoi{10.48550/arXiv.1705.10877}

\bibitem[{A.~V. {Filippenko}(1997){Filippenko}}]{fil97}
{Filippenko}, A.~V. 1997, \bibinfo{title}{{Optical Spectra of Supernovae},}
  Annual Review of Astronomy and Astrophysics, 35, 309,
  \dodoi{10.1146/annurev.astro.35.1.309}

\bibitem[{G. {Folatelli} {et~al.}(2025){Folatelli}, {Ferrari}, {Ertini},
  {Kuncarayakti}, \& {Maeda}}]{gaston_2025}
{Folatelli}, G., {Ferrari}, L., {Ertini}, K., {Kuncarayakti}, H., \& {Maeda},
  K. 2025, \bibinfo{title}{{SN 2023ixf: Interaction signatures in the spectrum
  at 445 days},} \aap, 698, A213, \dodoi{10.1051/0004-6361/202554128}

\bibitem[{O.~D. {Fox} {et~al.}(2010){Fox}, {Chevalier}, {Dwek}, {Skrutskie},
  {Sugerman}, \& {Leisenring}}]{fox_2010}
{Fox}, O.~D., {Chevalier}, R.~A., {Dwek}, E., {et~al.} 2010,
  \bibinfo{title}{{Disentangling the Origin and Heating Mechanism of Supernova
  Dust: Late-time Spitzer Spectroscopy of the Type IIn SN 2005ip},} \apj, 725,
  1768, \dodoi{10.1088/0004-637X/725/2/1768}

\bibitem[{O.~D. {Fox} {et~al.}(2011){Fox}, {Chevalier}, {Skrutskie},
  {Soderberg}, {Filippenko}, {Ganeshalingam}, {Silverman}, {Smith}, \&
  {Steele}}]{fox11}
{Fox}, O.~D., {Chevalier}, R.~A., {Skrutskie}, M.~F., {et~al.} 2011,
  \bibinfo{title}{{A Spitzer Survey for Dust in Type IIn Supernovae},} \apj,
  741, 7, \dodoi{10.1088/0004-637X/741/1/7}

\bibitem[{O.~D. Fox {et~al.}(2020)Fox, Fransson, Smith, Andrews,
  Azalee Bostroem, Brink, Bradley Cenko, Clayton, Filippenko, Fong,
  Gallagher, Kelly, Kilpatrick, Mauerhan, Miller, Montiel, Stritzinger, Szalai,
  \& Van Dyk}]{Fox2020}
Fox, O.~D., Fransson, C., Smith, N., {et~al.} 2020, \bibinfo{title}{The slow
  demise of the long-lived SN 2005ip,} \mnras, 498, 517,
  \dodoi{10.1093/mnras/staa2324}

\bibitem[{C. {Fransson} {et~al.}(2014){Fransson}, {Ergon}, {Challis},
  {Chevalier}, {France}, {Kirshner}, {Marion}, {Milisavljevic}, {Smith},
  {Bufano}, {Friedman}, {Kangas}, {Larsson}, {Mattila}, {Benetti}, {Chornock},
  {Czekala}, {Soderberg}, \& {Sollerman}}]{fra14}
{Fransson}, C., {Ergon}, M., {Challis}, P.~J., {et~al.} 2014,
  \bibinfo{title}{{High-density Circumstellar Interaction in the Luminous Type
  IIn SN 2010jl: The First 1100 Days},} \apj, 797, 118,
  \dodoi{10.1088/0004-637X/797/2/118}

\bibitem[{J. {Fuller} \& D. {Tsuna}(2024){Fuller} \& {Tsuna}}]{fuller_2024}
{Fuller}, J., \& {Tsuna}, D. 2024, \bibinfo{title}{{Boil-off of red
  supergiants: mass loss and type II-P supernovae},} The Open Journal of
  Astrophysics, 7, 47, \dodoi{10.33232/001c.120130}

\bibitem[{S. {Ghaziasgar} {et~al.}(2025){Ghaziasgar}, {Abdollahi}, {Javadi},
  {van Loon}, {McDonald}, {Oliveira}, {Masoudnezhad}, {Khosroshahi}, {Foing},
  \& {Fazel Hesar}}]{ghaziasgar_2025}
{Ghaziasgar}, S., {Abdollahi}, M., {Javadi}, A., {et~al.} 2025,
  \bibinfo{title}{{Dusty Stellar Source Classification by Implementing Machine
  Learning Methods Based on Spectroscopic Observations in the Magellanic
  Clouds},} \apj, 986, 168, \dodoi{10.3847/1538-4357/adceeb}

\bibitem[{D. {Gobrecht} {et~al.}(2016){Gobrecht}, {Cherchneff}, {Sarangi},
  {Plane}, \& {Bromley}}]{gob16}
{Gobrecht}, D., {Cherchneff}, I., {Sarangi}, A., {Plane}, J.~M.~C., \&
  {Bromley}, S.~T. 2016, \bibinfo{title}{{Dust formation in the oxygen-rich AGB
  star IK Tauri},} \aap, 585, A6, \dodoi{10.1051/0004-6361/201425363}

\bibitem[{M.~S. {Gordon} {et~al.}(2018){Gordon}, {Humphreys}, {Jones},
  {Shenoy}, {Gehrz}, {Helton}, {Marengo}, {Hinz}, \& {Hoffmann}}]{gordon_2018}
{Gordon}, M.~S., {Humphreys}, R.~M., {Jones}, T.~J., {et~al.} 2018,
  \bibinfo{title}{{Searching for Cool Dust. II. Infrared Imaging of The OH/IR
  Supergiants, NML Cyg, VX Sgr, S Per, and the Normal Red Supergiants RS Per
  and T Per},} \aj, 155, 212, \dodoi{10.3847/1538-3881/aab961}

\bibitem[{ {Gottlieb, C. A.} {et~al.}(2022){Gottlieb, C. A.}, {Decin, L.},
  {Richards, A. M. S.}, {De Ceuster, F.}, {Homan, W.}, {Wallström, S. H. J.},
  {Danilovich, T.}, {Millar, T. J.}, {Montargès, M.}, {Wong, K. T.},
  {McDonald, I.}, {Baudry, A.}, {Bolte, J.}, {Cannon, E.}, {De Beck, E.}, {de
  Koter, A.}, {El Mellah, I.}, {Etoka, S.}, {Gobrecht, D.}, {Gray, M.},
  {Herpin, F.}, {Jeste, M.}, {Kervella, P.}, {Khouri, T.}, {Lagadec, E.},
  {Maes, S.}, {Malfait, J.}, {Menten, K. M.}, {Müller, H. S. P.}, {Pimpanuwat,
  B.}, {Plane, J. M. C.}, {Sahai, R.}, {Van de Sande, M.}, {Waters, L. B. F.
  M.}, {Yates, J.}, \& {Zijlstra, A.}}]{Gottlieb2021}
{Gottlieb, C. A.}, {Decin, L.}, {Richards, A. M. S.}, {et~al.} 2022,
  \bibinfo{title}{ATOMIUM: ALMA tracing the origins of molecules in dust
  forming oxygen rich M-type stars - Motivation, sample, calibration, and
  initial results,} A\&A, 660, A94, \dodoi{10.1051/0004-6361/202140431}

\bibitem[{T.~P.~M. {Goumans} \& S.~T. {Bromley}(2012){Goumans} \&
  {Bromley}}]{gou12}
{Goumans}, T.~P.~M., \& {Bromley}, S.~T. 2012, \bibinfo{title}{{Efficient
  nucleation of stardust silicates via heteromolecular homogeneous
  condensation},} Monthly Notices of the Royal Astronomical Society, 420, 3344,
  \dodoi{10.1111/j.1365-2966.2011.20255.x}

\bibitem[{H. {Hassani} {et~al.}(2026){Hassani}, {Rosolowsky}, {Leroy},
  {Sandstrom}, {Boquien}, {Thilker}, {Whitmore}, {Anand}, {Barnes}, {Cao},
  {Chown}, {Congiu}, {Dale}, {Egorov}, {Gerasimov}, {Grasha}, {Indebetouw},
  {Lee}, {Liang}, {Maschmann}, {Meidt}, {Oakes}, {Pessa}, {Pety}, {Querejeta},
  {Ramambason}, {Rodr{\'\i}guez}, {Sarbadhicary}, {Sutter}, {{\'U}beda}, \&
  {Williams}}]{hassani_2026}
{Hassani}, H., {Rosolowsky}, E., {Leroy}, A.~K., {et~al.} 2026,
  \bibinfo{title}{{The Hidden Life of Stars: Embedded Beginnings to Asymptotic
  Giant Branch Endings in the PHANGS─JWST Sample. I. Catalog of Mid-infrared
  Sources},} \apjs, 284, 3, \dodoi{10.3847/1538-4365/ae4aa5}

\bibitem[{A. {Heger} {et~al.}(1997){Heger}, {Jeannin}, {Langer}, \&
  {Baraffe}}]{Heger1997}
{Heger}, A., {Jeannin}, L., {Langer}, N., \& {Baraffe}, I. 1997,
  \bibinfo{title}{{Pulsations in red supergiants with high L/M ratio.
  Implications for the stellar and circumstellar structure of supernova
  progenitors},} \aap, 327, 224, \dodoi{10.48550/arXiv.astro-ph/9705097}

\bibitem[{S. {Hillel} {et~al.}(2025){Hillel}, {Schreier}, \&
  {Soker}}]{hillel_2025}
{Hillel}, S., {Schreier}, R., \& {Soker}, N. 2025, \bibinfo{title}{{Forming a
  Clumpy Circumstellar Material in Energetic Pre-supernova Activity},} Research
  in Astronomy and Astrophysics, 25, 055014, \dodoi{10.1088/1674-4527/adcf87}

\bibitem[{S. {H{\"o}fner} \& H. {Olofsson}(2018){H{\"o}fner} \&
  {Olofsson}}]{hofner_2018}
{H{\"o}fner}, S., \& {Olofsson}, H. 2018, \bibinfo{title}{{Mass loss of stars
  on the asymptotic giant branch. Mechanisms, models and measurements},} \aapr,
  26, 1, \dodoi{10.1007/s00159-017-0106-5}

\bibitem[{A.~K. {Inoue} {et~al.}(2020){Inoue}, {Hashimoto}, {Chihara}, \&
  {Koike}}]{inoue_2020}
{Inoue}, A.~K., {Hashimoto}, T., {Chihara}, H., \& {Koike}, C. 2020,
  \bibinfo{title}{{Radiative equilibrium estimates of dust temperature and mass
  in high-redshift galaxies},} \mnras, 495, 1577,
  \dodoi{10.1093/mnras/staa1203}

\bibitem[{W.
  {Jacobson-Gal{\'a}n}(2025){Jacobson-Gal{\'a}n}}]{JacobsonGalan2025}
{Jacobson-Gal{\'a}n}, W. 2025, \bibinfo{title}{{SN 2023ixf: The Closest
  Supernova of the Decade},} Universe, 11, 231,
  \dodoi{10.3390/universe11070231}

\bibitem[{W.~V. {Jacobson-Gal{\'a}n} {et~al.}(2026){Jacobson-Gal{\'a}n},
  {Dessart}, \& {Vartanyan}}]{JacobsonGalan2026}
{Jacobson-Gal{\'a}n}, W.~V., {Dessart}, L., \& {Vartanyan}, D. 2026,
  \bibinfo{title}{{Mapping 3D Explosive Nucleosynthesis in the Type II
  Supernova 2024ggi with Infrared Emission Lines},} \apj, 1006, 66,
  \dodoi{10.3847/1538-4357/ae7d1b}

\bibitem[{A.~S. {Jermyn} {et~al.}(2023){Jermyn}, {Bauer}, {Schwab}, {Farmer},
  {Ball}, {Bellinger}, {Dotter}, {Joyce}, {Marchant}, {Mombarg}, {Wolf}, {Sunny
  Wong}, {Cinquegrana}, {Farrell}, {Smolec}, {Thoul}, {Cantiello}, {Herwig},
  {Toloza}, {Bildsten}, {Townsend}, \& {Timmes}}]{Jermyn2023}
{Jermyn}, A.~S., {Bauer}, E.~B., {Schwab}, J., {et~al.} 2023,
  \bibinfo{title}{{Modules for Experiments in Stellar Astrophysics (MESA):
  Time-dependent Convection, Energy Conservation, Automatic Differentiation,
  and Infrastructure},} \apjs, 265, 15, \dodoi{10.3847/1538-4365/acae8d}

\bibitem[{T. {Kami{\'n}ski}(2019){Kami{\'n}ski}}]{kaminski_2019}
{Kami{\'n}ski}, T. 2019, \bibinfo{title}{{Massive dust clumps in the envelope
  of the red supergiant VY Canis Majoris},} \aap, 627, A114,
  \dodoi{10.1051/0004-6361/201935408}

\bibitem[{T. {Kami{\'n}ski} {et~al.}(2013{\natexlab{a}}){Kami{\'n}ski},
  {Schmidt}, \& {Menten}}]{kaminski_2013a}
{Kami{\'n}ski}, T., {Schmidt}, M.~R., \& {Menten}, K.~M. 2013{\natexlab{a}},
  \bibinfo{title}{{Aluminium oxide in the optical spectrum of VY Canis
  Majoris},} \aap, 549, A6, \dodoi{10.1051/0004-6361/201220650}

\bibitem[{T. {Kami{\'n}ski} {et~al.}(2013{\natexlab{b}}){Kami{\'n}ski},
  {Gottlieb}, {Schmidt}, {Patel}, {Young}, {Menten}, {Br{\"u}nken},
  {M{\"u}ller}, {Winters}, \& {McCarthy}}]{kaminski_2013b}
{Kami{\'n}ski}, T., {Gottlieb}, C.~A., {Schmidt}, M.~R., {et~al.}
  2013{\natexlab{b}}, in EAS Publications Series, Vol.~60, EAS Publications
  Series, ed. P.~{Kervella}, T.~{Le Bertre}, \& G.~{Perrin} (EDP), 191--198,
  \dodoi{10.1051/eas/1360022}

\bibitem[{S. {Katsuda} {et~al.}(2014){Katsuda}, {Maeda}, {Nozawa}, {Pooley}, \&
  {Immler}}]{katsuda_2014}
{Katsuda}, S., {Maeda}, K., {Nozawa}, T., {Pooley}, D., \& {Immler}, S. 2014,
  \bibinfo{title}{{SN 2005ip: A Luminous Type IIn Supernova Emerging from a
  Dense Circumstellar Medium as Revealed by X-Ray Observations},} \apj, 780,
  184, \dodoi{10.1088/0004-637X/780/2/184}

\bibitem[{N.~D. {Kee} {et~al.}(2021){Kee}, {Sundqvist}, {Decin}, {de Koter}, \&
  {Sana}}]{kee_2021}
{Kee}, N.~D., {Sundqvist}, J.~O., {Decin}, L., {de Koter}, A., \& {Sana}, H.
  2021, \bibinfo{title}{{Analytic, dust-independent mass-loss rates for red
  supergiant winds initiated by turbulent pressure},} \aap, 646, A180,
  \dodoi{10.1051/0004-6361/202039224}

\bibitem[{C.~D. Kilpatrick {et~al.}(2023)Kilpatrick, Foley, Jacobson-Galán,
  Piro, Smartt, Drout, Gagliano, Gall, Hjorth, Jones, Mandel, Margutti,
  Ramirez-Ruiz, Ransome, Villar, Coulter, Gao, Matthews, Taggart, \&
  Zenati}]{Kilpatrick2023}
Kilpatrick, C.~D., Foley, R.~J., Jacobson-Galán, W.~V., {et~al.} 2023,
  \bibinfo{title}{SN 2023ixf in Messier 101: A Variable Red Supergiant as the
  Progenitor Candidate to a Type II Supernova,} The Astrophysical Journal
  Letters, 952, L23, \dodoi{10.3847/2041-8213/ace4ca}

\bibitem[{C.~D. Kilpatrick {et~al.}(2025)Kilpatrick, Suresh, Davis, Drout,
  Foley, Gagliano, Jacobson-Galán, Kaur, Taggart, \& Vazquez}]{Kilpatrick2025}
Kilpatrick, C.~D., Suresh, A., Davis, K.~W., {et~al.} 2025, \bibinfo{title}{The
  Type II SN 2025pht in NGC 1637: A Red Supergiant with Carbon-rich
  Circumstellar Dust as the First JWST Detection of a Supernova Progenitor
  Star,} The Astrophysical Journal Letters, 992, L10,
  \dodoi{10.3847/2041-8213/ae04de}

\bibitem[{C. {Koike} {et~al.}(1995){Koike}, {Kaito}, {Yamamoto}, {Shibai},
  {Kimura}, \& {Suto}}]{koike_1995}
{Koike}, C., {Kaito}, C., {Yamamoto}, T., {et~al.} 1995,
  \bibinfo{title}{{Extinction spectra of corundum in the wavelengths from UV to
  FIR.},} \icarus, 114, 203, \dodoi{10.1006/icar.1995.1055}

\bibitem[{A. {Kumar} \& A. {Sarangi}(2026){Kumar} \& {Sarangi}}]{aman_2026a}
{Kumar}, A., \& {Sarangi}, A. 2026, \bibinfo{title}{{Type Iax Supernovae as a
  Source of Iron-rich Silicate Dust},} \apj, 999, 234,
  \dodoi{10.3847/1538-4357/ae3d9e}

\bibitem[{R.~M. {Lau} {et~al.}(2022){Lau}, {Hankins}, {Han}, {Argyriou},
  {Corcoran}, {Eldridge}, {Endo}, {Fox}, {Garcia Marin}, {Gull}, {Jones},
  {Hamaguchi}, {Lamberts}, {Law}, {Madura}, {Marchenko}, {Matsuhara}, {Moffat},
  {Morris}, {Morris}, {Onaka}, {Ressler}, {Richardson}, {Russell},
  {Sanchez-Bermudez}, {Smith}, {Soulain}, {Stevens}, {Tuthill}, {Weigelt},
  {Williams}, \& {Yamaguchi}}]{lau_2022}
{Lau}, R.~M., {Hankins}, M.~J., {Han}, Y., {et~al.} 2022,
  \bibinfo{title}{{Nested dust shells around the Wolf-Rayet binary WR 140
  observed with JWST},} Nature Astronomy, \dodoi{10.1038/s41550-022-01812-x}

\bibitem[{E.~M. {Levesque} {et~al.}(2005){Levesque}, {Massey}, {Olsen}, {Plez},
  {Josselin}, {Maeder}, \& {Meynet}}]{Levesque2005ApJ}
{Levesque}, E.~M., {Massey}, P., {Olsen}, K.~A.~G., {et~al.} 2005,
  \bibinfo{title}{{The Effective Temperature Scale of Galactic Red Supergiants:
  Cool, but Not As Cool As We Thought},} \apj, 628, 973, \dodoi{10.1086/430901}

\bibitem[{E.~M. {Levesque} {et~al.}(2006){Levesque}, {Massey}, {Olsen}, {Plez},
  {Meynet}, \& {Maeder}}]{Levesque2006ApJ}
{Levesque}, E.~M., {Massey}, P., {Olsen}, K.~A.~G., {et~al.} 2006,
  \bibinfo{title}{{The Effective Temperatures and Physical Properties of
  Magellanic Cloud Red Supergiants: The Effects of Metallicity},} \apj, 645,
  1102, \dodoi{10.1086/504417}

\bibitem[{A. {Maeder} {et~al.}(2013){Maeder}, {Meynet}, {Lagarde}, \&
  {Charbonnel}}]{Maeder_2013}
{Maeder}, A., {Meynet}, G., {Lagarde}, N., \& {Charbonnel}, C. 2013,
  \bibinfo{title}{{The thermohaline, Richardson, Rayleigh-Taylor,
  Solberg-H{\o}iland, and GSF criteria in rotating stars},} \aap, 553, A1,
  \dodoi{10.1051/0004-6361/201220936}

\bibitem[{P. {Massey} {et~al.}(2021){Massey}, {Neugent}, {Levesque}, {Drout},
  \& {Courteau}}]{massey_2021}
{Massey}, P., {Neugent}, K.~F., {Levesque}, E.~M., {Drout}, M.~R., \&
  {Courteau}, S. 2021, \bibinfo{title}{{The Red Supergiant Content of M31 and
  M33},} \aj, 161, 79, \dodoi{10.3847/1538-3881/abd01f}

\bibitem[{M. {Matsuura} {et~al.}(2014){Matsuura}, {Yates}, {Barlow},
  {Swinyard}, {Royer}, {Cernicharo}, {Decin}, {Wesson}, {Polehampton},
  {Blommaert}, {Groenewegen}, {Van de Steene}, \& {van Hoof}}]{matsuura_2014}
{Matsuura}, M., {Yates}, J.~A., {Barlow}, M.~J., {et~al.} 2014,
  \bibinfo{title}{{Herschel SPIRE and PACS observations of the red supergiant
  VY CMa: analysis of the molecular line spectra},} \mnras, 437, 532,
  \dodoi{10.1093/mnras/stt1906}

\bibitem[{S. {Mattila} {et~al.}(2008){Mattila}, {Meikle}, {Lundqvist},
  {Pastorello}, {Kotak}, {Eldridge}, {Smartt}, {Adamson}, {Gerardy}, {Rizzi},
  {Stephens}, \& {van Dyk}}]{mattila_2008}
{Mattila}, S., {Meikle}, W.~P.~S., {Lundqvist}, P., {et~al.} 2008,
  \bibinfo{title}{{Massive stars exploding in a He-rich circumstellar medium -
  III. SN 2006jc: infrared echoes from new and old dust in the progenitor
  CSM},} \mnras, 389, 141, \dodoi{10.1111/j.1365-2966.2008.13516.x}

\bibitem[{N. Mauron \& E. Josselin(2011)Mauron \&
  Josselin}]{MauronJosselin2011}
Mauron, N., \& Josselin, E. 2011, \bibinfo{title}{The mass-loss rates of red
  supergiants and the de Jager prescription,} Astronomy \& Astrophysics, 526,
  A156, \dodoi{10.1051/0004-6361/201013993}

\bibitem[{C. {McCormick} {et~al.}(2023){McCormick}, {Majewski}, {Smith},
  {Hayes}, {Cunha}, {Masseron}, {Weiss}, {Shetrone}, {Almeida}, {Frinchaboy},
  {Garc{\'\i}a-Hern{\'a}ndez}, \& {Nitschelm}}]{McCormick2023MNRAS}
{McCormick}, C., {Majewski}, S.~R., {Smith}, V.~V., {et~al.} 2023,
  \bibinfo{title}{{An investigation of non-canonical mixing in red giant stars
  using APOGEE $^{12}$C/$^{13}$C ratios observed in open cluster stars},}
  \mnras, 524, 4418, \dodoi{10.1093/mnras/stad2156}

\bibitem[{S.~N. {Milam} {et~al.}(2009){Milam}, {Woolf}, \&
  {Ziurys}}]{milam_2009}
{Milam}, S.~N., {Woolf}, N.~J., \& {Ziurys}, L.~M. 2009,
  \bibinfo{title}{{Circumstellar $^{12}$C/$^{13}$C Isotope Ratios from
  Millimeter Observations of CN and CO: Mixing in Carbon- and Oxygen-Rich
  Stars},} \apj, 690, 837, \dodoi{10.1088/0004-637X/690/1/837}

\bibitem[{M. {Montarg{\`e}s} {et~al.}(2021){Montarg{\`e}s}, {Cannon},
  {Lagadec}, {de Koter}, {Kervella}, {Sanchez-Bermudez}, {Paladini},
  {Cantalloube}, {Decin}, {Scicluna}, {Kravchenko}, {Dupree}, {Ridgway},
  {Wittkowski}, {Anugu}, {Norris}, {Rau}, {Perrin}, {Chiavassa}, {Kraus},
  {Monnier}, {Millour}, {Le Bouquin}, {Haubois}, {Lopez}, {Stee}, \&
  {Danchi}}]{montar_2021}
{Montarg{\`e}s}, M., {Cannon}, E., {Lagadec}, E., {et~al.} 2021,
  \bibinfo{title}{{A dusty veil shading Betelgeuse during its Great Dimming},}
  \nat, 594, 365, \dodoi{10.1038/s41586-021-03546-8}

\bibitem[{T.~J. {Moriya} {et~al.}(2018){Moriya}, {F{\"o}rster}, {Yoon},
  {Gr{\"a}fener}, \& {Blinnikov}}]{Moriya2018}
{Moriya}, T.~J., {F{\"o}rster}, F., {Yoon}, S.-C., {Gr{\"a}fener}, G., \&
  {Blinnikov}, S.~I. 2018, \bibinfo{title}{{Type IIP supernova light curves
  affected by the acceleration of red supergiant winds},} \mnras, 476, 2840,
  \dodoi{10.1093/mnras/sty475}

\bibitem[{G. {Munoz-Sanchez} {et~al.}(2024){Munoz-Sanchez}, {Kalitsounaki}, {de
  Wit}, {Antoniadis}, {Bonanos}, {Zapartas}, {Boutsia}, {Christodoulou},
  {Maravelias}, {Soszynski}, \& {Udalski}}]{Munoz-Sanchez2024}
{Munoz-Sanchez}, G., {Kalitsounaki}, M., {de Wit}, S., {et~al.} 2024,
  \bibinfo{title}{{The dramatic transition of the extreme Red Supergiant WOH
  G64 to a Yellow Hypergiant},} arXiv e-prints, arXiv:2411.19329,
  \dodoi{10.48550/arXiv.2411.19329}

\bibitem[{T. {Nagao} {et~al.}(2026){Nagao}, {Kuncarayakti}, {Maeda}, {Mattila},
  {Kotak}, {Killestein}, {Humina}, {Steeghs}, \& {Jarvis}}]{nagao_2026}
{Nagao}, T., {Kuncarayakti}, H., {Maeda}, K., {et~al.} 2026,
  \bibinfo{title}{{Giant outbursts of clumpy material preceding Type II
  supernova 2024qiw},} \aap, 708, A294, \dodoi{10.1051/0004-6361/202558402}

\bibitem[{A.~J. {Nayana} {et~al.}(2025){Nayana}, {Margutti}, {Wiston},
  {Chornock}, {Campana}, {Laskar}, {Murase}, {Krips}, {Migliori}, {Tsuna},
  {Alexander}, {Chandra}, {Bietenholz}, {Berger}, {Chevalier}, {De Colle},
  {Dessart}, {Diesing}, {Grefenstette}, {Jacobson-Gal{\'a}n}, {Maeda},
  {Marcote}, {Matthews}, {Milisavljevic}, {Ray}, {Reguitti}, \&
  {Polzin}}]{Nayana2025ApJ...985...51N}
{Nayana}, A.~J., {Margutti}, R., {Wiston}, E., {et~al.} 2025,
  \bibinfo{title}{{Dinosaur in a Haystack: X-Ray View of the Entrails of SN
  2023ixf and the Radio Afterglow of Its Interaction with the Medium Spawned by
  the Progenitor Star (Paper I)},} \apj, 985, 51,
  \dodoi{10.3847/1538-4357/adc2fb}

\bibitem[{Z. {Niu} {et~al.}(2023){Niu}, {Sun}, {Maund}, {Zhang}, {Zhao}, \&
  {Liu}}]{niu_2023}
{Niu}, Z., {Sun}, N.-C., {Maund}, J.~R., {et~al.} 2023, \bibinfo{title}{{The
  Dusty Red Supergiant Progenitor and the Local Environment of the Type II SN
  2023ixf in M101},} \apjl, 955, L15, \dodoi{10.3847/2041-8213/acf4e3}

\bibitem[{T. {Nozawa} {et~al.}(2014){Nozawa}, {Yoon}, {Maeda}, {Kozasa},
  {Nomoto}, \& {Langer}}]{nozawa_2014}
{Nozawa}, T., {Yoon}, S.-C., {Maeda}, K., {et~al.} 2014, \bibinfo{title}{{Dust
  Production Factories in the Early Universe: Formation of Carbon Grains in
  Red-supergiant Winds of Very Massive Population III Stars},} \apjl, 787, L17,
  \dodoi{10.1088/2041-8205/787/2/L17}

\bibitem[{ {Ohnaka, K.} {et~al.}(2024){Ohnaka, K.}, {Hofmann, K.-H.}, {Weigelt,
  G.}, {van Loon, J. Th.}, {Schertl, D.}, \& {Goldman, S. R.}}]{Ohnaka2024}
{Ohnaka, K.}, {Hofmann, K.-H.}, {Weigelt, G.}, {et~al.} 2024,
  \bibinfo{title}{Imaging the innermost circumstellar environment of the red
  supergiant WOH G64 in the Large Magellanic Cloud⋆,} A\&A, 691, L15,
  \dodoi{10.1051/0004-6361/202451820}

\bibitem[{B. {Paxton} {et~al.}(2011){Paxton}, {Bildsten}, {Dotter}, {Herwig},
  {Lesaffre}, \& {Timmes}}]{Paxton2011}
{Paxton}, B., {Bildsten}, L., {Dotter}, A., {et~al.} 2011,
  \bibinfo{title}{{Modules for Experiments in Stellar Astrophysics (MESA)},}
  \apjs, 192, 3, \dodoi{10.1088/0067-0049/192/1/3}

\bibitem[{B. Paxton {et~al.}(2013)Paxton, Cantiello, Arras, Bildsten, Brown,
  Dotter, Mankovich, Montgomery, Stello, Timmes, \& Townsend}]{Paxton2013}
Paxton, B., Cantiello, M., Arras, P., {et~al.} 2013, \bibinfo{title}{MODULES
  FOR EXPERIMENTS IN STELLAR ASTROPHYSICS (MESA): PLANETS, OSCILLATIONS,
  ROTATION, AND MASSIVE STARS,} The Astrophysical Journal Supplement Series,
  208, 4, \dodoi{10.1088/0067-0049/208/1/4}

\bibitem[{B. {Paxton} {et~al.}(2015){Paxton}, {Marchant}, {Schwab}, {Bauer},
  {Bildsten}, {Cantiello}, {Dessart}, {Farmer}, {Hu}, {Langer}, {Townsend},
  {Townsley}, \& {Timmes}}]{Paxton2015}
{Paxton}, B., {Marchant}, P., {Schwab}, J., {et~al.} 2015,
  \bibinfo{title}{{Modules for Experiments in Stellar Astrophysics (MESA):
  Binaries, Pulsations, and Explosions},} \apjs, 220, 15,
  \dodoi{10.1088/0067-0049/220/1/15}

\bibitem[{B. {Paxton} {et~al.}(2018){Paxton}, {Schwab}, {Bauer}, {Bildsten},
  {Blinnikov}, {Duffell}, {Farmer}, {Goldberg}, {Marchant}, {Sorokina},
  {Thoul}, {Townsend}, \& {Timmes}}]{Paxton2018}
{Paxton}, B., {Schwab}, J., {Bauer}, E.~B., {et~al.} 2018,
  \bibinfo{title}{{Modules for Experiments in Stellar Astrophysics (MESA):
  Convective Boundaries, Element Diffusion, and Massive Star Explosions},}
  \apjs, 234, 34, \dodoi{10.3847/1538-4365/aaa5a8}

\bibitem[{B. {Paxton} {et~al.}(2019){Paxton}, {Smolec}, {Schwab}, {Gautschy},
  {Bildsten}, {Cantiello}, {Dotter}, {Farmer}, {Goldberg}, {Jermyn}, {Kanbur},
  {Marchant}, {Thoul}, {Townsend}, {Wolf}, {Zhang}, \& {Timmes}}]{Paxton2019}
{Paxton}, B., {Smolec}, R., {Schwab}, J., {et~al.} 2019,
  \bibinfo{title}{{Modules for Experiments in Stellar Astrophysics (MESA):
  Pulsating Variable Stars, Rotation, Convective Boundaries, and Energy
  Conservation},} \apjs, 243, 10, \dodoi{10.3847/1538-4365/ab2241}

\bibitem[{G. {Perrin} {et~al.}(2007){Perrin}, {Verhoelst}, {Ridgway}, {Cami},
  {Nguyen}, {Chesneau}, {Lopez}, {Leinert}, \& {Richichi}}]{perrin_2007}
{Perrin}, G., {Verhoelst}, T., {Ridgway}, S.~T., {et~al.} 2007,
  \bibinfo{title}{{The molecular and dusty composition of Betelgeuse's inner
  circumstellar environment},} \aap, 474, 599,
  \dodoi{10.1051/0004-6361:20077863}

\bibitem[{M. {Pozzo} {et~al.}(2004){Pozzo}, {Meikle}, {Fassia}, {Geballe},
  {Lundqvist}, {Chugai}, \& {Sollerman}}]{poz04}
{Pozzo}, M., {Meikle}, W.~P.~S., {Fassia}, A., {et~al.} 2004,
  \bibinfo{title}{{On the source of the late-time infrared luminosity of SN
  1998S and other Type II supernovae},} \mnras, 352, 457,
  \dodoi{10.1111/j.1365-2966.2004.07951.x}

\bibitem[{N. Prantzos {et~al.}(2018)Prantzos, Abia, Limongi, Chieffi, \&
  Cristallo}]{Praztzos2018}
Prantzos, N., Abia, C., Limongi, M., Chieffi, A., \& Cristallo, S. 2018,
  \bibinfo{title}{Chemical evolution with rotating massive star yields – I.
  The solar neighbourhood and the s-process elements,} Monthly Notices of the
  Royal Astronomical Society, 476, 3432, \dodoi{10.1093/mnras/sty316}

\bibitem[{Y.-J. Qin {et~al.}(2024)Qin, Zhang, Bloom, Sollerman, Zimmerman,
  Irani, Schulze, Gal-Yam, Kasliwal, Coughlin, Perley, Fremling, \&
  Kulkarni}]{Qin2024}
Qin, Y.-J., Zhang, K., Bloom, J., {et~al.} 2024, \bibinfo{title}{The progenitor
  star of SN 2023ixf: a massive red supergiant with enhanced, episodic
  pre-supernova mass loss,} Monthly Notices of the Royal Astronomical Society,
  534, 271, \dodoi{10.1093/mnras/stae2012}

\bibitem[{F. {Ragosta} {et~al.}(2026){Ragosta}, {Simongini}, {Ambrosino},
  {Imbrogno}, {Illiano}, {Piranomonte}, {Melandri}, {Di Palma}, {Papitto},
  {Ghedina}, {Cecconi}, {Leone}, {Gonz{\'a}lez}, {P{\'e}rez Ventura},
  {Hernandez Diaz}, \& {San Juan}}]{ragosta_2026}
{Ragosta}, F., {Simongini}, A., {Ambrosino}, F., {et~al.} 2026,
  \bibinfo{title}{{Unveiling the progenitor of SN 2023ixf: Circumstellar dust
  and its implications for the red supergiant problem},} \aap, 706, A320,
  \dodoi{10.1051/0004-6361/202557952}

\bibitem[{T. Rauscher {et~al.}(2002)Rauscher, Heger, Hoffman, \&
  Woosley}]{Rauscher_2002}
Rauscher, T., Heger, A., Hoffman, R.~D., \& Woosley, S.~E. 2002,
  \bibinfo{title}{Nucleosynthesis in Massive Stars with Improved Nuclear and
  Stellar Physics,} The Astrophysical Journal, 576, 323, \dodoi{10.1086/341728}

\bibitem[{A. {Sarangi}(2022){Sarangi}}]{sarangi_2022b}
{Sarangi}, A. 2022, \bibinfo{title}{{Formation, distribution, and IR emission
  of dust in the clumpy ejecta of Type II-P core-collapse supernovae, in
  isotropic and anisotropic scenarios},} \aap, 668, A57,
  \dodoi{10.1051/0004-6361/202244391}

\bibitem[{A. {Sarangi} \& I. {Cherchneff}(2013){Sarangi} \&
  {Cherchneff}}]{sar13}
{Sarangi}, A., \& {Cherchneff}, I. 2013, \bibinfo{title}{{The Chemically
  Controlled Synthesis of Dust in Type II-P Supernovae},} The Astrophysical
  Journal, 776, 107, \dodoi{10.1088/0004-637X/776/2/107}

\bibitem[{A. {Sarangi} \& I. {Cherchneff}(2015){Sarangi} \&
  {Cherchneff}}]{sar15}
{Sarangi}, A., \& {Cherchneff}, I. 2015, \bibinfo{title}{{Condensation of dust
  in the ejecta of Type II-P supernovae},} \aap, 575, A95,
  \dodoi{10.1051/0004-6361/201424969}

\bibitem[{A. {Sarangi} {et~al.}(2019){Sarangi}, {Dwek}, \&
  {Kazanas}}]{sarangi_2019}
{Sarangi}, A., {Dwek}, E., \& {Kazanas}, D. 2019, \bibinfo{title}{{Dust
  Formation in AGN Winds},} \apj, 885, 126, \dodoi{10.3847/1538-4357/ab46a9}

\bibitem[{A. {Sarangi} {et~al.}(2018){Sarangi}, {Matsuura}, \&
  {Micelotta}}]{sarangi2018book}
{Sarangi}, A., {Matsuura}, M., \& {Micelotta}, E.~R. 2018,
  \bibinfo{title}{{Dust in Supernovae and Supernova Remnants I: Formation
  Scenarios},} \ssr, 214, 63, \dodoi{10.1007/s11214-018-0492-7}

\bibitem[{A. {Sarangi} \& J.~D. {Slavin}(2022){Sarangi} \&
  {Slavin}}]{sarangi_2022a}
{Sarangi}, A., \& {Slavin}, J.~D. 2022, \bibinfo{title}{{Dust Production in a
  Thin Dense Shell in Supernovae with Early Circumstellar Interactions},} \apj,
  933, 89, \dodoi{10.3847/1538-4357/ac713d}

\bibitem[{A. {Sarangi} {et~al.}(2025){Sarangi}, {Zs{\'\i}ros}, {Szalai},
  {Martinez}, {Shahbandeh}, {Fox}, {Van Dyk}, {Filippenko}, {Bersten}, {De
  Looze}, {Ashall}, {Temim}, {Jencson}, {Rest}, {Milisavljevic}, {Dessart},
  {Dwek}, {Smith}, {Tinyanont}, {Brink}, {Zheng}, {Clayton}, \&
  {Andrews}}]{sarangi_2025a}
{Sarangi}, A., {Zs{\'\i}ros}, S., {Szalai}, T., {et~al.} 2025,
  \bibinfo{title}{{Two Decades of Dust Evolution in SN 2005af through JWST,
  Spitzer, and Chemical Modeling},} \apj, 993, 94,
  \dodoi{10.3847/1538-4357/ae0645}

\bibitem[{S.~K. {Sarbadhicary} {et~al.}(2025){Sarbadhicary}, {Thilker},
  {Leroy}, {Lee}, {Amiri}, {Anand}, {Barnes}, {Boquien}, {Dale}, {Dlamini},
  {Glover}, {Klessen}, {Larson}, {Maschmann}, {Pan}, {Sun}, {{\'U}beda},
  {Williams}, {Wofford}, \& {PHANGS Collaboration}}]{sarbadhicary_2025}
{Sarbadhicary}, S.~K., {Thilker}, D., {Leroy}, A.~K., {et~al.} 2025,
  \bibinfo{title}{{Evolved Supergiants in PHANGS I: Red Supergiants in 19
  Galaxies between 5-20 Mpc with HST and JWST},} arXiv e-prints,
  arXiv:2601.00055, \dodoi{10.48550/arXiv.2601.00055}

\bibitem[{S. {Sengupta} {et~al.}(2026){Sengupta}, {Sujit}, \&
  {Sarangi}}]{Sengupta_Sujit_Sarangi2026}
{Sengupta}, S., {Sujit}, D., \& {Sarangi}, A. 2026, \bibinfo{title}{{Dance to
  Demise{\textemdash}How Massive Stars May Form Dense Circumstellar Shells
  before Explosion},} \apj, 996, 18, \dodoi{10.3847/1538-4357/ae129c}

\bibitem[{M. {Shahbandeh} {et~al.}(2023){Shahbandeh}, {Sarangi}, {Temim},
  {Szalai}, {Fox}, {Tinyanont}, {Dwek}, {Dessart}, {Filippenko}, {Brink},
  {Foley}, {Jencson}, {Pierel}, {Zs{\'\i}ros}, {Rest}, {Zheng}, {Andrews},
  {Clayton}, {De}, {Engesser}, {Gezari}, {Gomez}, {Gonzaga}, {Johansson},
  {Kasliwal}, {Lau}, {De Looze}, {Marston}, {Milisavljevic}, {O'Steen},
  {Siebert}, {Skrutskie}, {Smith}, {Strolger}, {Van Dyk}, {Wang}, {Williams},
  {Williams}, {Xiao}, \& {Yang}}]{shahbandeh_2023}
{Shahbandeh}, M., {Sarangi}, A., {Temim}, T., {et~al.} 2023,
  \bibinfo{title}{{JWST observations of dust reservoirs in type IIP supernovae
  2004et and 2017eaw},} \mnras, 523, 6048, \dodoi{10.1093/mnras/stad1681}

\bibitem[{H. Shinnaga {et~al.}(2025)Shinnaga, Oyadomari, Imai, Oyama, Claussen,
  Shimojo, Yamamoto, Richards, Etoka, Gray, \& Suzuki}]{Shinnaga2025}
Shinnaga, H., Oyadomari, M., Imai, H., {et~al.} 2025, \bibinfo{title}{First
  VLBI imaging of SiO v = 0, J = 1→0 masers in VY Canis Majoris,}
  Publications of the Astronomical Society of Japan, 77, 459,
  \dodoi{10.1093/pasj/psaf013}

\bibitem[{A.~P. {Singh} {et~al.}(2022){Singh}, {Edwards}, \&
  {Ziurys}}]{singh_2022}
{Singh}, A.~P., {Edwards}, J.~L., \& {Ziurys}, L.~M. 2022, \bibinfo{title}{{The
  Arizona Radio Observatory 1 mm Spectral Survey of the Hypergiant Star NML
  Cygni (215-285 GHz)},} \aj, 164, 230, \dodoi{10.3847/1538-3881/ac8df0}

\bibitem[{S.~J. Smartt(2009)Smartt}]{Smartt2009}
Smartt, S.~J. 2009, \bibinfo{title}{The death of massive stars -- I.
  Observational constraints on the progenitors of Type II-P supernovae,}
  Monthly Notices of the Royal Astronomical Society, 395, 1409,
  \dodoi{10.1111/j.1365-2966.2009.14506.x}

\bibitem[{S.~J. Smartt(2015)Smartt}]{Smartt2015}
Smartt, S.~J. 2015, \bibinfo{title}{Observational constraints on the
  progenitors of core-collapse supernovae: the case for missing high mass
  stars,} Publications of the Astronomical Society of Australia, 32, e016,
  \dodoi{10.1017/pasa.2015.7}

\bibitem[{N. {Smith}(2017){Smith}}]{smith_2017}
{Smith}, N. 2017, {Interacting Supernovae: Types IIn and Ibn} (Springer), 403,
  \dodoi{10.1007/978-3-319-21846-5_38}

\bibitem[{N. Smith {et~al.}(2014)Smith, Mauerhan, \& Prieto}]{Smith2014}
Smith, N., Mauerhan, J.~C., \& Prieto, J.~L. 2014, \bibinfo{title}{SN 2009ip
  and SN 2010mc: core-collapse Type IIn supernovae arising from blue
  supergiants,} Monthly Notices of the Royal Astronomical Society, 438, 1191,
  \dodoi{10.1093/mnras/stt2269}

\bibitem[{N. Smith {et~al.}(2009)Smith, Silverman, Chornock, Filippenko, Wang,
  Li, Ganeshalingam, Foley, Rex, \& Steele}]{Smith_2009}
Smith, N., Silverman, J.~M., Chornock, R., {et~al.} 2009,
  \bibinfo{title}{{CORONAL} {LINES} {AND} {DUST} {FORMATION} {IN} {SN} 2005ip:
  {NOT} {THE} {BRIGHTEST}, {BUT} {THE} {HOTTEST} {TYPE} {IIn} {SUPERNOVA},} The
  Astrophysical Journal, 695, 1334, \dodoi{10.1088/0004-637x/695/2/1334}

\bibitem[{M.~D. Soraisam {et~al.}(2023)Soraisam, Szalai, Van~Dyk, Andrews,
  Srinivasan, Chun, Matheson, Scicluna, \& Vasquez-Torres}]{Soraisam2023}
Soraisam, M.~D., Szalai, T., Van~Dyk, S.~D., {et~al.} 2023, \bibinfo{title}{The
  SN 2023ixf Progenitor in M101. I. Infrared Variability,} The Astrophysical
  Journal, 957, 64, \dodoi{10.3847/1538-4357/acef22}

\bibitem[{T. {Sukhbold} {et~al.}(2016){Sukhbold}, {Ertl}, {Woosley}, {Brown},
  \& {Janka}}]{sukhbold_2016}
{Sukhbold}, T., {Ertl}, T., {Woosley}, S.~E., {Brown}, J.~M., \& {Janka}, H.~T.
  2016, \bibinfo{title}{{Core-collapse Supernovae from 9 to 120 Solar Masses
  Based on Neutrino-powered Explosions},} \apj, 821, 38,
  \dodoi{10.3847/0004-637X/821/1/38}

\bibitem[{T. {Szalai} \& S.~V. {Dyk}(2023){Szalai} \& {Dyk}}]{Szalai2023}
{Szalai}, T., \& {Dyk}, S.~V. 2023, \bibinfo{title}{{Spitzer constraints on
  pre-explosion variability of the SN 2023ixf progenitor},} The Astronomer's
  Telegram, 16042, 1

\bibitem[{F.~X. {Timmes}(1999){Timmes}}]{1999ApJS..124..241T}
{Timmes}, F.~X. 1999, \bibinfo{title}{{Integration of Nuclear Reaction Networks
  for Stellar Hydrodynamics},} \apjs, 124, 241, \dodoi{10.1086/313257}

\bibitem[{S.~D. {Van Dyk}(2025){Van Dyk}}]{VanDyk2025}
{Van Dyk}, S.~D. 2025, \bibinfo{title}{{Red Supergiants as Supernova
  Progenitors},} Galaxies, 13, 33, \dodoi{10.3390/galaxies13020033}

\bibitem[{S.~D. {Van Dyk} {et~al.}(2024){Van Dyk}, {Srinivasan}, {Andrews},
  {Soraisam}, {Szalai}, {Howell}, {Isaacson}, {Matheson}, {Petigura},
  {Scicluna}, {Stephens}, {Van Zandt}, {Zheng}, {Chun}, \&
  {Fillippenko}}]{vandyk_2024}
{Van Dyk}, S.~D., {Srinivasan}, S., {Andrews}, J.~E., {et~al.} 2024,
  \bibinfo{title}{{The SN 2023ixf Progenitor in M101. II. Properties},} \apj,
  968, 27, \dodoi{10.3847/1538-4357/ad414b}

\bibitem[{S.~D. {Van Dyk} {et~al.}(2026){Van Dyk}, {Szalai}, {Anand}, {Brink},
  {Zimmer}, {Milisavljevic}, {Fox}, {Jencson}, {Zheng}, \&
  {Filippenko}}]{VanDyk2026}
{Van Dyk}, S.~D., {Szalai}, T., {Anand}, G.~S., {et~al.} 2026,
  \bibinfo{title}{{The Progenitor of the Type II-Plateau SN 2025pht in NGC
  1637: The Dustiest, Most Luminous Red Supergiant So Far?},} arXiv e-prints,
  arXiv:2601.09087, \dodoi{10.48550/arXiv.2601.09087}

\bibitem[{J.~T. {van Loon}(2025){van Loon}}]{vanLoon_2025}
{van Loon}, J.~T. 2025, \bibinfo{title}{{Red Supergiant Mass Loss and Mass-Loss
  Rates},} Galaxies, 13, 72, \dodoi{10.3390/galaxies13040072}

\bibitem[{J.~T. {van Loon} {et~al.}(2005){van Loon}, {Cioni}, {Zijlstra}, \&
  {Loup}}]{vanloon2005}
{van Loon}, J.~T., {Cioni}, M.-R.~L., {Zijlstra}, A.~A., \& {Loup}, C. 2005,
  \bibinfo{title}{{An empirical formula for the mass-loss rates of
  dust-enshrouded red supergiants and oxygen-rich Asymptotic Giant Branch
  stars},} \aap, 438, 273, \dodoi{10.1051/0004-6361:20042555}

\bibitem[{J.~T. van Loon \& K. Ohnaka(2026)van Loon \&
  Ohnaka}]{VanLoonOhnaka2025}
van Loon, J.~T., \& Ohnaka, K. 2026, \bibinfo{title}{A phoenix rises from the
  ashes: WOH G64 is still a red supergiant, for now,} Monthly Notices of the
  Royal Astronomical Society, 546, stag012, \dodoi{10.1093/mnras/stag012}

\bibitem[{F. {V{\'a}rosi} \& E. {Dwek}(1999){V{\'a}rosi} \&
  {Dwek}}]{VarosiDwek1999}
{V{\'a}rosi}, F., \& {Dwek}, E. 1999, \bibinfo{title}{{Analytical
  Approximations for Calculating the Escape and Absorption of Radiation in
  Clumpy Dusty Environments},} \apj, 523, 265, \dodoi{10.1086/307729}

\bibitem[{T. {Verhoelst} {et~al.}(2009){Verhoelst}, {van der Zypen}, {Hony},
  {Decin}, {Cami}, \& {Eriksson}}]{verhoelst_2009}
{Verhoelst}, T., {van der Zypen}, N., {Hony}, S., {et~al.} 2009,
  \bibinfo{title}{{The dust condensation sequence in red supergiant stars},}
  \aap, 498, 127, \dodoi{10.1051/0004-6361/20079063}

\bibitem[{J.~J. {Walmswell} \& J.~J. {Eldridge}(2012){Walmswell} \&
  {Eldridge}}]{walmswell_2012}
{Walmswell}, J.~J., \& {Eldridge}, J.~J. 2012, \bibinfo{title}{{Circumstellar
  dust as a solution to the red supergiant supernova progenitor problem},}
  \mnras, 419, 2054, \dodoi{10.1111/j.1365-2966.2011.19860.x}

\bibitem[{K.~E. {Weil} {et~al.}(2020){Weil}, {Fesen}, {Patnaude}, {Raymond},
  {Chevalier}, {Milisavljevic}, \& {Gerardy}}]{weil_2020b}
{Weil}, K.~E., {Fesen}, R.~A., {Patnaude}, D.~J., {et~al.} 2020,
  \bibinfo{title}{{Detection of the Red Supergiant Wind from the Progenitor of
  Cassiopeia A},} \apj, 891, 116, \dodoi{10.3847/1538-4357/ab76bf}

\bibitem[{J. {Wen} {et~al.}(2024){Wen}, {Gao}, {Yang}, {Chen}, {Ren}, {Wang},
  \& {Jiang}}]{wen_2024}
{Wen}, J., {Gao}, J., {Yang}, M., {et~al.} 2024, \bibinfo{title}{{Evolved
  Massive Stars at Low Metallicity. VI. Mass-loss Rate of Red Supergiant Stars
  in the Large Magellanic Cloud},} \aj, 167, 51,
  \dodoi{10.3847/1538-3881/ad12bf}

\bibitem[{D. Xiang {et~al.}(2024)Xiang, Mo, Wang, Wang, Zhang, Lin, Chen, Song,
  Liu, Wang, \& Li}]{Xiang2024}
Xiang, D., Mo, J., Wang, X., {et~al.} 2024, \bibinfo{title}{The Red Supergiant
  Progenitor of Type II Supernova 2024ggi,} The Astrophysical Journal Letters,
  969, L15, \dodoi{10.3847/2041-8213/ad54b3}

\bibitem[{ {Yang, Ming} {et~al.}(2023){Yang, Ming}, {Bonanos, Alceste Z.},
  {Jiang, Biwei}, {Zapartas, Emmanouil}, {Gao, Jian}, {Ren, Yi}, {Lam, Man I.},
  {Wang, Tianding}, {Maravelias, Grigoris}, {Gavras, Panagiotis}, {Wang, Shu},
  {Chen, Xiaodian}, {Tramper, Frank}, {de Wit, Stephan}, {Chen, Bingqiu}, {Wen,
  Jing}, {Liu, Jiaming}, {Tian, Hao}, {Antoniadis, Konstantinos}, \& {Luo,
  Changqing}}]{Yang2023}
{Yang, Ming}, {Bonanos, Alceste Z.}, {Jiang, Biwei}, {et~al.} 2023,
  \bibinfo{title}{Evolved massive stars at low-metallicity - V. Mass-loss rate
  of red supergiant stars in the Small Magellanic Cloud⋆,} A\&A, 676, A84,
  \dodoi{10.1051/0004-6361/202244770}

\bibitem[{S.-C. {Yoon} \& M. {Cantiello}(2010){Yoon} \&
  {Cantiello}}]{YoonCantiello2010}
{Yoon}, S.-C., \& {Cantiello}, M. 2010, \bibinfo{title}{{Evolution of Massive
  Stars with Pulsation-driven Superwinds During the Red Supergiant Phase},}
  \apjl, 717, L62, \dodoi{10.1088/2041-8205/717/1/L62}

\bibitem[{L.~M. {Ziurys} \& A.~M.~S. {Richards}(2025){Ziurys} \&
  {Richards}}]{ziurys_2025}
{Ziurys}, L.~M., \& {Richards}, A. M.~S. 2025, \bibinfo{title}{{Molecules and
  Chemistry in Red Supergiants},} Galaxies, 13, 82,
  \dodoi{10.3390/galaxies13040082}

\bibitem[{L.~M. {Ziurys} {et~al.}(2009){Ziurys}, {Tenenbaum}, {Pulliam},
  {Woolf}, \& {Milam}}]{ziurys_2009}
{Ziurys}, L.~M., {Tenenbaum}, E.~D., {Pulliam}, R.~L., {Woolf}, N.~J., \&
  {Milam}, S.~N. 2009, \bibinfo{title}{{Carbon Chemistry in the Envelope of VY
  Canis Majoris: Implications for Oxygen-Rich Evolved Stars},} \apj, 695, 1604,
  \dodoi{10.1088/0004-637X/695/2/1604}

\bibitem[{V. {Zubko} {et~al.}(2004){Zubko}, {Li}, {Lim}, {Feuchtgruber}, \&
  {Harwit}}]{zubko_H2O_2004}
{Zubko}, V., {Li}, D., {Lim}, T., {Feuchtgruber}, H., \& {Harwit}, M. 2004,
  \bibinfo{title}{{Observations of Water Vapor Outflow from NML Cygnus},} \apj,
  610, 427, \dodoi{10.1086/421700}

\end{thebibliography}

\end{document}